\documentclass[aps, prd,superscriptaddress,A4paper,preprintnumbers,nofootinbib]{revtex4}
\usepackage{amssymb,amsmath,graphicx,epsfig,bm,color}
\usepackage{bm}
\usepackage{float}
\usepackage{slashed}
\usepackage{verbatim} 
\newcommand{\be}{\begin{eqnarray}}
\newcommand{\ee}{\end{eqnarray}}
\usepackage[colorlinks]{hyperref}
\usepackage{leftidx}
\usepackage{booktabs}
\usepackage{lmodern}
\usepackage{latexsym}
\usepackage{revsymb}
\usepackage{cancel}
\usepackage{multirow}
\usepackage{caption}
\usepackage{xcolor}
\usepackage{subcaption}
\usepackage[normalem]{ulem}
\newcommand{\stkout}[1]{\ifmmode\text{\sout{\ensuremath{#1}}}\else\sout{#1}\fi}

\hypersetup{
    colorlinks=true,
    citecolor=blue,
    filecolor=magenta,      
    urlcolor=cyan,
}

\renewcommand{\d}{\mathrm{d}}

\newcommand{\xB }{x_{\scriptscriptstyle B}}
\newcommand{\sT}{{\scriptscriptstyle T}}

\definecolor{amol_color}{RGB}{200, 0, 200}
\definecolor{khatiza_color}{RGB}{191, 0, 0}
\definecolor{asmita_color}{rgb}{0.0, 0.1, 0.9}
\definecolor{rajesh_color}{rgb}{0.93, 0.53, 0.18}

\begin{document}

\title{Unraveling Gluon TMDs in $J/\psi$ and Pion production at the EIC}

 \author{Khatiza Banu}
 \email{banu2765@iitb.ac.in}
 \affiliation{Department of Physics, Indian Institute of Technology Bombay, Mumbai-400076, India}
 \affiliation{Centre for Frontiers in Nuclear Science, Stony Brook University, Stony Brook, NY 11794-3800, USA}
 
 \author{Asmita Mukherjee}
 \email{asmita@phy.iitb.ac.in}
 \affiliation{Department of Physics, Indian Institute of Technology Bombay, Mumbai-400076, India}

\author{Amol Pawar}
 \email{194120018@iitb.ac.in}
 \affiliation{Department of Physics, Indian Institute of Technology Bombay, Mumbai-400076, India}
 
 \author{Sangem Rajesh}
\email{sangem.rajesh@vit.ac.in}
 \affiliation{Department of Physics, School of Advanced Sciences, Vellore Institute of Technology, Vellore,
Tamil Nadu 632014, India}

\date{\today}
\begin{abstract}
    We investigate the azimuthal asymmetries such as $\cos2{\phi_T}$ and Sivers symmetry for  $J/\psi$ and $\pi^\pm$ production in electron-proton scattering, {focusing on scenarios where} the $J/\psi$ and the pion are produced in {an} almost back-to-back configuration. The electron is unpolarized, while the proton can be unpolarized or transversely polarized.  For the $J/\psi$ formation, we use non-relativistic QCD (NRQCD), while $\pi^\pm$ is formed due to parton fragmentation. In this kinematics, we utilize the transverse momentum-dependent factorization framework to calculate the cross sections and asymmetries. We consider both quark and gluon-initiated processes and show that the gluon contribution dominates. In this work, we used the generalized parton model (GPM) for the TMD parametrizations and did not consider the effect of TMD evolution. We provide numerical estimates of the upper bounds on the azimuthal asymmetries, as well as employ a Gaussian parametrization for the gluon transverse momentum distributions (TMDs), within the kinematical region accessible by the upcoming Electron-Ion Collider (EIC).   
\end{abstract}
\maketitle
 
\section{Introduction}\label{sec1}
Unraveling the three-dimensional structure of a hadron is a fundamental quest in hadron physics, which has attracted a lot of interest both theoretically and experimentally. The 3D tomography of a hadron is encoded in the transverse momentum-dependent parton distribution functions (TMD PDFs), in short TMDs \cite{Mulders:1995dh}. Generally,  TMDs are non-perturbative functions, hence they have to be extracted from the experimental data which come from processes like semi-inclusive deep inelastic scattering (SIDIS) and Drell-Yan (DY) \cite{Boer:1997nt, Bacchetta:2006tn, Tangerman:1994eh, Arnold:2008kf}. 
Unlike collinear PDFs, TMDs depend both on the longitudinal momentum fraction of partons ($x_a$) and on their intrinsic transverse momentum ($k_{\perp a}$).
Moreover, TMDs are process-dependent because their operator definition contains Wilson lines \cite{Mulders:2001pj}. The quark TMD operator requires a single Wilson line to ensure local gauge invariance, while two Wilson lines are needed for the gluon TMD operator. These Wilson lines encapsulate information about the processes under consideration, whether future-pointing (representing final-state interaction) or past-pointing (indicating initial-state interaction). Final-state interactions arise from the interaction between the proton remnant and final-state particles, whereas initial-state interactions involve the interaction between the proton remnant and initial-state partons. 

At the leading twist, there are eight quark and gluon TMDs. In recent years, significant attention has been drawn to certain TMDs, for instance, the Boer-Mulders function, $h_{1}^{\perp q}$ \cite{Mulders:2001pj}, and the Sivers function, $f_{1T}^{\perp q}$  \cite{Sivers:1989cc, Sivers:1990fh}. Similarly, efforts in both the experimental and theoretical realms have brought recognition to TMDs, particularly the quark Sivers function, as evidenced by studies such as \cite{ Anselmino:2016uie, Boglione:2018dqd, Bury:2021sue}. However, the understanding of gluon TMDs remains relatively limited. 
The linearly polarized gluon distribution was introduced in \cite{Mulders:2001pj}, followed by its calculation in a model presented in \cite{Meissner:2007}. The linearly polarized gluon distribution describes the linearly polarized gluons within an unpolarized proton. In particular, $h_{1}^{\perp g}$ is both time-reversal ($T$) and chiral-even, thus retaining non-zero values even in the absence of initial-state interactions or final-state interactions \cite{Mulders:2001pj}. Linearly polarized gluon TMD could be probed by studying $\cos2\phi$ azimuthal asymmetry. 
The gluon Sivers TMD is also in the spotlight, as it correlates the hadron spin with the intrinsic transverse momentum of the gluon. It is interpreted as the probability of finding an unpolarized gluon within a transversely polarized hadron\cite{ElkePhysRevD.98.034011, Boer2015Cristian,ZengPhysRevD.106.094039,agrawal2024spinflip}.
Unlike the linearly polarized TMD, the Sivers  TMD depends on the initial and final state interactions. These interactions contribute to the Sivers asymmetry \cite{Sivers:1989cc, Sivers:1990fh}. This asymmetry also offers insight into the spin crisis as well \cite{Ji:2002xn}. The first transverse moment of the Sivers function is related to the twist-three Qiu-Sterman function \cite{Qiu:1998ia, Qiu:1991pp}. The $f_{1T}^{\perp g}$, $T$-odd, exhibits a sign reversal between the SIDIS and the DY processes \cite{Collins:2002kn,GambergPhysRevLett.110.232301,KangPhysRevD.83.094001}.   Generally, the gluon Sivers  TMD can be expressed in terms of two independent TMDs known as $f$-type and $d$-type \cite{Brodsky:2002cx, Collins:2002kn, Belitsky:2002sm, Ji:2002aa, Boer:2003cm}. The $f$-type gluon Sivers TMD involves a $(++$ or $--$) Wilson line and is often referred to as the Weizsäcker-Williams (WW) gluon distribution in the small-$x$ physics literature. Conversely, the $d$-type gluon Sivers  TMD incorporates a $(+-)$ Wilson line and is termed the dipole-type gluon distribution. While the non-zero quark Sivers function has been extracted from experiments like HERMES \cite{Airapetian:1994, HERMES:1999ryv} and COMPASS \cite{COMPASS:2005csq, ADOLPH2012383}, the gluon Sivers function remains elusive. Despite initial attempts \cite{DAlesio:2015fwo, DAlesio:2018rnv} to extract the gluon Sivers TMD from the RHIC data \cite{PHENIX:2013wle} in the mid-rapidity region, it still remains to be understood.

The $J/\psi$ meson is the lightest meson consisting of charm and anti-charm quarks. Because of its small mass, it can be produced abundantly in the collider experiments. $J/\psi$ has similar quantum numbers as the photon and can decay into $l^+l^-$ with a branching ratio of about $6\%$. 
Three models have been widely used to describe the $J/\psi$ formation, namely color-singlet model (CSM) \cite{Braaten_1996}, non-relativistic QCD (NRQCD) \cite {BodwinPhysRevD.51.1125} and color evaporation model (CEM)\cite{AMUNDSON1997323}. The most widely used framework is the NRQCD approach for $J/\psi$ production, which is an effective field theory that offers a systematic approach to understand the heavy-quarkonium production and its decay. In the NRQCD framework, a heavy-quark pair is produced initially either in color singlet (CS) or color octet (CO) configuration with a definite quantum number. Later, the heavy-quark pair hadronizes into physical $J/\psi$ via exchanging soft gluons through a non-perturbative mechanism, which is encoded in the so-called long-distance matrix elements (LDMEs) \cite{KniehlPhysRevLett.106.022003}. Calculation within NRQCD typically involves a double expansion, both in terms of the strong coupling constant $\alpha_s$ and the average velocity $v$ of the heavy quark within the quarkonium rest frame. This expansion scheme is particularly effective for the charmonium and bottomonium states, where $v^2$ is approximately $0.3$ and $0.1$, respectively. 
Semi-inclusive $J/\psi$ production is known to be a prominent channel to probe gluon TMDs \cite{GRMAMARVPhysRevD.85.094013,RSKRMAPhysRevD.98.014007,Mukherjee2017}. In the large transverse momentum region, $J/\psi$ production is described in the collinear factorization framework.
In the low transverse momentum region, TMD factorization is expected to hold. In the intermediate region, the results from these two formalisms should match. It has been shown that the TMD factorized description of the process needs the inclusion of smearing effects in the form of TMD shape functions, a perturbative tail of which can be calculated through a matching procedure \cite{Fleming2020, Echevarria2019, Boer2023}. In \cite{Echevarria:2024idp}, the TMD factorization has been proven for the semi-inclusive electroproduction of $J/\psi$ in small transverse momentum region, with the incorporation of TMD shape functions.  In addition, gluon TMDs can also be probed in back-to-back production of  $J/\psi$ and photon \cite{Chakrabarti:2022rjr}, $J/\psi$-jet production \cite{DAlesio:2019qpk, Kishore:2022ddb, Kishore:2019fzb,maxia2024azimuthal}, and the production of heavy-quark pairs or dijets \cite{Boer:2011, Pisano:2013cya, Efremov:2017iwh} at the Electron-Ion Collider (EIC). These investigations focus on measuring the transverse momentum imbalance of the produced pairs, basically, in these processes, the transverse momentum of the pair ($q_T$) is typically smaller than the individual transverse momentum ($K_{\perp}$). By varying the total invariant mass of the final state pair, the gluon TMDs can be probed at different scales.  Azimuthal asymmetries have been explored in different processes, including $J/\psi$ production \cite{Kishore:2018ugo, Kishore:2021vsm, DAlesio:2021yws, DAlesio:2020eqo, Rajesh:2018qks}, photon pair production \cite{Qiu:2011}, and Higgs boson-jet production  \cite{Boer:2014lka} at the Large Hadron Collider (LHC). As discussed in Ref.\cite{delCastillo:2020omr,Kang:2020xgk}, a new soft function is required for TMD factorized cross-section for both dijet and heavy-meson pair production in DIS process. 

In this article, we present a calculation of azimuthal asymmetries in the almost back-to-back electroproduction of $J/\psi$ and $\pi^{\pm}$ in the process of $e+p\rightarrow e+ J/\psi+\pi^{\pm}+X$. TMD factorization is applicable in processes where two scales are involved.  In this process, the hard scale is provided by the virtuality of the photon, and the soft scale is provided by the total transverse momentum of the $J/\psi$-pion pair, which is small, as they are almost back-to-back \cite{Kishore:2019fzb}. Another advantage is that only the WW-type gluon TMDs contribute to this process \cite{DAlesio:2019qpk}. We present the analysis using a TMD factorization framework, considering both unpolarized and transversely polarized proton targets. We use NRQCD to estimate the $J/\psi$ production. In the back-to-back $J/\psi$-pion kinematical configuration, the quarkonium is produced with large transverse momentum. We use NRQCD factorization in terms of LDMEs, and did not include the effect of initial and final state radiations \cite{Echevarria:2024idp}; which is a topic for a separate publication. In fact, for such processes, TMD shape functions have not been calculated yet by a matching procedure. Our aim here is to provide numerical estimates of the asymmetries at leading order, in the kinematics of the upcoming EIC.  We primarily focus on the calculation of azimuthal asymmetries like $\cos2\phi_T$, $\cos2(\phi_T-\phi_{\perp})$, and $\sin(\phi_S-\phi_T)$. These asymmetries help us to probe the linearly polarized and Sivers gluon TMDs.  As described above, the $J/\psi$ and $\pi^\pm$ have almost equal and opposite transverse momenta in the transverse plane. In this kinematics, the total transverse momentum of the system $q_T$ is much smaller than the individual transverse momentum of the final particle $K_\perp$ $i.e.~|\bm q_T| \ll |\bm K_\perp|$.  This article is organized as follows: in section \ref{sec2}, we outline the kinematics and formalism of $J/\psi$ and $\pi^\pm$ production.
Section \ref{sec3} describes the azimuthal asymmetries along with upper bounds and  Gaussian parametrizations of the gluon TMDs. Subsequently, section \ref{sec4}, {presents} the numerical results. Finally, we conclude in section \ref{sec5}. 

\section{Kinematics and Formalism}\label{sec2}
We investigate $J/\psi$ and pion production in the electron-proton scattering process, represented as: 
\be
e(l)+p(P)\rightarrow e(l^\prime)+ J/\psi(P_\psi)+\pi^{\pm}(P_\pi)+X\,,
\ee
where the momenta of the particles are denoted within the brackets. Our analysis encompasses both unpolarized and transversely polarized protons.  {Our} study, {considers} the photon-proton center-of-mass (cm) frame, where the photon and proton travel along the positive and negative $z$ axes, respectively. In this work, we consider electro-production, where the virtuality of the photon $Q^2$  is large. In photoproduction, when  $Q^2$ is close to zero, one has to consider the contribution from the resolved photon as well, where the photon acts as a bunch of quarks or gluons that contribute to the hard process \cite{Kniehl1999}.

The 4-momenta of the colliding proton  $P$, virtual photon $q$ and the incoming lepton $l$  can be written as,
\be\label{pq4m}
P^\mu&=&n^\mu_-+\frac{M_p^2}{2}n^\mu_+\approx n^\mu_-\,,\nonumber\\
q^\mu&=&-\xB n^\mu_-+\frac{Q^2}{2\xB}n^\mu_+\approx -\xB P^\mu+(P\cdot q)n^\mu_+\,,\nonumber\\
l^\mu &=&\frac{1-y}{y}\xB n^\mu_-+\frac{1}{y}\frac{Q^2}{2\xB }n^\mu_
++\frac{\sqrt{1-y}}{y}Q\hat{l}^\mu_\perp\,,
\ee

where $M_p$ denotes the mass of the proton, and it has been neglected throughout the study. The Bjorken variable $\xB$ is an important kinematic quantity defined as $\xB=\frac{Q^2}{2P\cdot q}$, where $Q^2$ is the squared invariant mass of the virtual photon. Additionally, the inelasticity ($y$) is defined as the fraction of the energy of the electron transfer to the photon, given by $y = \frac{P\cdot q}{P\cdot l}$.  These kinematic variables are related to the square of the center-of-mass (cm) energy   $S$ of the electron-proton system and invariant mass of the virtual photon-proton system $W_{\gamma p}$ through the following relations:
\begin{eqnarray}
Q^2\approx \xB y S~~~~~~~~  W^2_{\gamma p}=\frac{Q^2(1-\xB )}{\xB } \approx yS-Q^2.
\end{eqnarray}
\par 
The dominant partonic subprocess contributing to our investigated process is $ \gamma^*(q) +a(k) \rightarrow J/\psi(P_\psi) + a(P_a)$, where $a$ is a parton (quark or gluon). In this process, the virtual photon interacts with the parton inside the proton, producing  $J/\psi$ meson and a final-state parton.  This parton further fragments to form a pion.  In this work, we take into account the contributions that come from both the gluon and quark channels and estimate their relative contribution in the relevant kinematics. The 4-momenta of the $J/\psi (P_{\psi})$ meson, initial parton ($k$), and final parton ($P_a$) can be written as \cite{DAlesio:2019qpk,Mukherjee2017,Bacchetta2020},

\be\label{fv:pc}
k^\mu &\approx& x_a P^\mu +k^\mu_{\perp a}, \,\nonumber \\
P_{\psi}^\mu&=&z_{\psi}(P\cdot q)n_+^\mu + \frac{M_{\psi}^2+\bm {P}_{{\psi}\perp}^2} {2z_{\psi} P\cdot q} P^\mu+ P_{{\psi}\perp}^\mu\,\nonumber \\
P_{a}^\mu&=&z_a(P\cdot q)n_+^\mu + \frac{\bm {P}_{a\perp}^2}{2z_a P\cdot q} P^\mu+ P_{a\perp}^\mu\,,
\ee
where $a=q, g$ refers to both quark and gluon.
The variables $x_a$ and $k_{\perp a}$ denote the light-cone momentum fraction and the intrinsic transverse momentum of the incoming parton relative to the direction of the parent proton, respectively. Furthermore, we consider the momentum fractions $z_{\psi}=\frac{P \cdot P_{\psi}}{P\cdot q}$ and $z_a=\frac{P \cdot P_{a}}{P\cdot q}$ which represent the momentum fractions carried by the $J/\psi$ and final parton with respect to the virtual photon. The $M_{\psi}$ is mass of $J/\psi$. The ${P}_{\psi\perp}$ and ${P}_{a\perp}$ are transverse momenta of $J/\psi$ and final parton, respectively. As
the $J/\psi$ and the pion in the final state are observed in almost back-to-back configurations, the transverse momentum of each of them is large. Compared to this, the transverse momentum of the pion relative to the fragmenting quark is small.   So we assume a collinear fragmentation function for the pion in the final state. The four-momentum of the pion, denoted as $P_{\pi}$, can be expressed in terms of lightlike vectors as follows:
\be\label{fv:pc1}
P_{\pi}^\mu=z_{\pi}(P\cdot q)n_+^\mu + \frac{M_{\pi}^2+\bm {P}_{{\pi}\perp}^2} {2z_{\pi} P\cdot q} P^\mu+ P_{{\pi}\perp}^\mu \,,
\ee
where $ z_{\pi}=\frac{P \cdot P_{\pi}}{P\cdot q}$ is the  momentum fraction carried by pion $(\pi^\pm)$ from the virtual photon and its mass is represented with $M_{\pi}$. The 4-momentum of the final parton, $P_a$, as expressed in Eq.\eqref{fv:pc}, can be parametrized using the momentum fraction $z$ as,
\be\label{fv:pc2}
P_a^\mu=\frac{1}{z}P^\mu_{\pi}- \frac{1}{2P\cdot P_\pi}\left( \frac{M_\pi^2}{z}\right)P^\mu.\,
\ee
Here $z=\frac{P\cdot P_\pi}{P\cdot P_a}=\frac{z_\pi}{z_a}$ represents the momentum fraction of the pion in the parton frame. Using Eqs.\eqref{fv:pc} and \eqref{fv:pc2}, the transverse momentum of the parton and the fragmented pion are related by the following equation \cite{Koike:2011ns}:,
\be\label{eq:pcpd}
P_{a\perp}^\mu=\frac{1}{z} P_{\pi}^\mu.
\ee

For defining the four momenta of all the particles, we have used the light-like vector $n_+$ and, $n_-$, satisfying $n_+^2=n_-^2=0$ and $n_+\cdot n_-=1$. Out of these two light-like vectors, one of them aligns with the proton's direction, denoted as $P^\mu=n_{-}^{\mu}$.

\begin{figure}[H]
\begin{center} 
\includegraphics[height=4cm,width=14cm]{ 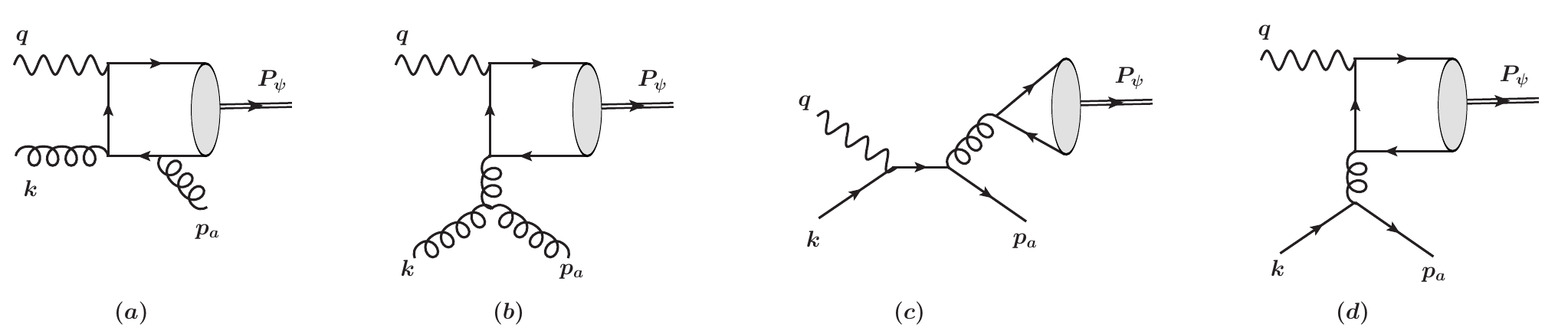}
\end{center}
\caption{\label{fig_feyn} Feynman diagrams for partonic subprocess $ \gamma^\ast(q) +a(k) \rightarrow J/\psi(P_\psi) + a(P_a)$. The gluon channel of types (a) and (b) contain six and two Feynman diagrams respectively. The quark channel of type (c) and (d) has two Feynman diagrams.}
\end{figure}

As discussed in the introduction, in the kinematical region when the outgoing $J/\psi$ and the pion are observed in almost back-to-back configuration, TMD factorization is expected to hold. To compute the cross-section for the electron-proton scattering process $e+p\to e+ J/\psi(P_\psi)+a(P_a)+X$, we adopt the TMD factorization framework, production of $J/\psi$ is calculated in (NRQCD).  We can write the 
 total differential scattering cross-section as a convolution of leptonic tensor, a soft gluon correlator, and a hard part \cite{Pisano:2013cya,BKMAPARSPhysRevD.108.034005},
\begin{equation}\label{eq:1}
 \begin{aligned}
 \d\sigma^{ep\to e+ J/\psi(P_\psi)+a(P_a)+X}={}&\frac{1}{2S}\frac{\d^3{\bm l^\prime}}{(2\pi)^32E_{l^\prime}}
\frac{\d^3{\bm P}_\psi}{(2\pi)^32E_{P_\psi}}\frac{\d^3{\bm P}_{a}}{(2\pi)^32E_{a}}\int \d x_a \,\d^2 {\bm k}_{\perp a} \, (2\pi)^4 \,\delta^4(q+k-P_{\psi}-P_a)\\
& \times\frac{1}{Q^4}L^{\mu\mu'}(l,q)\,\Phi^{\alpha\alpha'}_a(x,{\bm k}_{\perp a})\, \mathcal{M}_{\mu\alpha}^{\gamma^\ast a \to J/\psi+a}\mathcal{M}_{\mu'\alpha'}^{\dagger;\gamma^\ast a \to J/\psi+a}\, ,
\end{aligned}
\end{equation}
where $S=(l+P)^2\approx 2 l\cdot P$ and $Q^2 \equiv -q^2 > 0$.
The leptonic tensor $L^{\mu\mu'}$ takes the standard form given by:
\begin{eqnarray}\label{lep:ep}
L^{\mu\mu'}&=&e^2 Q^2\left(-g^{\mu\mu'}+\frac{2}{Q^2}(l^\mu l^{\prime\mu'}+l^{\mu'} l^{\prime\mu})\right).
\label{eq:lep1}
\end{eqnarray}
Here $e$ represents the electronic charge, and the spins of the initial lepton are averaged over.  The 4-momentum of the final scattered lepton is denoted as $l^\prime=l-q$. By using Eq.\eqref{pq4m}, the leptonic tensor can be reformulated as,
\begin{equation}\label{lep:ep1}
 \begin{aligned}
L^{\mu\mu'}={}&e^2 \frac{Q^2}{y^2}\Big[-(1+(1-y)^2)g_T^{\mu\mu'}+4(1-y)\epsilon_{ L}^\mu \epsilon_{ L}^{\mu'}+4(1-y)\left(\hat{l}^\mu_\perp \hat{l}^{\mu'}_\perp +\frac{1}{2} g_T^{\mu{\mu'}} \right)\\
&+2(2-y)\sqrt{1-y}\left(\epsilon_{ L}^\mu \hat{l}^{\mu'}_\perp + \epsilon_{ L}^{\mu'} \hat{l}^\mu_\perp \right)\Big]\,,
\end{aligned}
\end{equation}
where the transverse metric tensor is defined as $
g_{T}^{\mu{\mu'}}=g^{\mu{\mu'}}-n_+^{\mu}n_-^{\mu'}-n_+^{\mu'}n_-^{\mu}$. The light-like vectors $n_-^\mu$ and $n_+^\mu$ are expresses as:
\be\label{ll_1}
n_-^\mu=P^\mu\,,~~~~~~~~n_+^\mu=\frac{1}{P\cdot q}(q^\mu+x_B P^\mu)\,.
\ee
Additionally, the longitudinal polarization vector of the virtual photon is given by,
\be
\epsilon_{ L}^\mu(q) =\frac{1}{Q}\left(q^\mu +\frac{Q^2}{P\cdot q} P^\mu\right)\,,
\ee 
 which satisfy the relations $\epsilon_{ L}^2(q)=1$ and $\epsilon_{ L}^\mu(q) q_\mu=0$. The factor $\mathcal{M}$ in Eq.\eqref{eq:1} corresponds to the scattering amplitude of partonic process $\gamma^\ast(q) +a(k) \rightarrow J/\psi(P_\psi) + a(P_a)\,$; as depicted in Feynman diagrams shown in Fig.~\ref{fig_feyn}.

The Mandelstam variables for the partonic subprocess $ \gamma^\ast(q) +a(k) \rightarrow J/\psi(P_\psi) + a(P_a)$ are defined as,
\be
s=(q+k)^2=-Q^2+2q\cdot k=Q^2\left(\frac{x_a-\xB}{\xB}\right), ~~~ \,\nonumber \\
u=({P}_{\psi}-k)^2=M_{\psi}^2-2{P}_{\psi}\cdot k=M_{\psi}^2-\frac{x_a {z_{\psi}} Q^2}{\xB},~~~  \,\nonumber \\
t=(q-{P}_{\psi})^2=M_{\psi}^2-Q^2-2q\cdot {P}_{\psi}=M_{\psi}^2-Q^2(1-z_{\psi})-\frac{M_{\psi}^2+\bm {P}_{{\psi}\perp}^2}{z_{\psi}}\,.
\ee

The gluon and quark correlators contain the parton dynamics inside the proton and are non-perturbative quantities. For an unpolarized proton, the gluon correlator is given as:

\be\label{gc:un}
\Phi_{U}^{\alpha\alpha'}(x_g,{\bm k}_{\perp g})=\frac{1}{2x_g}\Bigg\{-g_{T}^{\alpha\alpha'}f_1^g(x_g,{\bm k}^2_{\perp g})+\left(\frac{k_{\perp g}^{\alpha} k_{\perp g}^{\alpha'}}{M_p^2}+g_{T}^{\alpha\alpha'}\frac{{\bm 
k}^2_{\perp g}}{2M_p^2}\right)h^{\perp g}_1(x_g,{\bm k}^2_{\perp g})\Bigg\}\,,
\ee
where $f_{1}^{g}$ and $h_{1}^{\perp g}$ are the unpolarized and linearly polarized gluon TMDs, which depend on both the momentum fraction $x_g$ and intrinsic transverse momentum of partons.
For a polarized proton, the gluon correlator is given by
\be
\begin{aligned}\label{gc:T}
\Phi_T^{\alpha\alpha'}(x_g,{\bm k}_{\perp g} )={}& 
\frac{1}{2x_g}\bigg \{-g^{\alpha\alpha'}_T
    \frac{ \epsilon^{\rho\sigma}_T k_{\perp g \rho} S_{T\sigma}}{M_p}
f_{1T}^{\perp\,g}(x_g, {\bm k}_{\perp g}^2) + i \epsilon_T^{\alpha\alpha'}
    \frac{ k_{\perp g} \cdot  S_T}{M_p} g_{1T}^{g}(x_g, {\bm k}_{\perp g}^2) \\
&+  
\frac{k_{\perp g \rho}\epsilon_{T}^{\rho\{ \alpha}k_{\perp g}^{\alpha'\}}} 
{2M_p^2} \frac{k_{\perp g }\cdot S_T}{M_p} h_{1T}^{\perp g}(x_g, {\bm k}_{\perp g}^2)
-\frac{k_{\perp g \rho}\epsilon_T^{\rho \{\alpha}S_T^{\alpha'\}}+ 
S_{T\rho}\epsilon_T^{\rho\{\alpha}k_{\perp g}^{\alpha'\}}}{4M_p}h_{1T}^{g}(x_g, {\bm k}_{\perp g}^2)  
 \bigg \}\,,
\end{aligned}
\ee

where $\epsilon_T^{\alpha\alpha^\prime}$ is the transverse metric tensor, $S_T$ represents the transverse spin vector of the proton. 
 The Sivers function, $f_{1T}^{\perp g}$,  describes the density of unpolarized gluons, while $h_{1T}^{\perp g}$ and $h_{1T}^g$ are linearly polarized gluon densities of a transversely polarized proton. The  $T$-even TMD, {$g_{1T}^g$} is the distribution of circularly polarized gluons in a transversely polarized proton, which does not contribute here since it is in the antisymmetric part of the correlator. The quark correlator for unpolarized proton is given by
\be
\Phi_{U}(x_q,{\bm k}_{\perp q})= \frac{1}{2x_q} \Bigg\{ \slashed{n}_+f_1^q(x_q,{\bm k}^2_{\perp q})+\frac{i \Big[\slashed{k}^2_{\perp q},\slashed{n}_+\Big]}{2M_p} h_1^{\perp q}(x_q,{\bm k}^2_{\perp q}) \Bigg\},
\ee
Here, $f^q_1$ and $h_1^{\perp q}$ are unpolarized quark TMD and Boer-Mulders quark TMD respectively. $f^q_1$ represents the probability of finding an unpolarized quark inside the unpolarized proton, and $h_1^{\perp q}$, T-odd function,  represents the probability of getting transversely polarized quarks in an unpolarized proton. The quark correlator for transversely polarized proton can be written  as   
\be
\begin{aligned}\label{qc:T}
\Phi_{T}(x_q,{\bm k}_{\perp q}) ={}& \frac{1}{2x_q} \Bigg\{ \frac{\epsilon_{\mu\nu\rho\sigma}\gamma^\mu n_+^\nu k_{\perp q}^\rho S_T^\sigma}{M_p} f_{1T}^{\perp\,q}(x_q, {\bm k}_{\perp q}^2)+ \frac{{\bm k}_{\perp q} \cdot {\bm S}_T}{M_p}\gamma_5\slashed{n}_+ g_{1T}^{q}(x_q, {\bm k}_{\perp q}^2) + \frac{\gamma_5 [\slashed{S}_T,\slashed{n}_+]}{2}h_{1T}^q (x_q, {\bm k}_{\perp q}^2)\\
&\qquad +\frac{{\bm k}_{\perp q} \cdot {\bm S}_T}{M_p} \frac{\gamma_5 \Big[\slashed{k}^2_{\perp q},\slashed{n}_+\Big]}{2M_p}  h_{1T}^{\perp q} (x_q, {\bm k}_{\perp q}^2) \Bigg\},
\end{aligned}
\ee
where, $f_{1T}^{\perp\,q}$ is quark Sivers function which is a probability of finding unpolarized quarks in a transversely polarized proton, $g_{1T}^{q}(x_q, {\bm k}_{\perp q}^2)$ is the probability of finding longitudinally polarized quarks in a transversely polarized proton. $h_{1T}^q,~h_{1T}^q$ are transversity TMDs of quark. 

The differential cross section given in Eq.\eqref{eq:1}  can be written as

\begin{equation}\label{eq:2}
 \begin{aligned}
 \d\sigma^{ep\to e+ J/\psi+\pi^\pm+X}={}&\frac{1}{2S}\frac{\d^3{\bm l^\prime}}{(2\pi)^32E_{l^\prime}}
\frac{\d^3{\bm P}_\psi}{(2\pi)^32E_{\psi}}\frac{\d^3{\bm P}_{\pi}}{(2\pi)^32E_{\pi}}\int \d x_a \,\d^2 {\bm k}_{\perp a} \, \d z\,(2\pi)^4 \,\delta^4(q+k-P_{\psi}-P_a)\\
& \times\frac{1}{Q^4}L^{\mu\mu'}(l,q)\,\Phi^{\alpha\alpha'}_a(x,{\bm k}_{\perp a})\, \mathcal{M}_{\mu\alpha}^{\gamma^\ast a \to J/\psi+a}\mathcal{M}_{\mu'\alpha'}^{\ast;\gamma^\ast a \to J/\psi+a}\, D(z)J(z).
\end{aligned}
\end{equation}
The phase space of the produced final parton is related to the phase space of the outgoing pion through the Jacobian $J$ as
\begin{eqnarray}
\frac{\d^3{\bm P}_a}{E_a}=J(z)\frac{\d ^3{\bm P}_\pi}{E_\pi}\,
~~~~~~~~{\rm with}~~~~~~~~
J=\frac{1}{z^3}\frac{E_\pi}{E_g}.
\end{eqnarray} \\
The momentum conservation delta function, given in  Eq.\eqref{eq:1},  can be decomposed as follows

\begin{equation}\label{dfun1}
\begin{aligned}
 \delta^{4}\bigl(q+k-P_{\psi}-P_a\bigr)
 & =\frac{2}{yS}\delta\bigl(1-z_\psi-z_a\bigr)\delta\left(x_a-\frac{z_a(M^2_{\psi}+ \bm{P}_{\psi\perp}^2)+z_\psi(\bm{P}^{2}_{a\perp}) + z_\psi z_a Q^2}{z_\psi z_a yS}\right )\delta^{2}\bigl(\bm{k}_{\perp a}- \bm{P}_{\psi\perp}- \bm{P}_{a\perp}\bigr)\;,
\end{aligned}
\end{equation}
After substituting Eq.\eqref{eq:pcpd} in Eq.\eqref{dfun1} we get,
\begin{equation}\label{dfun2}
\begin{aligned}
 \delta^{4}\bigl(q+k-P_\psi-P_a\bigr)
 & =\frac{2}{yS}\delta\bigl(1-z_\psi-\frac{z_\pi}{z}\bigr)\delta\left(x_a-\frac{z_a(M^2_{\psi}+ \bm{P}_{\psi\perp}^2)+z_\psi({\bm{P}^{2}_{\pi\perp}/}{z^2}) + z_\psi z_a Q^2}{z_\psi z_a yS}\right )\delta^{2}\bigl(\bm{k}_{\perp a}- \bm{P}_{\psi\perp}- \frac{\bm{P}_{\pi\perp}}{z}\bigr)\;.
\end{aligned}
\end{equation}
The phase space of the outgoing particles is given by
\be\label{felectron}
 \frac{\d^3{\bm l^\prime}}{(2\pi)^32E_{l^\prime}}&=&\frac{1}{16\pi^2}\d Q^2\d y\,, \quad 
 \frac{\d^3{\bm P}_\pi}{(2\pi)^32E_\pi}=\frac{\d^2{\bm P}_{\pi\perp}\d z_{\pi}}{(2\pi)^3 2z_\pi}\,, \quad \frac{\d^3{\bm P}_\psi}{(2\pi)^32E_\psi}=\frac{\d^2{\bm P}_{\psi\perp}\d z_{\psi}}{(2\pi)^3 2z_\psi}.
\ee

In our analysis, we consider a scenario in which the final parton and the $J/\psi$ are oriented back-to-back in the transverse plane, {which means} the sum of their transverse momenta is significantly smaller than the individual transverse momenta of the $J/\psi$ and final parton.  This condition allows us to use the TMD factorization for the scattering cross-section. We define the sum and difference of their transverse momentum as follows,
\be\label{eq:newfv}
\bm q_T={\bm P}_{\psi\perp}+\frac{{\bm P}_{\pi\perp}}{z}, \quad
\bm K_\perp= \frac{{\bm P}_{\psi\perp}-\frac{{\bm P}_{\pi\perp}}{z}}{2}\,.
\ee
 The Eq.\eqref{eq:newfv} shows that the transverse momentum of {the final} parton and $J/\psi$ are almost equal in magnitude but point in approximately opposite directions $i.e.$, $|\bm q_T| \ll |\bm K_\perp|$.  
From Eq.\eqref{dfun2}, we have $z_{\psi}=1-\frac{z_\pi}{z}$ and $\bm q_T=\bm{k}_{\perp a}$.{ Further, the transformation given in Eq.\eqref{eq:newfv} leads us to the transverse phase space given below,}
 \be\label{eq:newfq}
\d^2{\bm  P}_{\psi\perp} \d^2 {\bm  P}_{\pi\perp}=z \d^2{\bm  q}_{T} \d^2 {\bm  K}_{\perp}.
\ee
After performing the integration over $z_\pi$, $x_a$ and ${\bm k}_{\perp a}$  in Eq.\eqref{eq:2}, we get
\begin{equation}\label{eq:ff3}
 \begin{aligned}
 \frac{\d\sigma^{ep\to e+ J/\psi+\pi^\pm+X}}{\d Q^2 \d y \d z_\psi \d^2{\bm  q}_{T} \d^2 {\bm  K}_{\perp} }={}&\frac{1}{yS^2}\frac{1}{16(2\pi)^4}
\int \d z\, D(z)\,\\
& \times\frac{1}{Q^4}L^{\mu\mu^\prime}(l,q)\,\Phi_{\alpha\alpha^\prime}^a(x_a,{\bm q}_{T})\, {\mathcal M}_{\mu\alpha}^{\gamma^\ast a \to c\bar{c}+a}{\mathcal M}_{\mu^\prime\alpha^\prime}^{\dagger;\gamma^\ast a \to c\bar{c}+a}\, \,\frac{zJ(z)}{z_\psi(1- z_\psi)}\,.
\end{aligned}
\end{equation}

\section{Azimuthal asymmetries }\label{sec3}
The cross-section for $J/\psi$ and $\pi^\pm$ production at NLO, where both the $J/\psi$ and $\pi^\pm$ are almost back-to-back in the transverse plane, can be written as the sum of unpolarized and transversely polarized cross-sections \cite{Pisano:2013cya},  

\begin{equation}
 \frac{\d\sigma}{\d Q^2 \d y \d z_\psi \d^2{\bm  q}_{T} \d^2 {\bm  K}_{\perp} }  \equiv \d\sigma (\phi_S, \phi_\sT) =    \d\sigma^U(\phi_\sT,\phi_\perp)  +  \d\sigma^T (\phi_S, \phi_\sT)  \,.
\label{eq:cs}
\end{equation}

\begin{figure}[H]
\begin{center} 
\includegraphics[height=4cm,width=8cm]{ 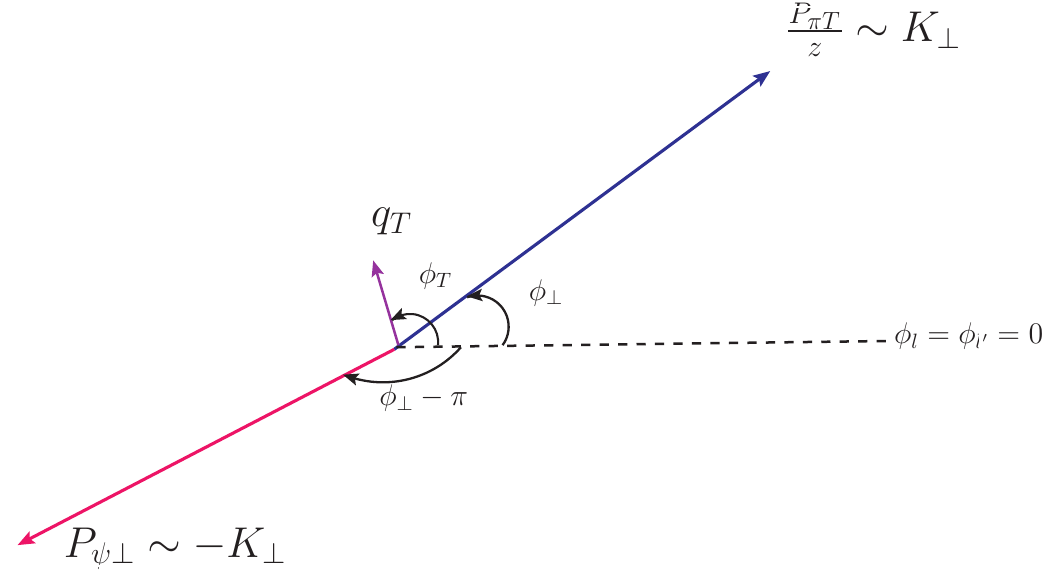}
\end{center}
\caption{\label{fig_angles} Azimuthal angles of $J/\psi$ and $\pi^{\pm}$ in the transverse plane.}
\end{figure}

For an unpolarized proton, the differential cross-section can be written in terms of different azimuthal modulations associated with the $f_{1}^{g}(x_g,\bm{q}_{\sT}^{2})$ and $h_{1}^{\perp\, g} (x_g,\bm{q}_{\sT}^2)$,  along with  the fragmentation function,
\begin{align}\label{eq:Un}
\d\sigma^{U} & =\mathcal{N}\int \d z\bigg[\bigl(\mathcal{A}_{0}+\mathcal{A}_{1} \cos\phi_{\perp}+\mathcal{A}_{2} \cos2\phi_{\perp}\bigr)f_{1}^{g}(x_g,\bm{q}_{\sT}^{2})+\bigl(\mathcal{B}_{0}  \cos2\phi_{\sT}+\mathcal{B}_{1}  \cos(2\phi_{\sT}-\phi_{\perp})\nonumber\\
 & \qquad\quad+\mathcal{B}_{2}  \cos2(\phi_{\sT}-\phi_{\perp})+\mathcal{B}_{3}  \cos(2\phi_{\sT}-3\phi_{\perp})+\mathcal{B}_{4}  \cos(2\phi_{\sT}-4\phi_{\perp})\bigr)\frac{ \bm q_\sT^2  }{M_p^2} \,h_{1}^{\perp\, g} (x_g,\bm{q}_{\sT}^2)\bigg]\, D(z)\,.
\end{align}

where $\mathcal{N}$ is the normalization factor given as,
\begin{equation}
\mathcal{N}=\frac{\alpha^2 \alpha_s e_c^2}{\pi y^3 S^2 x_g}. 
\end{equation}

while the transversely polarized scattering cross-section is given by,
\begin{align}\label{eq:Tr}
\d\sigma^T
  & =  \mathcal{N}{\vert \bm S_\sT\vert}\,\int \d z \bigg[\sin(\phi_S -\phi_\sT) \bigl( \mathcal{A} _0 + \mathcal{A} _1 \cos \phi_\perp  + \mathcal{A} _2 \cos 2 \phi_\perp \bigr ) \frac{ |\bm q_\sT | }{M_p}f_{1T}^{\perp\, g} (x_g,\bm{q}_{\sT}^2)\nonumber \\
 & + \cos(\phi_S-\phi_\sT) \bigl( \mathcal{B} _0 \sin 2 \phi_\sT  +  \mathcal{B} _1 \sin(2\phi_\sT-\phi_\perp) + \mathcal{B} _2 \sin 2 (\phi_\sT-\phi_\perp) \nonumber \\
 &\quad+ \mathcal{B} _3 \sin( 2\phi_\sT-3\phi_\perp)  + \mathcal{B} _4\sin(2\phi_\sT- 4\phi_\perp)  \bigr) \frac{ |\bm q_\sT |^3  }{M_p^3} \,h_{1T}^{\perp\, g} (x_g,\bm{q}_{\sT}^2) \nonumber\\
& + \bigl (\mathcal{B}_0  \sin(\phi_S+\phi_\sT) + \mathcal{B}_1  \sin(\phi_S + \phi_T-\phi_\perp) + \mathcal{B}_2  \sin(\phi_S+\phi_\sT-2\phi_\perp) \nonumber \\
&  \quad  +  \mathcal{B}_3  \sin(\phi_S+\phi_\sT-3\phi_\perp) + \mathcal{B}_4  \sin(\phi_S+\phi_\sT-4 \phi_\perp) \bigr)\frac{ |\bm q_\sT | }{M_p} h_{1T}^{g} (x_g,\bm{q}_{\sT}^2) \bigg ]\, D(z)\,,
\end{align}
where  $\phi_S$, $\phi_\sT$ and $\phi_\perp$  are  the azimuthal angles of the three-vectors $\bm S_\sT$,  $\bm q_\sT$ and  $\bm K_\perp$ respectively, measured w.r.t. the lepton plane      
 ($\phi_{\ell}=\phi_{\ell^\prime}=0)$ as shown in Fig.~\ref{fig_angles}. The expressions for the coefficients of various angular modulations, denoted as $\mathcal{A}_i$ for $i=0$ and $\mathcal{B}_j$ for $j=0,1,2$, are calculated for $^{3}{S}_{1}^{(1,8)}$,  $^{1}{S}_{0}^{8}$ and $^{3}{P}_{J}^{8}$ states in NRQCD. The S-wave amplitude is obtained by Taylor expanding the amplitude, $\mathcal{M}^{\gamma^\ast a\to c\bar{c} +a}$, around the zero relative momentum of the heavy quark(anti-quark) in the quarkonium rest frame \cite{Rajesh:2018qks,Mukherjee:2016qxa}. The second term in the Taylor expansion, the first-order derivative term, gives P-wave amplitude.  The analytical expressions of the $\mathcal{A}_0$, $\mathcal{B}_0$, and $\mathcal{B}_2$ are given in Appendix \ref{appen} for  $S$-waves.   The expressions for the $P$-wave contributions are too lengthy to be given in the appendix but are available upon request. The analytical expressions of  $\mathcal{A}_i$  and $\mathcal{B}_j$ match with Ref.\cite{DAlesio:2019qpk}, except the $^{3}P_0$ state  contribution, which differs by a negative sign.   
 \par
In case we do not measure the azimuthal angle of the final scattered lepton, then only one modulation term in Eq.\eqref{eq:Un} is defined, and the cross-section is expressed as,
\begin{align}\label{eq:Un1}
\d\sigma^{U} & =\mathcal{N}\int \d z\bigg[\mathcal{A}_{0}f_{1}^{g}(x_g,\bm{q}_{\sT}^{2})+\mathcal{B}_{2}  \cos2(\phi_{\sT}-\phi_{\perp})\frac{ \bm q_\sT^2  }{M_p^2} \,h_{1}^{g} (x_g,\bm{q}_{\sT}^2)\bigg]\,D(z)\,,
\end{align}

The weighted azimuthal asymmetry gives the ratio of the specific gluon TMD over unpolarized $f_1^g$ and is defined as \cite{DAlesio:2019qpk},
\begin{align}
A^{W(\phi_S,\phi_\sT)} & \equiv 2\,\frac {\int\d \phi_S\, \d \phi_\sT \,\d\phi_\perp\, W(\phi_S,\phi_\sT)\,\d\sigma(\phi_S,\,\phi_\sT,\,\phi_\perp)}{\int \d \phi_S\,\d\phi_\sT \,\d\phi_\perp\,\d\sigma(\phi_S,\phi_\sT,\phi_\perp)} \,,
\label{eq:mom}
\end{align}
where the denominator is given by
\begin{align}\label{eq:f1}
\int \d\phi_S\,\d \phi_\sT \,\d\phi_\perp\,\d\sigma(\phi_S,\phi_\sT,\phi_\perp)  & =\int \d\phi_S\,\d \phi_\sT \,\d\phi_\perp\,\d\sigma^U(\phi_\sT,\phi_\perp)=  (2\pi)^3 \mathcal{N}\int \d z {\cal A}_0 f_{1}^{g}(x_g,\bm{q}_{\sT}^{2})\,D(z)\;.
\end{align}
By integrating over the azimuthal angle $\phi_\perp$, the transversely polarized cross-section, Eq.~(\ref{eq:Tr}), can be simplified further as,
\begin{align}\label{eq:sivers}
\int\d\phi_\perp\d\sigma^T & =  2\pi {\vert \bm S_\sT\vert}\, \frac{ |\bm q_\sT | }{M_p}\int  \d z\left [\mathcal{A} _0 \sin(\phi_S -\phi_\sT)    f_{1T}^{\perp\, g} (x_g,\bm{q}_{\sT}^2) -\frac{1}{2} \mathcal{B} _0\sin(\phi_S-3\phi_\sT)   \frac{ |\bm q_\sT |^2 }{M_p^2} \,h_{1T}^{\perp\, g} (x_g,\bm{q}_{\sT}^2) \right . \nonumber\\*
  &\qquad+\mathcal{B}_0  \sin(\phi_S+\phi_\sT)  h_{1}^{g} (x_g,\bm{q}_{\sT}^2) \bigg ]\,D(z)\,,
\end{align}
where we have used the relation 
\begin{equation}
h_1^g \equiv h_{1T}^g +\frac{\bm p_\sT^2}{2 M_p^2}\,  h_{1T}^{\perp\,g}\,
\label{eq:h1}
\end{equation}
where $h_1^g$ ($T$-odd), is the helicity flip gluon distribution which is chiral-even and vanishes upon integration of transverse momentum \cite{Boer:2016fqd}. In contrast, the quark distribution is chiral-odd {($T$-even)} and survives even after the transverse momentum integration.
The $h_1^{\perp\, g}$ gluon TMD could be extracted by studying the following two azimuthal asymmetries\cite{DAlesio:2019qpk},
\begin{align}
A^{\cos 2 \phi_\sT}&= \frac{\bm q_\sT^2}{M_p^2}
\,\frac{\int \d z\,  {\cal B}_0 \,D(z)\, h_1^{\perp\, g}(x_g,\bm q_\sT^2 )}{\int \d z \, {\cal A}_0\, D(z)\, f_1^{g}(x_g,\bm q_\sT^2 )}\,,
\label{eq:cos2phiT}
\end{align}
and
 \begin{align}
A^{\cos 2 (\phi_\sT-\phi_\perp)}&= \frac{\bm q_\sT^2}{M_p^2}\, \frac{\int \d z\,  {\cal B}_2\, D(z)\,h_1^{\perp\, g}(x_g,\bm q_\sT^2 )}{\int \d z\,  {\cal A}_0\, D(z)\, f_1^{g}(x_g,\bm q_\sT^2 )}\,.
\label{eq:cos2phiT2phiP}
\end{align}
Using Eq.\eqref{eq:sivers} with ${\vert \bm S_\sT\vert}=1$ , one could utilize the following asymmetries to extract the $f_{1T}^{\perp g}$, $h_1^g$ and $h_{1T}^{\perp g}$ TMDs,
\begin{align}\label{eq:f1t_sivers}
A^{\sin(\phi_S-\phi_\sT)} & =  \frac{\vert \bm q_\sT\vert}{M_p}\, \frac{\int \d z\,  {\cal A}_0\, D(z)\, f_{1T}^{\perp\,g}(x_g,\bm q_\sT^2) }{\int \d z\,  {\cal A}_0\, D(z)\,f_1^g(x_g,\bm q_\sT^2)}\,,
\end{align}
\begin{align} \label{eq:h1g-asy}
A^{\sin(\phi_S+\phi_\sT)}  & =  \frac{\vert \bm q_\sT\vert}{M_p}\, \frac{\int \d z\,  {\cal B}_0\, D(z)\, h_{1 }^{g}(x_g,\bm q_\sT^2) }{\int \d z\,  {\cal A}_0\, D(z)\,f_1^g(x_g,\bm q_\sT^2)}\,,
 \end{align}
 and
\begin{align}\label{eq:h1Tg-asy}
A^{\sin(\phi_S-3\phi_\sT)}  & =   -  \frac{\vert \bm q_\sT\vert ^3}{2M_p^3}\,  \frac{\int \d z\,  {\cal B}_0\, D(z)\, h_{1T}^{\perp\,g}(x_g,\bm q_\sT^2)}{\int \d z\,  {\cal A}_0\, D(z)\, f_1^g(x_g,\bm q_\sT^2)} \,.
\end{align}

\subsection{Upper bounds}

The upper bounds of the azimuthal asymmetries as defined in Eqs.\eqref{eq:cos2phiT}-\eqref{eq:h1Tg-asy}, can be determined by saturating the positivity bounds on the gluon TMDs, which are model-independent constraints \cite{Bacchetta:1999kz, Mulders:2001pj}; given below :

\begin{align}
\frac{\vert \bm q_\sT \vert }{M_p}\, \vert f_{1T}^{\perp \,g}(x_g,\bm q_\sT^2) \vert & \le   f_1^g(x_g,\bm q_\sT^2)\,,\nonumber \\
\frac{ \bm q^2_\sT }{2 M_p^2}\, \vert h_{1}^{\perp\,g}(x_g,\bm q_\sT^2) \vert & \le     f_1^g(x_g,\bm q_\sT^2)\,,\nonumber \\
\frac{\vert \bm q_\sT \vert }{M_p}\, \vert h_{1}^g(x_g,\bm q_\sT^2) \vert & \le   f_1^g(x_g,\bm q_\sT^2)\,,\nonumber \\
\frac{\vert \bm q_\sT \vert^3}{2 M_p^3}\, \vert h_{1T}^{\perp \,g}(x_g,\bm q_\sT^2) \vert & \le   f_1^g(x_g,\bm q_\sT^2)\,.
\label{eq:bound}
\end{align}
  The bounds, as stated in Eq. \eqref{eq:bound}, essentially ensure that the Sivers functions as well as the linearly polarized gluon distribution do not dominate the unpolarized gluon distribution $f_1^g(x_g,\bm{q}_T^2)$. Upper bounds on the asymmetries are also model-independent. By incorporating these positivity bounds into our analysis, we can derive upper limits on azimuthal asymmetries such as $A^{\cos 2 \phi_T}$ and $A^{\cos 2(\phi_T-\phi_\perp)}$ as follows:
\begin{align}
\vert A^{\cos 2 \phi_\sT}\vert = 2\frac{ |{\cal B}_0|  }{{\cal A}_0}\,, \quad  \vert A^{\cos 2 (\phi_\sT-\phi_\perp)}\vert = 2\frac{ |{\cal B}_2|  }{{\cal A}_0}\,,
\end{align}
 The gluon TMDs influence these asymmetries, and the bounds provide a ceiling on how large these asymmetries can be for a given kinematic condition. Additionally, the upper bound for the Sivers asymmetry, $A^{\sin(\phi_S-\phi_T)}$, reaches unity.  While the upper bounds for the other asymmetries are related as below 
\begin{align}\label{eq:ub1}
\vert A^{\sin(\phi_S+\phi_\sT)}\vert = \frac12 \,\vert A^{\cos 2 \phi_\sT}\vert \,, \quad  \vert A^{\sin(\phi_S-3\phi_\sT)}\vert =\frac12\,\vert A^{\cos 2 (\phi_\sT-\phi_\perp)}\vert\,.
\end{align}

\subsection{Gaussian Parametrization of the TMDs}
The asymmetries are dependent on the parametrizations of the gluon TMDs.  In our analysis, we employ a widely used  Gaussian parameterization for the TMDs. This parametrization involves factorizing the TMDs into collinear PDFs and an exponential factor that depends on the transverse momentum of the parton. For the unpolarized TMD $f_1^a(x_a,\bm{q}_\sT^2)$, the parametrization is given by :
\begin{eqnarray}\label{eq:gauss_f1}
f_1^a(x_a,\bm{q}_\sT^2)=f_1^a(x_a,\mu)\frac{e^{-\bm{q}_\sT^2/\langle q_\sT^2\rangle}}{\pi\langle q_\sT^2\rangle}\,.
\end{eqnarray}

Here, $f_1^a(x_a,\mu)$ represents the collinear gluon and quark PDFs at the probing scale $\mu=\sqrt{M^2_\psi+Q^2}$ \cite{Kniehl:2006mw}.  In this work, we have used the Gaussian width $\langle q_\sT^2\rangle=1~\mathrm{GeV}^2$ for gluons \cite{Alesio:2017um} and $\langle q_\sT^2\rangle=0.25~\mathrm{GeV}^2$ for quarks.
%
For the linearly polarized TMD $h_1^{\perp g}(x_g,\bm{q}_\sT^2)$, we adopted the Gaussian parameterization proposed in Ref. \cite{Boer:2011kf, Boer:2012bt}:

\begin{eqnarray}
			\label{eq:gauss_h1p}
			h_1^{\perp g}(x_g,\bm{q}_\sT^2)=\frac{M_p^2f_1^g(x_g,\mu)}{\pi\langle q_\sT^2\rangle^2}\frac{2(1-r)}{r}e^{1-\frac{\bm{q}_\sT^2}{r\langle q_\sT^2\rangle}},
\end{eqnarray}

where $M_p$ is the proton mass, $r$ (with $0<r<1$), and the average intrinsic transverse momentum width of the incoming gluon, $\langle q_\sT^2\rangle$, are parameters of this model. In our numerical estimation, we take $r=1/3$ and 
$\langle q_\sT^2\rangle=1~\text{GeV}^2$ for gluons {in line with Ref.\cite{Boer:2012bt}}.  

For the numerical estimates of Sivers asymmetry, we utilized Gaussian {parameterization} for both the gluon and quark Sivers functions, denoted as $f_{1T}^{\perp g}$ and $f_{1T}^{\perp q}$, respectively.
Starting with the gluon Sivers function $f_{1T}^{\perp g}$, we employed the Gaussian parameterization as defined in Refs. \cite{Bacchetta:2004jz, DAlesio:2018rnv, Anselmino:2005ea}. The parameterization is given by:
\begin{eqnarray}\label{eq:Sivers:par}
\Delta^{N} f_{g / p^{\uparrow}}\left(x_g, q_\sT\right)= \left(-\frac{2|\bm{q}_\sT|}{M_p}\right)f_{1T}^{\perp g}\left(x_g, q_\sT\right)=2\frac{\sqrt{2 e}}{\pi} \mathcal{N}_{g}\left(x_g\right) f_{g / p}\left(x_g\right)  \sqrt{\frac{1-\rho}{\rho}} q_\sT \frac{e^{-\bm{q}_\sT^2 / \rho\left\langle q_\sT^2\right\rangle}}{\left\langle q_\sT^2\right\rangle^{3 / 2}}\,,
\end{eqnarray}

Here, $\rho$ is a parameter with $0<\rho<1$, and the $x$-dependence is encapsulated in the function $\mathcal{N}_{g}\left(x_a\right)$, defined as:
\begin{eqnarray}\label{eq:Sivers:Ng}
\mathcal{N}_{g}\left(x_a\right)=N_{g} x_a^{\alpha}\left(1-x_a\right)^{\beta} \frac{(\alpha+\beta)^{(\alpha+\beta)}}{\alpha^{\alpha} \beta^{\beta}}.\,
\end{eqnarray}

The parameters $N_g$, $\alpha$, and $\beta$ are determined from fits to experimental data on single spin asymmetries (SSAs) in inclusive hadron production processes \cite{DAlesio:2018rnv}. For our numerical estimations at $\left\langle q_\sT^2\right\rangle=1~\text{GeV}^2$, we used the values $ N_g=0.25$, $\alpha=0.6$, $\beta=0.6$, and $\rho=0.1$.

For the quark Sivers function $f_{1T}^{\perp q}$, we have used the Gaussian parameterization introduced in Ref. \cite{Boglione:2021aha},
\begin{eqnarray}\label{eq:qSivers:par}
\Delta^{N} f_{q / p^{\uparrow}}\left(x_q, q_\sT\right)= \left(-\frac{2|\bm{q}_\sT|}{M_P}\right)f_{1T}^{\perp q}\left(x_q, q_\sT\right)= {4 {M_p} q_\sT}\Delta^{N} f_{q / p^{\uparrow}}^{(1)}(x_q) \frac{e^{-\bm{q}_\sT^2 / \left\langle q_\sT^2\right\rangle}}{\pi\left\langle q_\sT^2\right\rangle^{2}}\,,
\end{eqnarray}

\begin{eqnarray}
\Delta^{N} f_{q / p^{\uparrow}}\left(x_q, q_\sT\right) = {4 {M_p} q_\sT}\Delta^{N} f_{q / p^{\uparrow}}^{(1)}(x_q) \frac{e^{-\bm{q}_\sT^2 / \left\langle q_\sT^2\right\rangle}}{\pi\left\langle q_\sT^2\right\rangle^{2}}.
\end{eqnarray}

Here $q=u,d$ and $\Delta^{N} f_{q / p^{\uparrow}}^{(1)}(x_q)$  represents the Sivers first $q_\sT$-moment and is parameterized as:
\begin{eqnarray}\label{eq:Sivers:Nq}
\Delta^{N} f_{q / p^{\uparrow}}^{(1)}(x_q)&=&\int d^{2}q_{\sT} \frac{q_\sT}{4 M_p}\Delta^{N} f_{q / p^{\uparrow}}\left(x_q, q_\sT\right)\equiv -f_{1T}^{\perp (1) q}(x_q)\\
&&=N_q(1-x_q)^{\beta_q}.\,
\end{eqnarray}

The parameters $N_u=0.36$, $N_d=-0.55$, $\beta_u=4.98$, $\beta_d=6.45$, and $\left\langle q_\sT^2\right\rangle =0.28 ~\rm{GeV}^2$ were determined through reweighting the Sivers function from SIDIS-jet data in the GPM formalism \cite{Boglione:2021aha} compared with STAR measurements \cite{adam2021measurement}. 

There are recent parametrizations of gluon TMDs, based on spectator model,  that do not assume a factorization of the $x$ and transverse momentum dependence \cite{Bacchetta:2020vty, Chakrabarti:2023djs}. We also comment on the results in such models.


\section{Numerical Results} \label{sec4}

In the section, we present numerical results for  $J/\psi$ and $\pi^\pm$ production in electron-proton collisions. The process involves the interaction of a virtual photon with the partonic content of the proton, resulting in the creation of a pair of charm and anti-charm quarks along with a final-state parton. The charm and anti-charm quark pair subsequently hadronize into a $J/\psi$ meson through the NRQCD mechanism, while the produced parton fragments to form a pion in the final state. The process considered is $\gamma^\ast + a \rightarrow J/\psi + a \rightarrow J/\psi + \pi^\pm$. The heavy quark pair can be produced in different color and spin configurations, depending on which, contribution to $J/\psi$ production comes from the states  $^{3}{S}_{1}^{(1,8)}$,  $^{1}{S}_{0}^{8}$ and $^{3}{P}_{J}^{8}$, with  $J=0,1,2$. The hadronization of heavy-quark pair into the quarkonium is encoded in the LDMEs. Even though LDMEs are assumed to be universal, there are several sets of LDMEs in the literature that are distinct from each other. This discrepancy among the LDMEs is related to the fact that the order of perturbative QCD considered and the second one, the transverse momentum cut imposed in the fitting. Although the choice of the LDME set strongly influences the asymmetries, we considered two LDME sets, labeled with CMSWZ \cite{Chao:2012iv}  and SV \cite{Sharma:2012dy}, to obtain maximum asymmetries. As discussed earlier, for this process, 
an analytic form of the smearing effect coming from the TMD shape functions is still not available, as matching the TMD factorized result with the collinear framework at the intermediate transverse momentum region is still to be done. So in this work, we did not include any smearing effect. The effect  of the shape function on the asymmetry is expected to be small, as the $J/\psi$ is produced with relatively large transverse momentum. However, a careful consideration of these effects  is needed, possibly in  a future publication. For the pion production, we consider a collinear fragmentation process, that is we assume that the transverse momentum of the pion with respect to the fragmenting quark is small compared to the transverse momentum of the pion itself.  For the pion fragmentation, we use the neural networks fragmentation function (NNFF) \cite{Bertone:2017tyb}. The CT18NLO \cite{HouPhysRevD.103.014013} is employed for collinear PDFs. The PDFs and fragmentation function are probed at the scale $\mu=\sqrt{M^2_\psi+Q^2}$. The mass of the $J/\psi$ is taken as $M_{\psi}=3.1$ GeV.  We consider kinematics where the  $J/\psi$ and $\pi^\pm$ are produced almost back-to-back in the transverse plane, wherein the relative transverse momentum, $q_T$,  is significantly smaller than the transverse momentum, $K_{\perp}$, of the $J/\psi$,  i.e., $|\bm q_T| \ll |\bm K_{\perp}|$. To ensure this, we consider the following kinematical cuts:  the transverse momentum imbalance $q_T$ and $z$, the momentum fraction carried by pion from partons,  are integrated in the range [0,1]. The momentum fraction carried by $J/\psi$ from virtual photon $z_\psi$ is considered in the range $0.3<z_\psi<0.7$. The lower cut here is chosen to eliminate the resolved photon contribution, while the uppercut avoids the collinear divergences as $z_\psi \to 1$.

\subsection{Upper bounds}
\begin{figure}[H]
	\begin{center} 
		\begin{subfigure}{0.49\textwidth}
		\includegraphics[height=6cm,width=7.5cm]{ 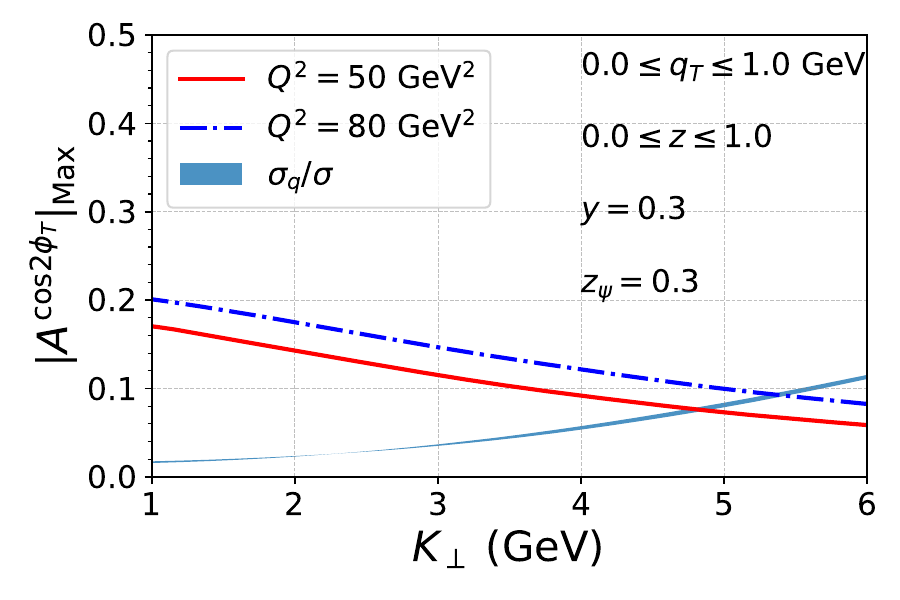}
			\caption{}
		\end{subfigure}
	    \begin{subfigure}{0.49\textwidth}
	    \includegraphics[height=6cm,width=7.5cm]{ 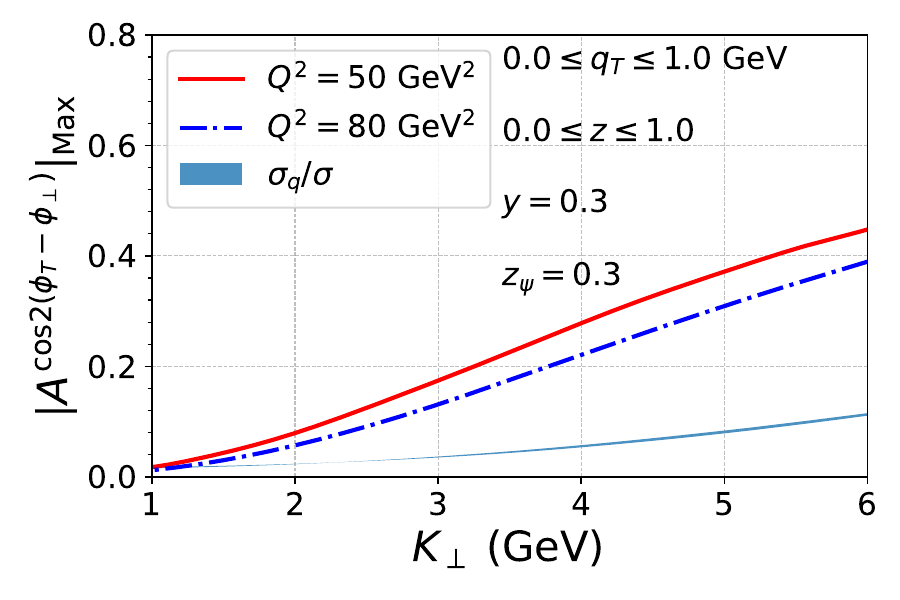}
	        \caption{}
	    \end{subfigure}
	\end{center}
 \caption{\label{fig: KP_VAR_fix_z1_y} The upper bounds for the $A^{\cos2\phi_T}$ (left panel) and $A^{\cos2(\phi_{T}-\phi_\perp)}$ (right panel) azimuthal asymmetries are shown as function of $K_\perp$ in the process $ep\rightarrow e+ J/\psi+\pi^\pm+X$ at  $\sqrt{S}=140$ GeV using CMSWZ \cite{Chao:2012iv}  LDME set. The fixed parameters include $y=0.3$, $z_\psi=0.3$ and $Q^2=50$ and $80$  GeV$^2$. Kinematic variables $z$ and $q_T$ are integrated over the interval $[0,1]$. The ratio of quark contribution to the total scattering cross-section is shown as a band, obtained by varying $Q^2$ from 50 to 80 GeV$^2$.}
 \end{figure}

\begin{figure}[H]
	\begin{center} 
		\begin{subfigure}{0.49\textwidth}
		\includegraphics[height=6cm,width=7.5cm]{ 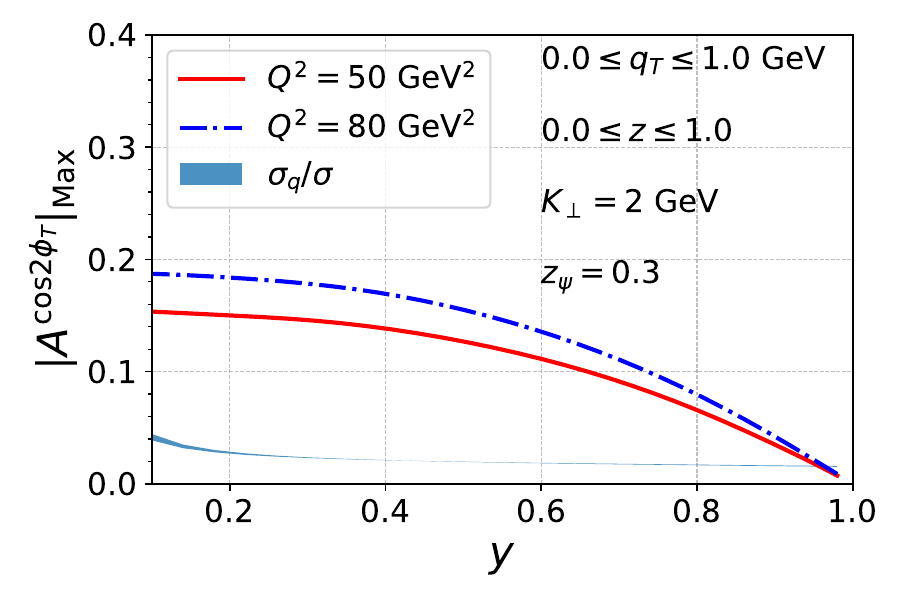}
			\caption{}
        \end{subfigure}
	    \begin{subfigure}{0.49\textwidth}
	    \includegraphics[height=6cm,width=7.5cm]{ 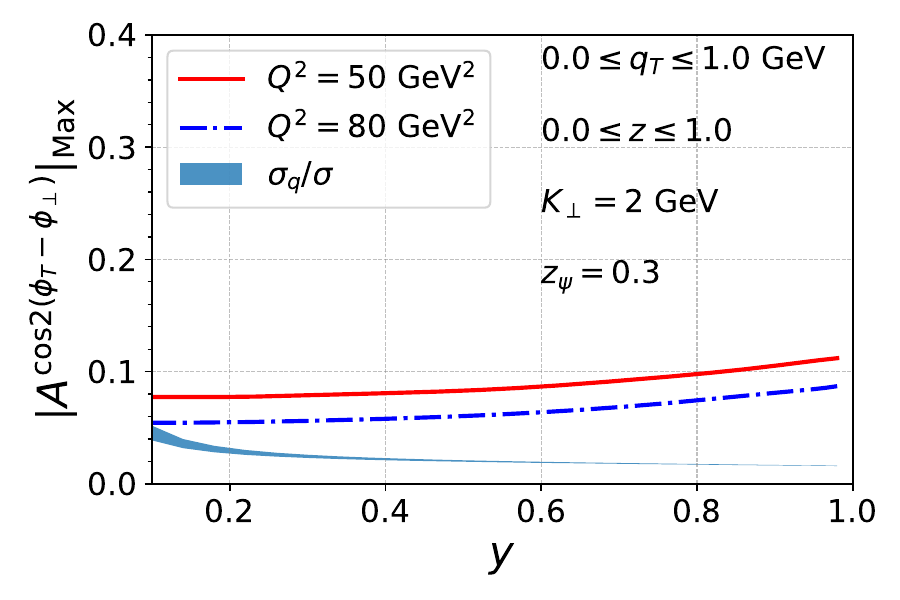}
	        \caption{}
	    \end{subfigure}
	\end{center}
 \caption{\label{fig:y_VAR_fix_z1_KP} The upper bounds for the $A^{\cos2\phi_T}$ (left panel) and $A^{\cos2(\phi_{T}-\phi_\perp)}$ (right panel) azimuthal asymmetries are shown as function of $y$ in the process $ep\rightarrow e+ J/\psi+\pi^\pm+X$ at  $\sqrt{S}=140$ GeV using CMSWZ \cite{Chao:2012iv}  LDME set. The fixed parameters include $K_\perp=2.0$ GeV, $z_\psi=0.3$ and $Q^2=50$  and $80$ GeV$^2$. Kinematic variables $z$ and $q_T$ are integrated over the interval $[0,1]$. The ratio of quark contribution to the total scattering cross-section is shown as a band, which is obtained by varying $Q^2$ from 50 to 80 GeV$^2$.} 
 \end{figure}

\begin{figure}[H]
	\begin{center} 
		\begin{subfigure}{0.49\textwidth}
		\includegraphics[height=6cm,width=7.5cm]{ 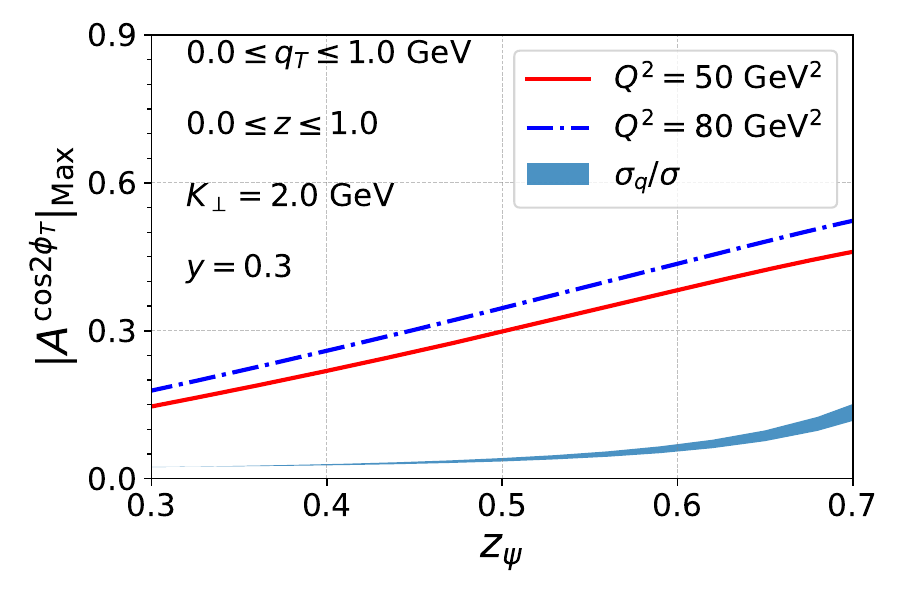}
			\caption{}
		\end{subfigure}
	    \begin{subfigure}{0.49\textwidth}
	    \includegraphics[height=6cm,width=7.5cm]{ 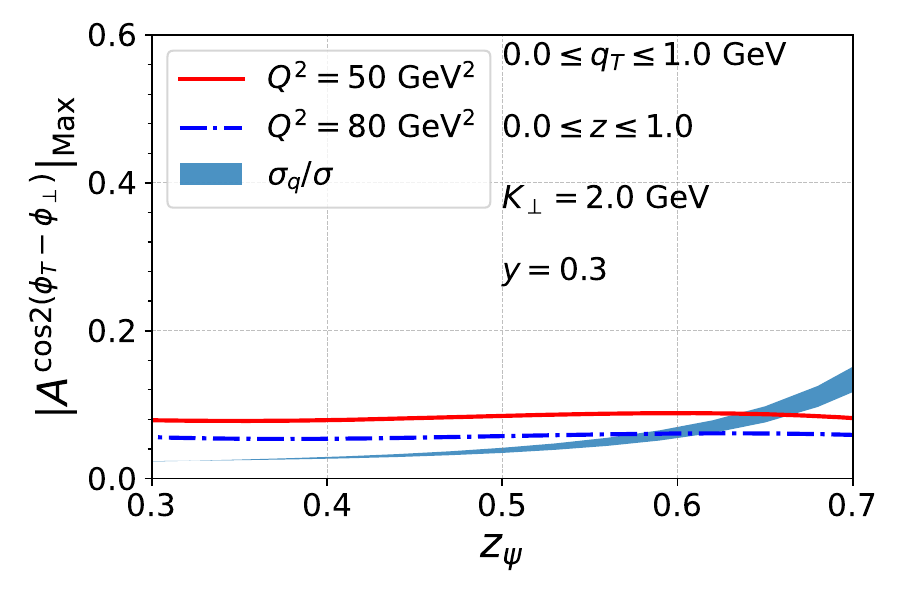}
	        \caption{}
	    \end{subfigure}
     \caption{\label{fig:z1_VAR_fix_KP_y} The upper bounds for the $A^{\cos2\phi_T}$ (left panel) and $A^{\cos2(\phi_{T}-\phi_\perp)}$ (right panel) azimuthal asymmetries are shown as function of $z_\psi$ in the process $ep\rightarrow e+ J/\psi+\pi^\pm+X$ at  $\sqrt{S}=140$ GeV using CMSWZ \cite{Chao:2012iv}  LDME set. The fixed parameters include $K_\perp=2.0$ GeV, $y=0.3$ and $Q^2=50$  and $80$ GeV$^2$. Kinematic variables $z$ and $q_T$ are integrated over the interval $[0,1]$. The ratio of quark contribution to the total scattering cross-section is shown as a band obtained by varying $Q^2$ from 50 to 80 GeV$^2$.}
     \end{center},
\end{figure}

This subsection focuses on presenting the numerical estimates of the upper bounds of azimuthal asymmetries, specifically ${\cos2\phi_T}$  and ${\cos2(\phi_{T}-\phi_\perp)}$, for $J/\psi$ and pion production. These azimuthal asymmetries are evaluated for two different virtualities of photons: $Q^2=50$ and $80$ GeV$^2$. The parameters are chosen to maximize the azimuthal asymmetries for probing the gluon TMDs such as $h_{1}^{\perp g}$. We set the cm energy to  $\sqrt{S}=140$ GeV. This is accessible at EIC; at such higher energies, the momentum fraction carried by the parton $x_a$ becomes small, leading to a significant contribution from gluons. The kinematic variables $z$ and $q_T$ are integrated over the intervals $[0, 1]$. The asymmetries receive contributions both from quark and gluon TMDs. In Figs. \ref{fig: KP_VAR_fix_z1_y}-\ref{fig:z1_VAR_fix_KP_y}, the quark channel contribution to the total unpolarized cross-section is shown as a band at a cm energy $\sqrt{S}=140$ GeV, which can be accessible at the upcoming EIC The band is obtained by varying the square of the invariant mass of the virtual photon in the range  $Q^2=[50, 80]$ GeV$^2$. From the plots, one can see that the quark contribution is significantly smaller compared to the gluon contribution in most of the kinematical region, where the asymmetries can be sizeable. Therefore, the azimuthal asymmetry in this region will mainly probe the gluon TMDs.
The estimated azimuthal asymmetries using the CMSWZ LDME set are shown in Fig.~\ref{fig: KP_VAR_fix_z1_y}-\ref{fig:z1_VAR_fix_KP_y}. 
The  upper bounds on  ${\cos2\phi_T}$  and ${\cos2(\phi_{T}-\phi_\perp)}$ as  functions of $K_{\perp}$ are shown in Fig.~\ref{fig: KP_VAR_fix_z1_y}. The bounds are evaluated at the fixed values of $y=0.3$ and $z_{\psi}=0.3$ for which the bounds are maximal. In the left panel of the figure, gluon contribution to the ${\cos2\phi_T}$ azimuthal asymmetry decreases as $K_{\perp}$ increases. This decrease is primarily due to the increase in the momentum fraction carried by the parton $x_a$ at high $K_{\perp}$, resulting in a reduction of the ${\cos2\phi_T}$ azimuthal asymmetry for gluons. The quark contribution to the cross-section, on the other hand, increases as $K_{\perp}$ increases.
In the right panel of Fig.~\ref{fig: KP_VAR_fix_z1_y}, the ${\cos2(\phi_{T}-\phi_\perp)}$ azimuthal asymmetry is depicted as a function of $K_{\perp}$. As $K_{\perp}$   increases, the gluon contribution to the ${\cos2(\phi_{T}-\phi_\perp)}$ asymmetry increases for both values of $Q^2$. The quark contribution to the cross-section also increases, however, is always less compared to the gluon contribution in the kinematical region considered. 

Fig.~\ref{fig:y_VAR_fix_z1_KP} illustrates the azimuthal asymmetries as a function of the energy fraction carried by the photon, ($y$), for fixed values of $K_{\perp}=2$ GeV and $z_{\psi}=0.3$ of $J/\psi$ for two different values of $Q^2$.
In the left panel of the figure, we observe that the ${\cos2\phi_T}$ asymmetry is quite large for small values of $y$, which decreases as  $y$ increases and eventually vanishes at $y=1$. This behavior is attributed to the coefficient ${\cal B}_0$ in the numerator of the ${\cos2\phi_T}$ azimuthal asymmetry, which vanishes at $y=1$. At $y=0.1$, the quark contribution constitutes approximately $5\%$ of the total scattering cross-section, but this contribution becomes negligible with increasing $y$. In the right panel of the figure, the ${\cos2(\phi_{T}-\phi_\perp)}$ azimuthal asymmetry exhibits less dependence on $y$ compared to ${\cos2\phi_T}$. It slightly increases for larger values of $y$. Similar to the variation with $K_{\perp}$, the ${\cos2(\phi_{T}-\phi_\perp)}$ azimuthal asymmetry is larger  for smaller values of $Q^2$.

In Fig.~\ref{fig:z1_VAR_fix_KP_y}, we show the azimuthal asymmetries as functions of the momentum fraction carried by the $J/\psi$ from the virtual photon, denoted as $z_{\psi}$, while keeping the transverse momentum fixed at $K_{\perp}=2$ GeV and the energy fraction carried by the photon fixed at $y=0.3$.
The ${\cos2\phi_T}$ azimuthal asymmetry for gluons strongly depends on $z_{\psi}$. As $z_{\psi}$ increases, the ${\cos2\phi_T}$ azimuthal asymmetry also significantly increases. For instance, it reaches approximately around $50\%$ and $45\%$ for $Q^2=80$ and 50 GeV$^2$ respectively. Moreover, the ${\cos2\phi_T}$ azimuthal asymmetry tends to increase with increasing $Q^2$. In contrast, the ${\cos2(\phi_{T}-\phi_\perp)}$ azimuthal asymmetry remains almost independent of $z_{\psi}$. As $z_{\psi}$ increases, this azimuthal asymmetry remains constant for a given value of $Q^2$.
Similar to the variations observed with $K_{\perp}$ and $y$, the ${\cos2(\phi_{T}-\phi_\perp)}$ azimuthal asymmetry is higher for lower values of $Q^2$. We also estimated the asymmetries in the recently proposed spectator model \cite{ChakrabartiPhysRevD.108.014009}, wherein the parametrization of linearly polarized and unpolarized gluon TMDs are given. For the above kinematical conditions, the estimated azimuthal asymmetry using the spectator model coincides with the upper bound as shown in Figs. \ref{fig: KP_VAR_fix_z1_y} - \ref{fig:z1_VAR_fix_KP_y}. \par
\begin{figure}[H]
	\begin{center} 
		\begin{subfigure}{0.49\textwidth}
		\includegraphics[height=6cm,width=7.5cm]{ 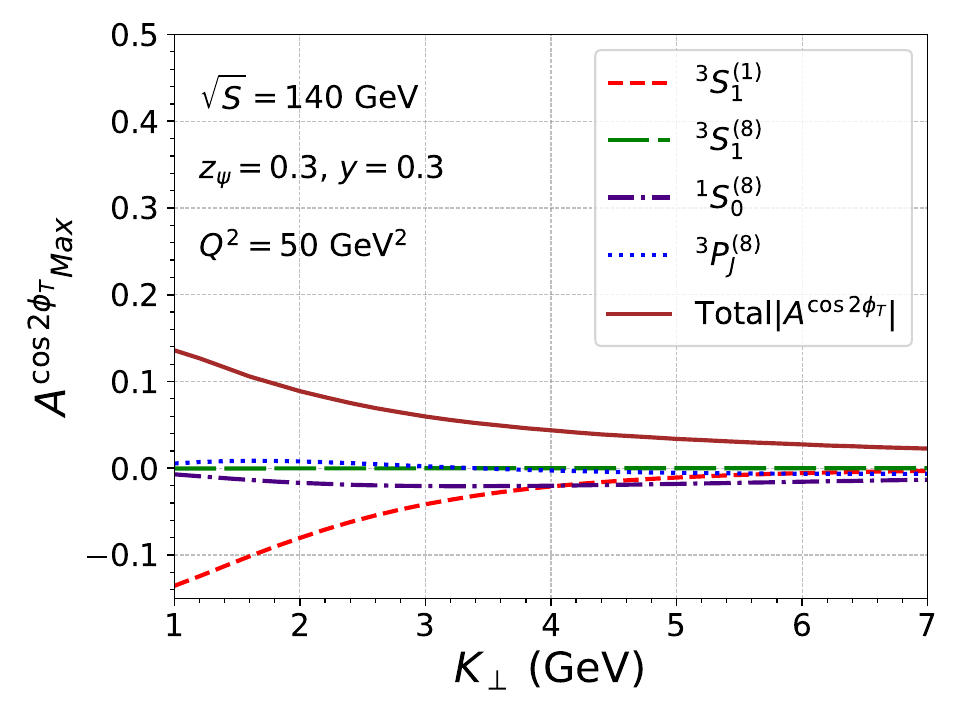}
			\caption{}
		\end{subfigure}
	    \begin{subfigure}{0.49\textwidth}
	    \includegraphics[height=6cm,width=7.5cm]{ 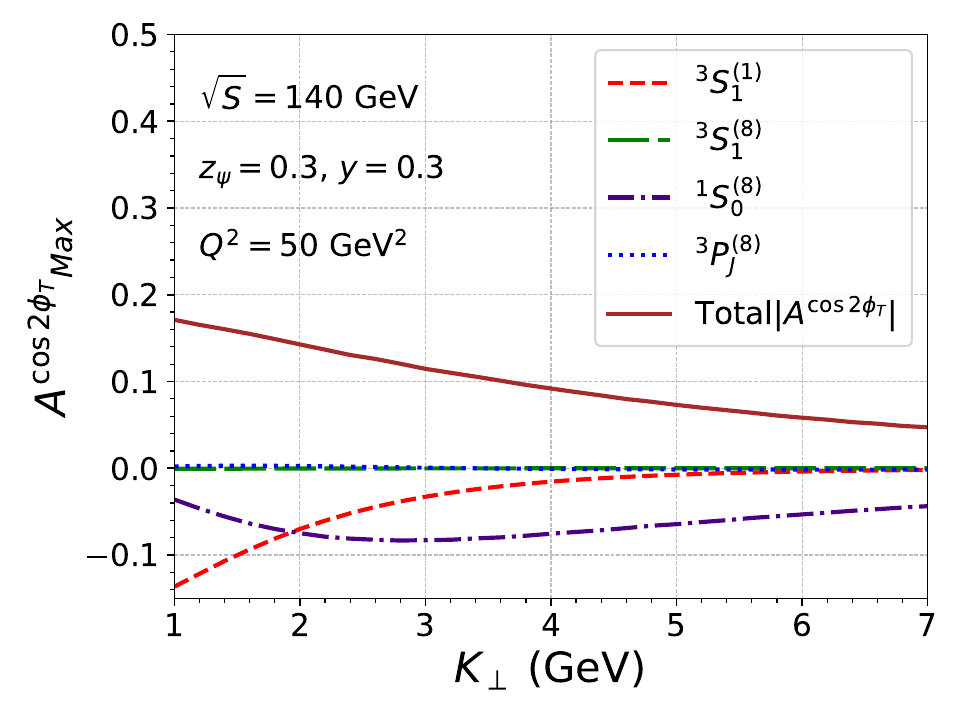}
	        \caption{}
	    \end{subfigure}
     \caption{\label{fig: UB_LDMEs} The upper bound of $A^{\cos2\phi_T}$ is shown as function of $K_\perp$ for $^{3}{S}_{1}^{(1,8)}$,  $^{1}{S}_{0}^{8}$ and $^{3}{P}_{J}^{8}$ states using LDME sets CMSWZ (left panel)\cite{Chao:2012iv} and SV (right panel)\cite{Sharma:2012dy} in the process $ep\rightarrow e+ J/\psi+\pi^\pm+X$ at $\sqrt{S}=140$ GeV. The fixed parameters include  $y=0.3$,  $z_{\psi}=0.3$ and $Q^2=50$ GeV$^2$. The variables $z$ and $q_T$ are integrated over the interval $[0,1]$.}
     \end{center}
\end{figure}

In Fig.~\ref{fig: UB_LDMEs}, we show the contribution of individual states, such as $^{3}{S}_{1}^{(1,8)}$, $^{1}{S}_{0}^{8}$, and $^{3}{P}_{J}^{8}$,  to the upper bound of ${\cos2\phi_T}$ azimuthal asymmetry for two different sets of LDMEs, namely,   CMSWZ (left)\cite{Chao:2012iv} and SV (right)\cite{Sharma:2012dy}. Among all the states, mainly $^{3}{S}_{1}^{1}$ and $^{1}{S}_{0}^{8}$ states contribute significantly to the symmetry. As seen in the plot, for the CMSWZ set, the main contribution to the asymmetry comes from $^{1}{S}_{0}^{8}$ state, while for the SV set, the main contribution to the asymmetry comes from $^{3}{S}_{1}^{1}$ state.

\subsection{Gaussian Parametrizations}

\begin{figure}[H]
	\begin{center} 
		\begin{subfigure}{0.49\textwidth}
		\includegraphics[height=6cm,width=7.5cm]{ 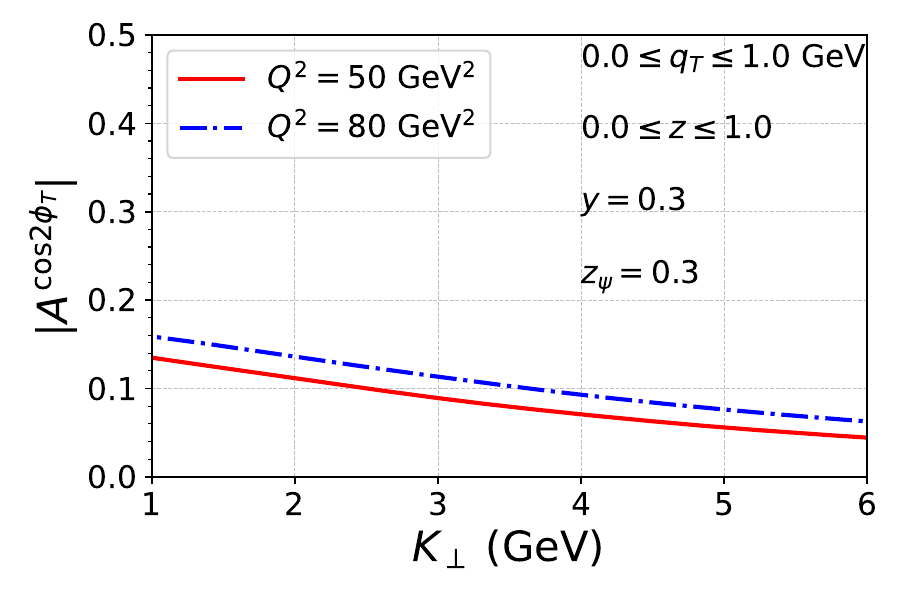}
			\caption{}
		\end{subfigure}
	    \begin{subfigure}{0.49\textwidth}
	    \includegraphics[height=6cm,width=7.5cm]{ 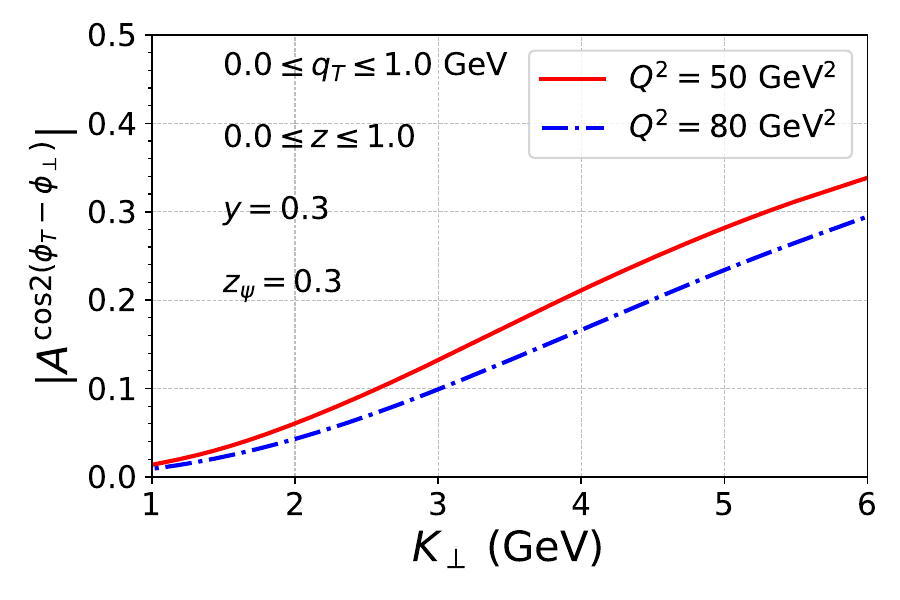}
	        \caption{}
	    \end{subfigure}
	\end{center}
 \caption{\label{fig:Gau_KP_VAR_fix_z1_y} $A^{\cos2\phi_T}$ (left panel) and $A^{\cos2(\phi_{T}-\phi_\perp)}$ (right panel) azimuthal asymmetries are shown as function of $K_\perp$ at $\sqrt{S}=140$ GeV using CMSWZ \cite{Chao:2012iv}  LDME set and Gaussian parametrization of the TMDs.  The fixed parameters include $y=0.3$, $z_\psi=0.3$ and $Q^2=50$ and $80$ GeV$^2$. Kinematic variables $z$ and $q_T$ are integrated over the interval $[0,1]$.}
 \end{figure}

\begin{figure}[H]
	\begin{center} 
		\begin{subfigure}{0.49\textwidth}
		\includegraphics[height=6cm,width=7.5cm]{ 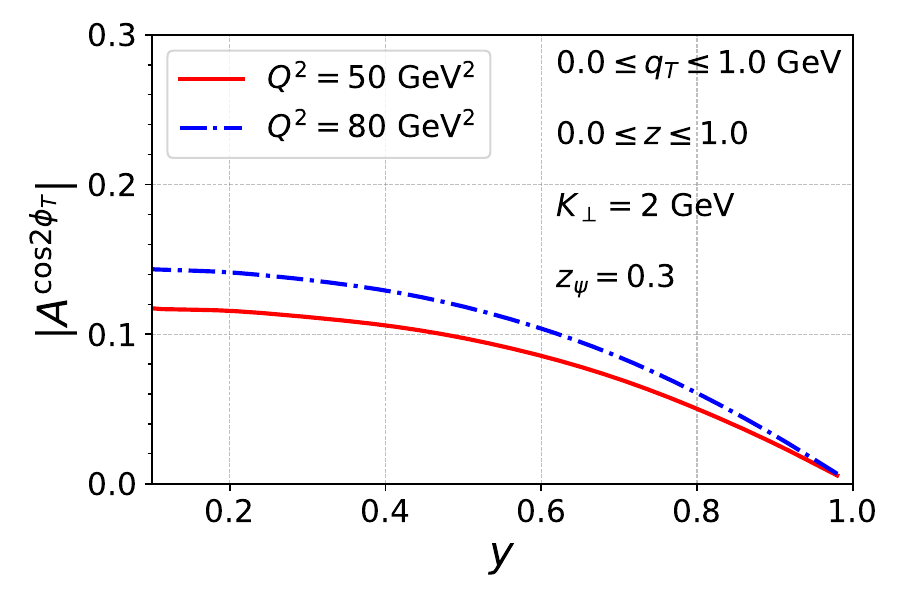}
			\caption{}
        \end{subfigure}
	    \begin{subfigure}{0.49\textwidth}
	    \includegraphics[height=6cm,width=7.5cm]{ 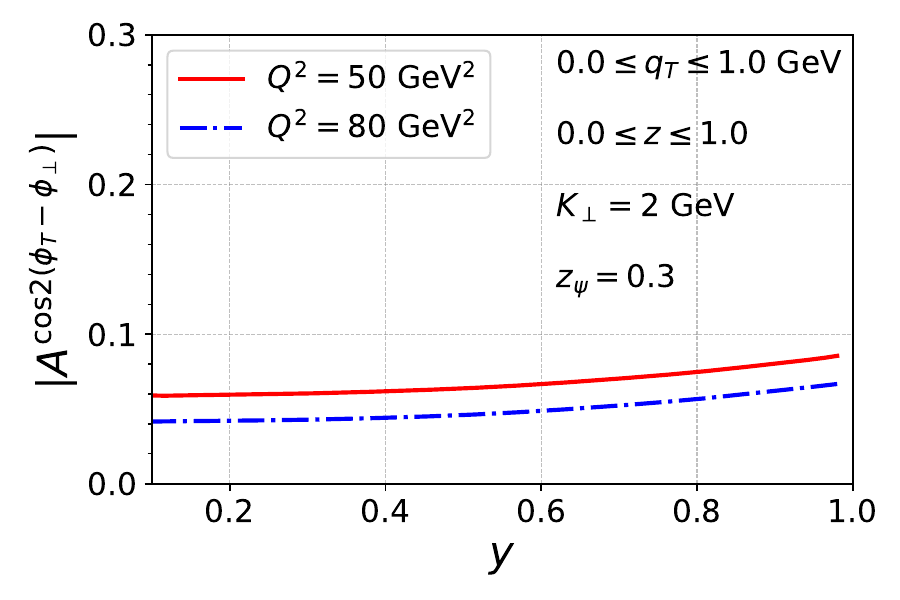}
	        \caption{}
	    \end{subfigure}
	\end{center}
 \caption{\label{fig:Gau_y_VAR_fix_z1_KP} $A^{\cos2\phi_T}$ (left panel) and $A^{\cos2(\phi_{T}-\phi_\perp)}$ (right panel) azimuthal asymmetries are shown as function of $y$ at $\sqrt{S}=140$ GeV using CMSWZ \cite{Chao:2012iv}  LDME set and Gaussian parametrization of the TMDs. The fixed parameters include $K_\perp=2.0$ GeV, $z_\psi=0.3$ and $Q^2=50$ and $80$ GeV$^2$. Kinematic variables $z$ and $q_T$ are integrated over the interval $[0,1]$.} 
  \end{figure}

\begin{figure}[H]
	\begin{center} 
		\begin{subfigure}{0.49\textwidth}
		\includegraphics[height=6cm,width=7.5cm]{ 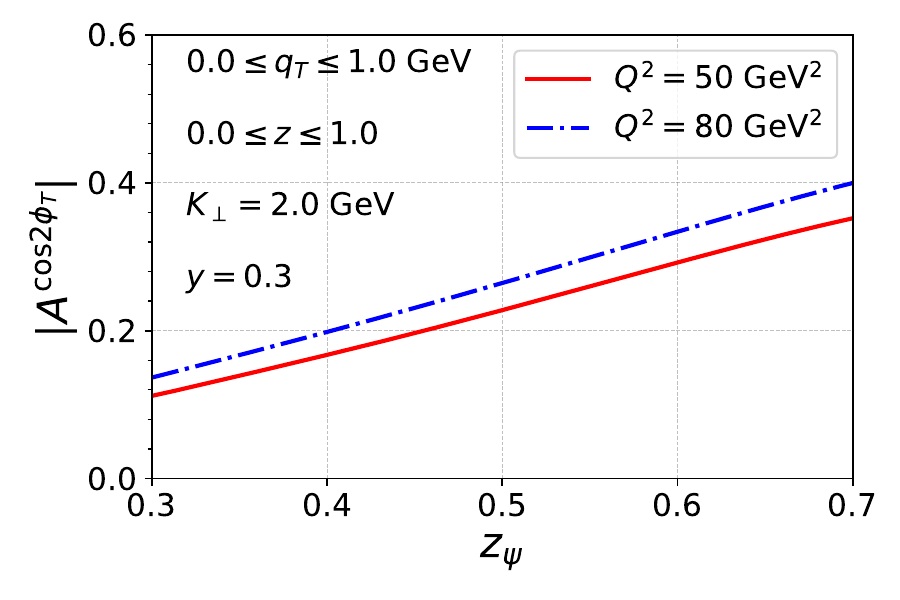}
			\caption{}
		\end{subfigure}
	    \begin{subfigure}{0.49\textwidth}
	    \includegraphics[height=6cm,width=7.5cm]{ 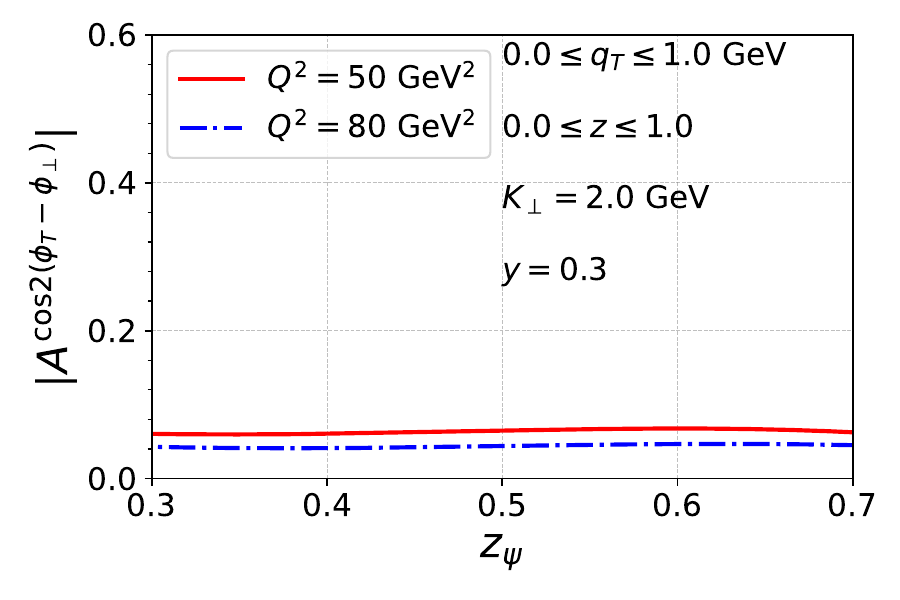}
	        \caption{}
	    \end{subfigure}
     \caption{\label{fig:Gau_z1_VAR_fix_KP_y} $A^{\cos2\phi_T}$ (left panel) and $A^{\cos2(\phi_{T}-\phi_\perp)}$ (right panel) azimuthal asymmetries are shown as a function of $z_\psi$ at $\sqrt{S}=140$ GeV using CMSWZ \cite{Chao:2012iv}  LDME set and Gaussian parametrization of the TMDs. The fixed parameters include $K_\perp=2.0$ GeV, $y=0.3$ and $Q^2=50$ and $80$ GeV$^2$. Kinematic variables $z$ and $q_T$ are integrated over the interval $[0,1]$.}
     \end{center}
\end{figure}

In this section, we esimate the ${\cos2\phi_T}$ and ${\cos2(\phi_T - \phi_\perp)}$ azimuthal asymmetries using Gaussian parametrization of the TMDs. We choose the kinematical variables to ensure that the quark contribution remains minimal compared to the gluon contribution. In Figs. \ref{fig:Gau_KP_VAR_fix_z1_y}-\ref{fig:Gau_z1_VAR_fix_KP_y}, we have shown the ${\cos2\phi_T}$ and ${\cos2(\phi_T - \phi_\perp)}$ azimuthal asymmetries as functions of $K_{\perp}$, $y$, and $z_{\psi}$, respectively. From the plots, it is seen that the qualitative behavior of the asymmetries remains the same as that in the upper-bound plots, however, the magnitude is lower than the upper bound 

\begin{figure}[H]
	\begin{center} 
		\begin{subfigure}{0.49\textwidth}
		\includegraphics[height=6cm,width=7.5cm]{ 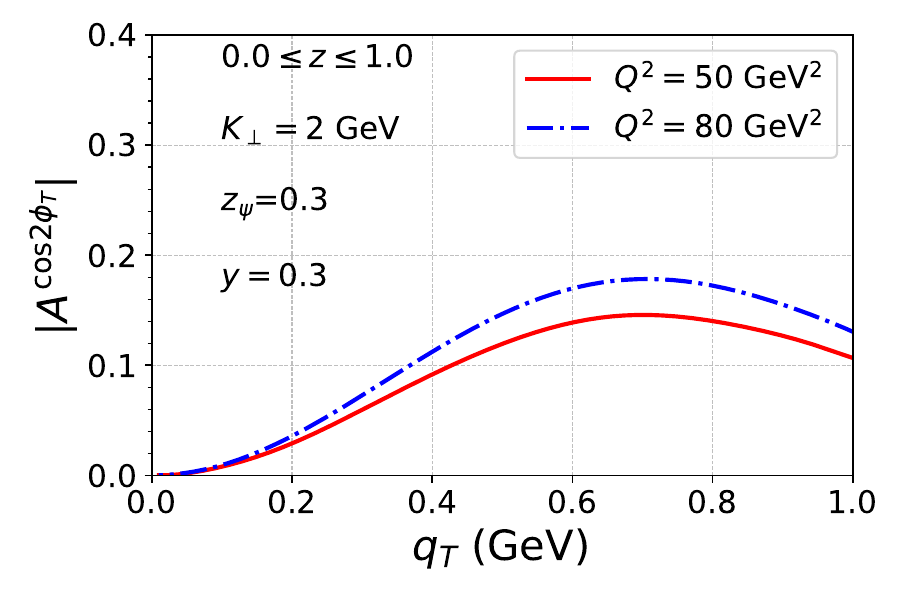}
			\caption{}
		\end{subfigure}
	    \begin{subfigure}{0.49\textwidth}
	    \includegraphics[height=6cm,width=7.5cm]{ 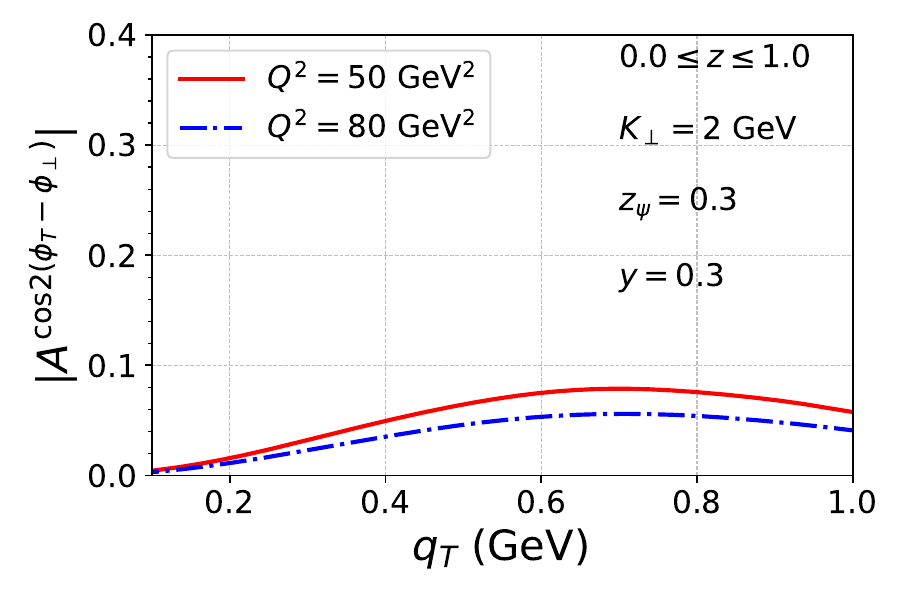}
	        \caption{}
	    \end{subfigure}
     \caption{\label{fig:qt_VAR_fix_KP_y}  $A^{\cos2\phi_T}$ (left panel) and $A^{\cos2(\phi_{T}-\phi_\perp)}$ (right panel) azimuthal asymmetries are shown as function of $q_T$ at $\sqrt{S}=140$ GeV using CMSWZ \cite{Chao:2012iv}  LDME set and Gaussian parametrization of the TMDs. The fixed parameters include $K_\perp=2.0$ GeV, $y=0.3$,  $z_{\psi}=0.3$ and $Q^2=50$ and $80$ GeV$^2$. The variable $z$ is integrated over the interval $[0,1]$.}
     \end{center}
\end{figure}

In Fig. \ref{fig:qt_VAR_fix_KP_y}, the ${\cos2\phi_T}$ and ${\cos2(\phi_{T}-\phi_\perp)}$ azimuthal asymmetries are plotted to explore their behavior as a function of transverse momentum imbalance of the final state particles,  $q_T$, which, through momentum conservation,  is related to the intrinsic transverse momentum of the initial gluon as given in Eq.\eqref{dfun2}. These azimuthal asymmetries are calculated for two virtualities of the photon, namely $Q^2=50, 80$ GeV$^{2}$, while keeping $K_{\perp}=2$ GeV, $z_{\psi}=0.3$ and $y=0.3$ fixed.  The azimuthal asymmetries first increase {as} $q_T$ increases, reach a maximum, and then decrease. The peak occurs at $q_T=0.7$ for both the azimuthal asymmetries in the kinematics chosen. The ${\cos2\phi_T}$ asymmetry is about $18\%$ for $Q^2=80$ GeV$^{2}$ at the peak, and the magnitude is larger for higher  $Q^2$. On the other hand, the peak of ${\cos2(\phi_{T}-\phi_\perp)}$, is approximately $7\%$, and this asymmetry is larger for lower values of $Q^2$.

\begin{figure}[H]
	\begin{center} 
		\begin{subfigure}{0.49\textwidth}
		\includegraphics[height=6cm,width=7.5cm]{ 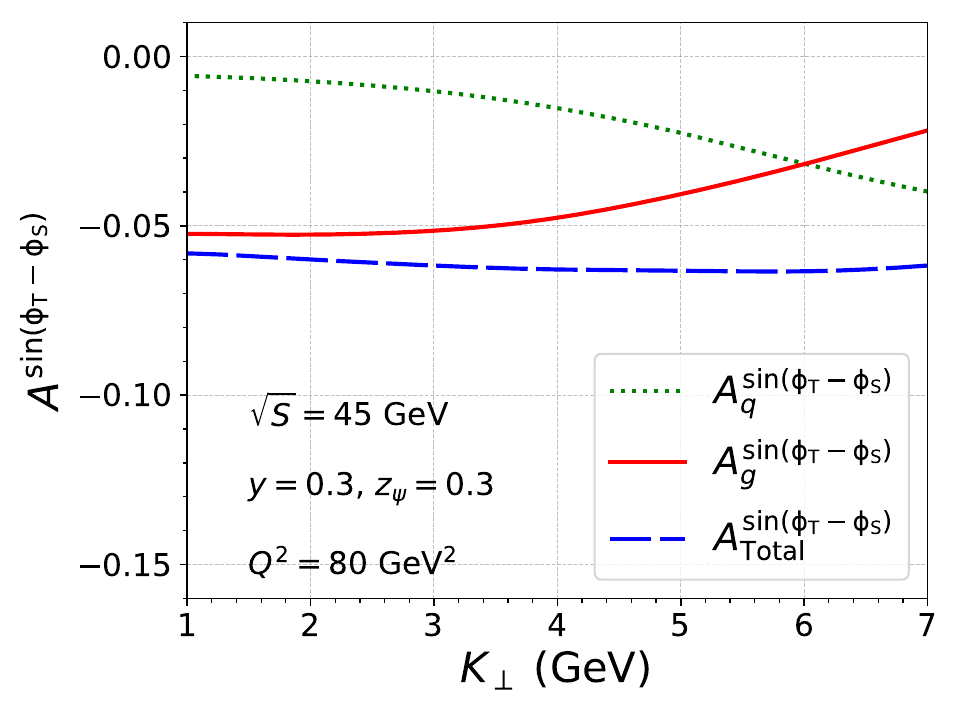}
			\caption{}
		\end{subfigure}
	    \begin{subfigure}{0.49\textwidth}
	    \includegraphics[height=6cm,width=7.5cm]{ 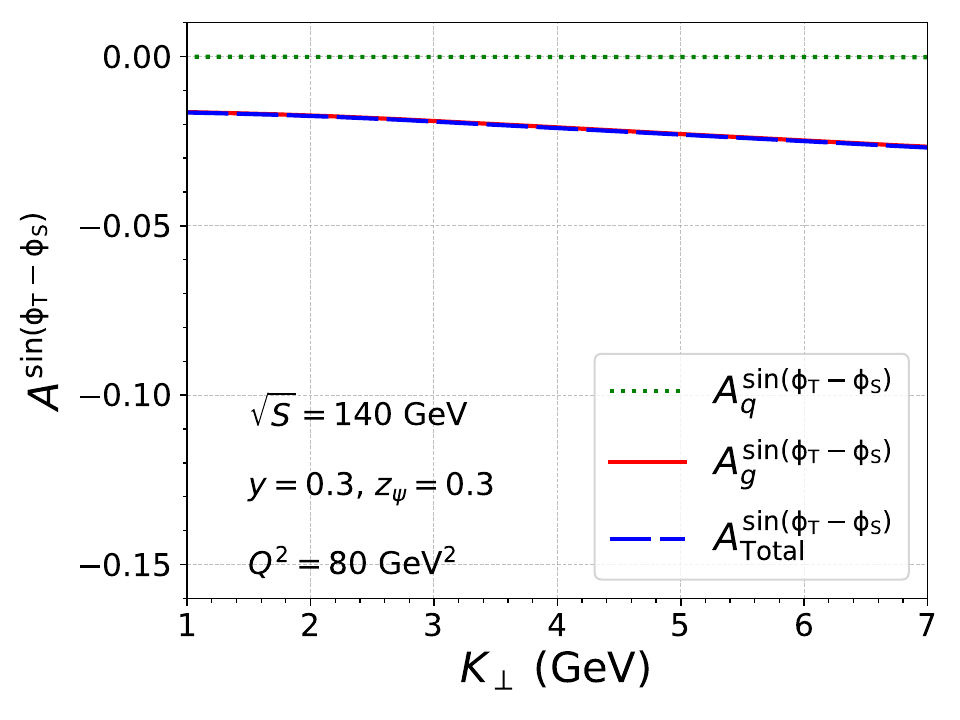}
	        \caption{}
	    \end{subfigure}
     \caption{\label{fig: Siv_KP_VAR_fix_z1_y} Sivers asymmetry is shown as function of $K_\perp$ in the process $ep\rightarrow e+ J/\psi+\pi^\pm+X$ at $\sqrt{S}=45$ GeV (left panel), $\sqrt{S}=140$ GeV (right panel) using CMSWZ \cite{Chao:2012iv}  LDME set. The fixed parameters include $y=0.3$,  $z_{\psi}=0.3$ and $Q^2=80$ GeV$^2$. The variables $z,~ q_T$ are integrated over the interval $[0,1]$.}
     \end{center}
\end{figure}

\begin{figure}[H]
	\begin{center} 
		\begin{subfigure}{0.49\textwidth}
		\includegraphics[height=6cm,width=7.5cm]{ 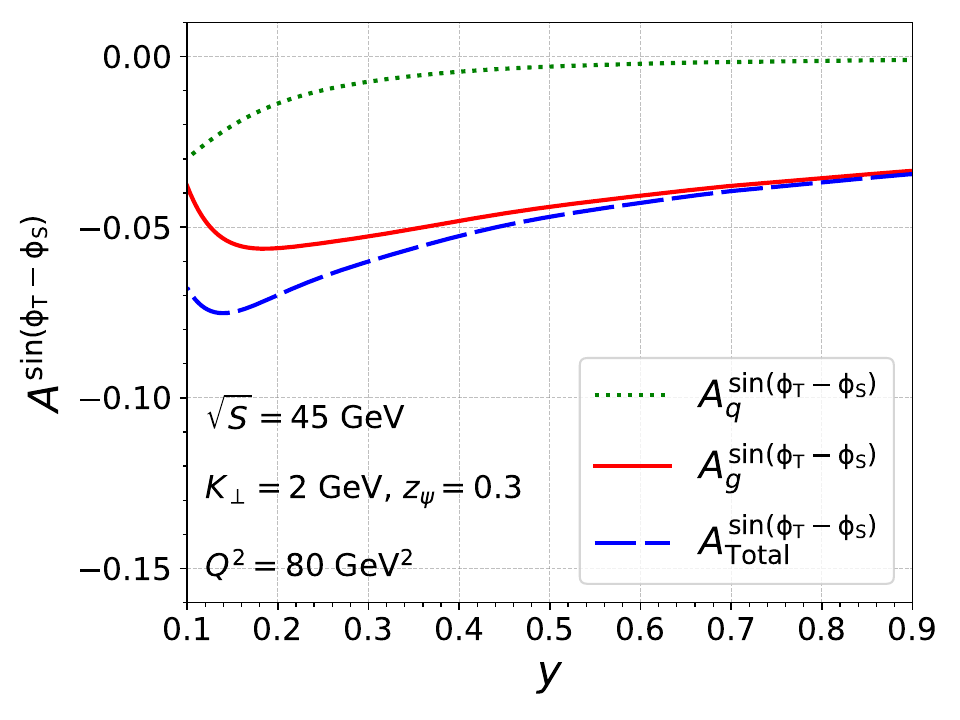}
			\caption{}
		\end{subfigure}
	    \begin{subfigure}{0.49\textwidth}
	    \includegraphics[height=6cm,width=7.5cm]{ 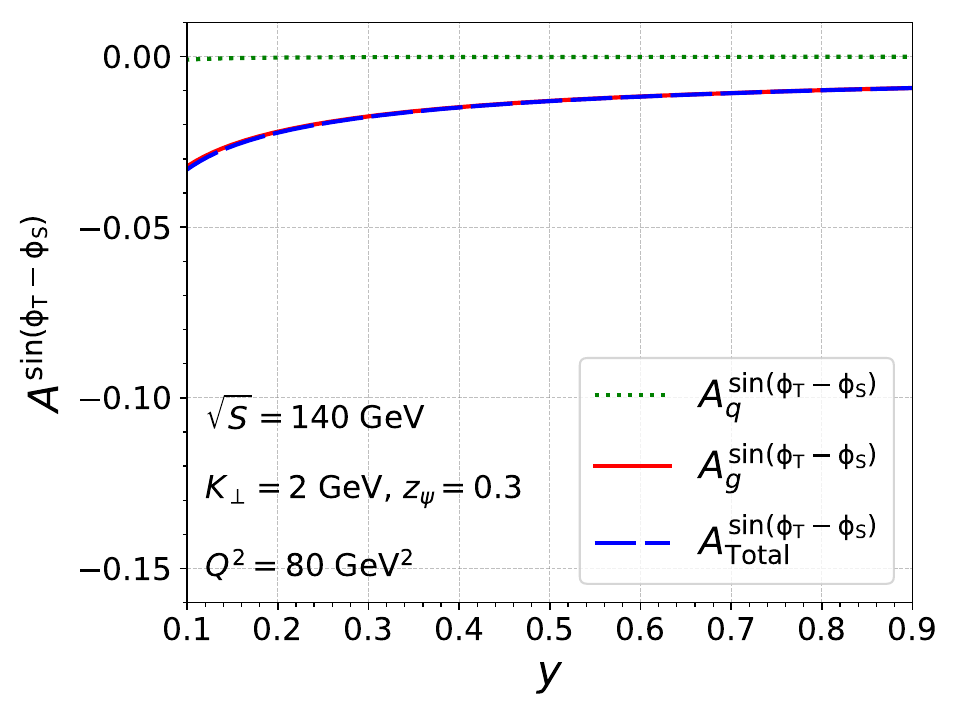}
	        \caption{}
	    \end{subfigure}
     \caption{\label{fig: Siv_y_VAR_fix_KP_z1} Sivers asymmetry is shown as function of $y$ in the process $ep\rightarrow e+ J/\psi+\pi^\pm+X$ at $\sqrt{S}=45$ GeV (left panel), $\sqrt{S}=140$ GeV (right panel) using CMSWZ \cite{Chao:2012iv}  LDME set. The fixed parameters include $K_\perp=2.0$ GeV,  $z_{\psi}=0.3$ and $Q^2=80$ GeV$^2$. The variables $z,~ q_T$ are integrated over the interval $[0,1]$.}
     \end{center}
\end{figure}

\begin{figure}[H]
	\begin{center} 
		\begin{subfigure}{0.49\textwidth}
		\includegraphics[height=6cm,width=7.5cm]{ 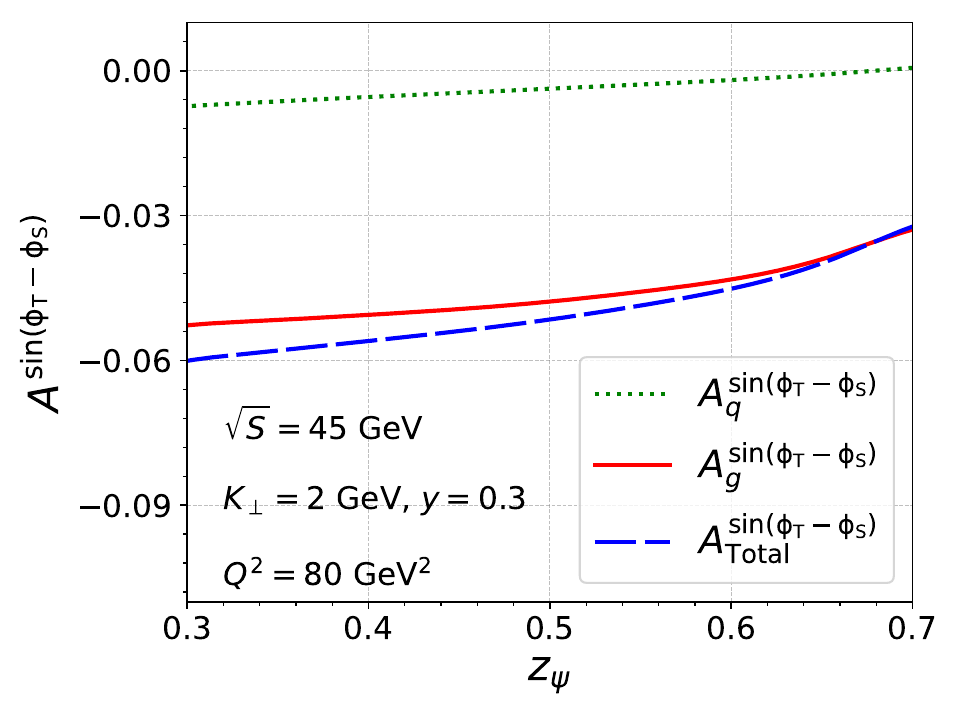}
			\caption{}
		\end{subfigure}
	    \begin{subfigure}{0.49\textwidth}
	    \includegraphics[height=6cm,width=7.5cm]{ 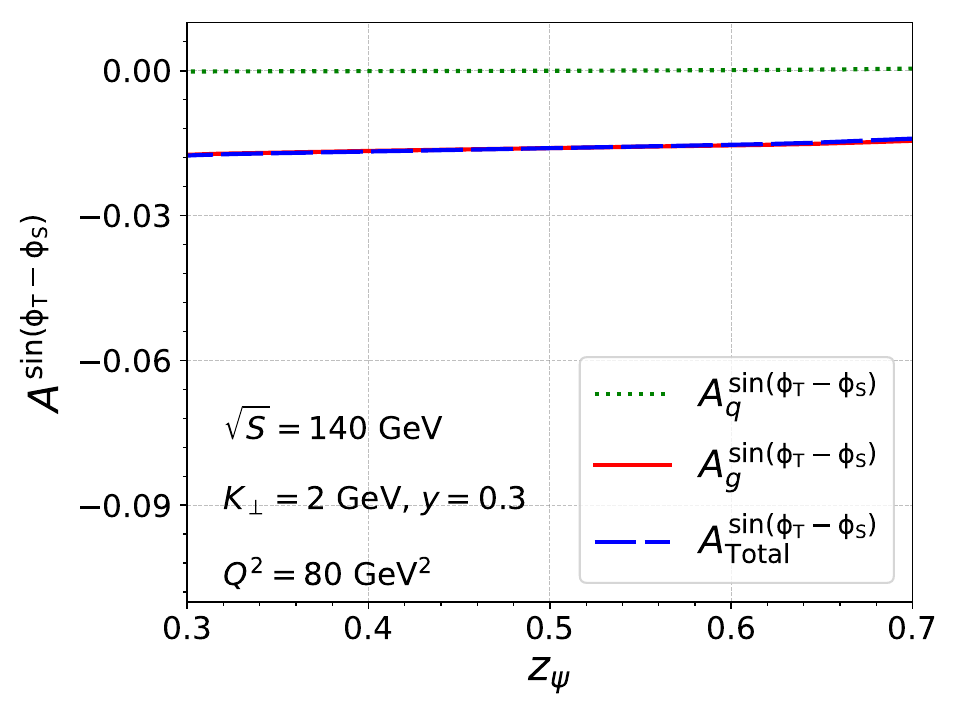}
	        \caption{}
	    \end{subfigure}
     \caption{\label{fig: Siv_z1_VAR_fix_KP_y} Sivers asymmetry is shown as function of $z_{\psi}$ in the process $ep\rightarrow e+ J/\psi+\pi^\pm+X$ at $\sqrt{S}=45$ GeV (left panel), $\sqrt{S}=140$ GeV (right panel) using CMSWZ \cite{Chao:2012iv}  LDME set. The fixed parameters include $K_\perp=2.0$ GeV,  $y=0.3$ and $Q^2=80$ GeV$^2$. The variables $z,~ q_T$ are integrated over the interval $[0,1]$.}
     \end{center}
\end{figure}

\begin{figure}[H]
	\begin{center} 
		\begin{subfigure}{0.49\textwidth}
		\includegraphics[height=6cm,width=7.5cm]{ 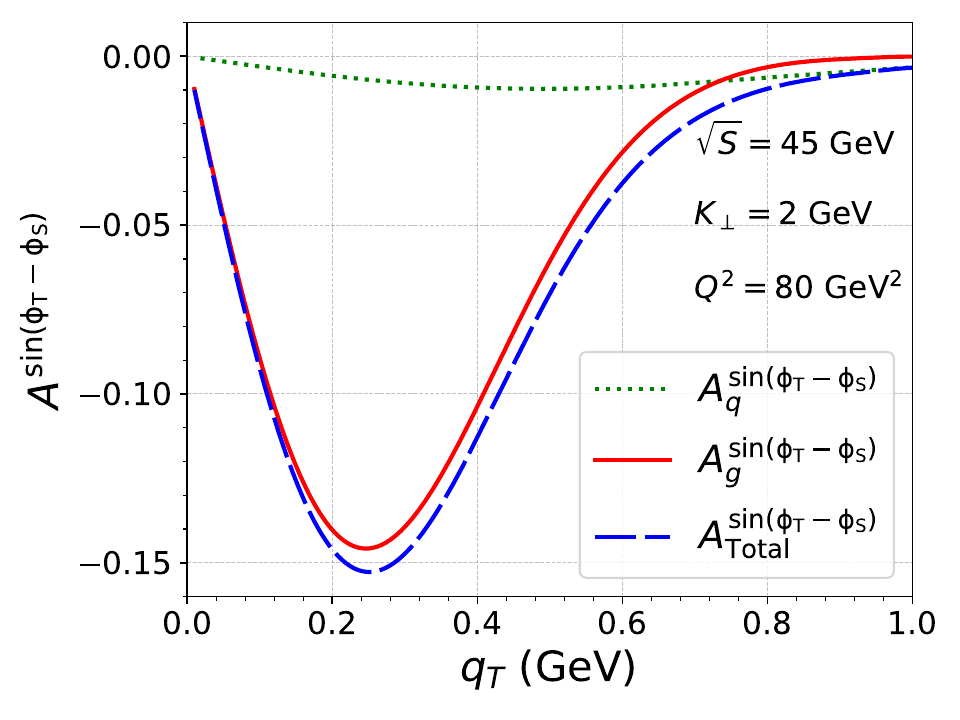}
			\caption{}
		\end{subfigure}
	    \begin{subfigure}{0.49\textwidth}
	    \includegraphics[height=6cm,width=7.5cm]{ 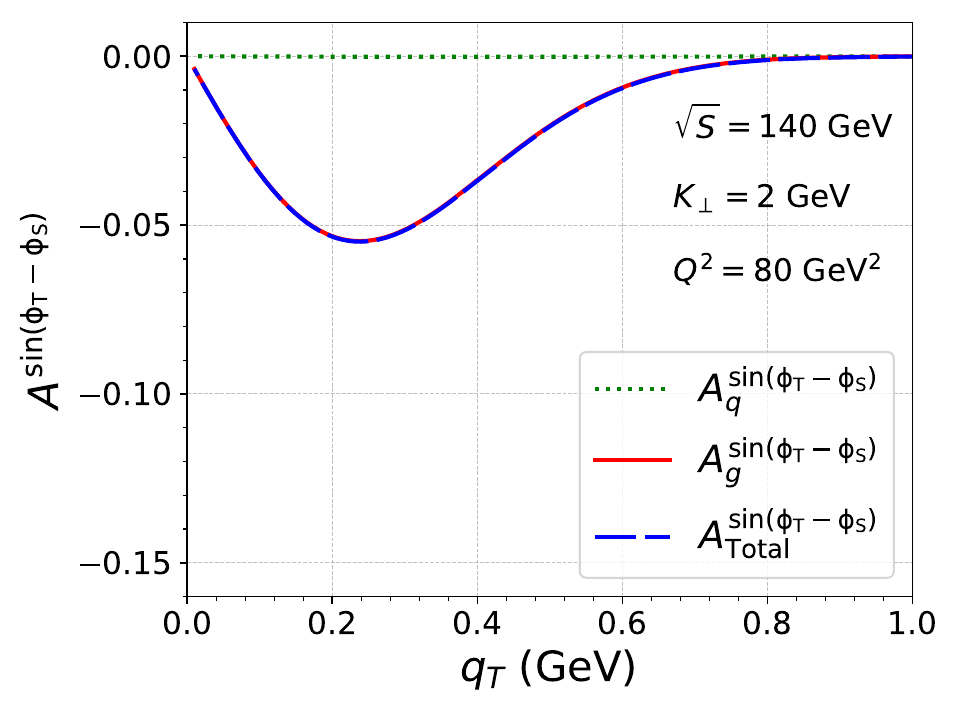}
	        \caption{}
	    \end{subfigure}
     \caption{\label{fig: Siv_qt_VAR_fix_KP_y} Sivers asymmetry is shown as function of $q_T$ in the process $ep\rightarrow e+ J/\psi+\pi^\pm+X$ at $\sqrt{S}=45$ GeV (left panel), $\sqrt{S}=140$ GeV (right panel) using CMSWZ \cite{Chao:2012iv}  LDME set. The fixed parameters include $K_\perp=2.0$ GeV, $y=0.3$,  $z_{\psi}=0.3$ and $Q^2=80$ GeV$^2$. The variable $z$ is integrated over the interval $[0,1]$.}
     \end{center}
\end{figure}

In Figs. \ref{fig: Siv_KP_VAR_fix_z1_y}-\ref{fig: Siv_qt_VAR_fix_KP_y}, we present the Sivers asymmetry, $A^{\sin(\phi_S-\phi_T)}$, for two different cm energies, specifically $\sqrt{S}=45$ GeV (left panel) and $\sqrt{S}=140$ GeV (right panel), which can be accessed at the upcoming EIC. {We have used the Gaussian parameterizations for the gluon and quark Sivers functions given in Eq.\eqref{eq:Sivers:par} and Eq.\eqref{eq:qSivers:par}.}The asymmetries are evaluated using CMSWZ \cite{Chao:2012iv} sets of LDMEs at fixed values of $Q^2=80$ GeV$^2$, $K_\perp=2$ GeV $y=0.3$, and $z_{\psi}=0.3$, while $q_T$ is integrated in the range  [0,1] GeV. The dotted and solid lines represent quark and gluon contributions, respectively, while the dashed line indicates the sum of quark and gluon contributions to the asymmetry. In all plots, the quark channel contribution is practically insignificant compared to the gluon channel. For higher cm energy, $\sqrt{S}=140$ GeV, the quark contribution is effectively zero.  {
The Sivers asymmetry is observed to be negative and slightly more pronounced at $\sqrt{S}=45$ GeV compared to $\sqrt{S}=140$ GeV. This behavior is primarily attributed to the $\mathcal{N}_g(x)$ term in the gluon Sivers function, as described in  Eq.\eqref{eq:Sivers:Ng}.
The $\mathcal{N}_g(x)$ term inversely depends on the cm energy through $x_g$ as given in Eq.\eqref{dfun2}.
As $\sqrt{S}$ increases, $x_g$ decreases, which in turn reduces  $\mathcal{N}_g(x)$ in the gluon Sivers function. Consequently, this reduction in $\mathcal{N}_g(x)$ leads to a decrease in the Sivers asymmetry at higher 
$\sqrt{S}$ values. In Fig. \ref{fig: Siv_KP_VAR_fix_z1_y} , for $\sqrt{S}=45$ GeV, as $K_{\perp}$ increases, the value of $x_a$ also increases as given in Eq.\eqref{dfun2}. This results in a decrease in the gluon contribution and a corresponding increase in the quark contribution. Consequently, the total Sivers asymmetry remains relatively constant with variations in $K_{\perp}$.} Furthermore, at  $\sqrt{S}=140$ GeV the dependence of the asymmetry on the kinematical variables is less pronounced, especially for $K_\perp$, $y$, and $z_{\psi}$.   Sivers asymmetry is mainly dominated by the gluon channel in the low $K_\perp$ region, particularly $K_\perp \ll 6$ GeV. In Fig. \ref{fig: Siv_y_VAR_fix_KP_z1}, Sivers asymmetry as a function of $y$ is presented, which decreases with increasing $y$ and it reaches a maximum of about 7$\%$ at low $y$ region. Sivers asymmetry as a function of $z_\psi$ is shown in Fig. \ref{fig: Siv_z1_VAR_fix_KP_y}. Finally, in Fig. \ref{fig: Siv_qt_VAR_fix_KP_y}  the asymmetry is plotted against $q_T$. Here, the asymmetry at first increases in magnitude reaches a maximum, and then {decreases}.   The maximum value of the asymmetry occurs at $q_T=0.24$ GeV; notably, this peak remains consistent across different cm energies, showing independence from energy variations.

\section{Conclusion}\label{sec5}
We have presented a calculation of the azimuthal asymmetries, particularly $\cos2\phi_T$, $\cos2(\phi_T-\phi_{\perp})$, and Sivers asymmetries, in almost back-to-back production of $J/\psi$ and $\pi^\pm$ in the process $e+p \rightarrow e+J/\psi + \pi^\pm +X$.  We considered the kinematics of the upcoming EIC.  In this kinematical region, TMD factorization is expected to be valid. We calculated the $J/\psi$ production using NRQCD factorization in terms of LDMEs and ignoring the effects of  TMD shape functions  and soft functions, at leading order. We considered the formation of the pion through fragmentation of the final state parton.  We estimated the  {contribution} to the cross-section and asymmetries both from quark and gluon channels and showed that in the kinematics to be accessed by EIC, {the} contribution from virtual photon-gluon fusion is dominant.  We presented numerical estimates of the upper bounds on the asymmetries by saturating the positivity bound. 
We also calculated them using a Gaussian parametrization of the gluon TMDs and a more recent spectator model-based parametrization.   The asymmetries in the Gaussian parameterization satisfy the positivity bound, but the spectator model saturates the bound. The asymmetries depend on the LDME sets used in NRQCD.  For an unpolarized proton, the estimated weighted  $\cos2\phi_T$ and $\cos2(\phi_T-\phi_{\perp})$  asymmetries can reach a maximum of about $40\%$ within the considered kinematical region.  The estimated Sivers asymmetry is about $14\%$ and $5\%$ for $\sqrt{S}=45~$GeV and $\sqrt{S}=140~$GeV respectively when the proton is transversely polarized. Thus, an almost back-to-back $J/\psi$ and $\pi^\pm$ production at EIC will be  a potential channel to probe the gluon TMDs
\section*{Acknowledgments}
A part of this work was done at the International Centre for Theoretical Sciences (ICTS) during the  International School and Workshop on Probing Hadron Structure at the Electron-Ion Collider (ICTS/QEICIII2024/01).  We also thank A. Bacchetta, C. Pisano, A. Signori, M. Radici and L. Maxia for helpful discussions during this workshop.  A.M. would like to thank SERB MATRICS (MTR/2021/000103) for funding. The work of S.R. is supported by a SEED grant provided by VIT University, Vellore, Grant No. SG20230031. 

\appendix
\section{Amplitude modulations }\label{appen}
Analytical expressions of different modulations are given below:
\begin{align} \mathcal{A}_0=\langle0|\mathcal{O}_{J/\psi}^{^3S_1[1]}|0\rangle~\mathcal{A}_0^{^3S_1[1]} + \langle0|\mathcal{O}_{J/\psi}^{^3S_1[8]}|0\rangle~\mathcal{A}_0^{^3S_1[8]} +\langle0|\mathcal{O}_{J/\psi}^{^1S_0[8]}|0\rangle~\mathcal{A}_0^{^1S_0[8]} + \langle0|\mathcal{O}_{J/\psi}^{^3P_0[8]}|0\rangle~\mathcal{A}_0^{^3P_J[8]}\\
\mathcal{B}_{0,2}=\langle0|\mathcal{O}_{J/\psi}^{^3S_1[1]}|0\rangle~\mathcal{B}_{0,2}^{^3S_1[1]} + \langle0|\mathcal{O}_{J/\psi}^{^3S_1[8]}|0\rangle~\mathcal{B}_{0,2}^{^3S_1[8]} + \langle0|\mathcal{O}_{J/\psi}^{^1S_0[8]}|0\rangle~\mathcal{B}_{0,2}^{^1S_0[8]} + \langle0|\mathcal{O}_{J/\psi}^{^3P_0[8]}|0\rangle~\mathcal{B}_{0,2}^{^3P_J[8]}
\end{align}

\subsection{Gluon channel}
\begin{align}
    \mathcal{A}_0^{^3S_1[1]} &=\frac{1024 Q^2}{y^2 (Q^2+s)^2 (Q^2+s+t)^2 (2 Q^2+s+u)^2 \left(Q^2+t+u\right)^2}  \Bigg\{2 Q^2 (y-1) \Big[u^2 \left(Q^2+s\right) (t^2 \left(Q^2+s\right) 
    +\left(Q^2+s\right)^3\nonumber\\
        &+4 t^3) +2 u \left(t^2 \left(Q^2+s\right)^2+2 t^3 \left(2 Q^2+s\right)+Q^2 \left(Q^2+s\right)^3\right)\left(Q^2+s+t\right)+(t^2 \left(Q^2+s\right)^2+Q^4 \left(Q^2+s\right)^2\nonumber\\
        &+4 Q^2 t^3)(Q^2+s+t)^2\big]-((y-2) y+2) \Big[Q^{10} \left(2 \left(5 s^2+14 s t+6 t^2\right)+14 u (s+t)+3 u^2\right)+Q^8 (s^2 (43 t+26 u)\nonumber\\
    &+10 s^3+13 s (t+u) (3 t+u)+(t+u) \left(8 t^2+7 t u+u^2\right))+Q^6 (u \left(73 s^2 t+24 s^3+44 s t^2+7 t^3\right)+47 s^2 t^2\nonumber\\
    &+31 s^3 t
    +5 s^4+19 s t^3+u^3 (4 s+t)+u^2 (s+t) (22 s+5 t)+4 t^4)+Q^4 \Bigg(s^3 (26 t^2+47 t u+18 u^2)\nonumber\\
    &+s^2 (3 t+2 u) (5 t^2+12 t u
    +3 u^2)+s^4 (10 t+11 u)+s^5+s t \left(13 t^2 u+4 t^3+13 t u^2+3 u^3\right)\nonumber\\
    &+t^2 (t+u) \left(2 t^2+2 t u+u^2\right)\Bigg)+Q^2 s (s^3 (7 t^2+13 t u+7 u^2)+s^2 (20 t^2 u+5 t^3+17 t u^2+4 u^3)\nonumber\\
    &+s^4 (t+2 u)+s t u \left(7 t^2+11 t u+3 u^2\right)+2 t^2 u (t+u)^2)+Q^{12} (5 s+7 t+3 u)+Q^{14}\nonumber\\
    &+s^2 (s+t+u)(s^2 (t^2 +t u+u^2)+s t u (t+u)+t^2 u^2)\Big]\Bigg\}
    \end{align}
   \begin{align}
    \mathcal{A}_0^{^3S_1[8]} = \frac{15}{8}    \mathcal{A}_0^{^3S_1[1]} 
      \end{align}
    \begin{align}
        \mathcal{A}_0^{^1S_0[8]}&=\frac{512 Q^2 }{y^2 \left(Q^2+s\right)^2 \left(Q^2+s+t\right)^2 \left(2 Q^2+s+u\right)^2 \left(Q^2+t+u\right)^2}\Bigg\{4 Q^2 (y-1) \Bigg(u^2 \Big(t^2 \left(8 Q^2 s+7 Q^4+3 s^2\right)+4 t (4 Q^2 s\nonumber\\
        &+6 Q^4+s^2) \left(Q^2+s\right)+\left(8 Q^2 s+19 Q^4+s^2\right) \left(Q^2+s\right)^2\Big)+2 u \left(Q^2+s+t\right) \Big(Q^4 \left(6 s^2+11 s t+2 t^2\right)+Q^2 s (s^2+\nonumber\\
        &6 s t+3 t^2)+Q^6 (13 s+8 t)+8 Q^8+s^2 t (s+t)\Big)+\left(Q^2+s+t\right)^2 (Q^4 \left(s^2+6 s t+t^2\right)+4 Q^6 (s+t)+2 Q^2 s t (s+t)\nonumber\\
        &+5 Q^8+s^2 t^2)+2 u^3 \left(Q^2+s\right) \left(3 Q^2 (2 s+t)+5 Q^4+s (s+t)\right)+2 u^4 \left(Q^2+s\right)^2\Bigg)-\frac{((y-2) y+2) }{t}\Bigg[(Q^2 (s+t\nonumber\\
        &+u)+Q^4+s u) \Bigg(Q^8 \left(60 s^2+62 s t+60 s u+18 t^2+34 t u+15 u^2\right)+Q^6 \Big(78 s^2 (t+u)+54 s^3+s (42 t^2+86 t u+\nonumber\\
        &42 u^2)+22 t^2 u+6 t^3+20 t u^2+6 u^3\Big)+Q^4 \Big(s^2 \left(41 t^2+82 t u+42 u^2\right)+50 s^3 (t+u)+28 s^4+2 s (t+u)^2 (6 t+7 u)\nonumber\\
        &+7 t^2 u^2+4 t^3 u+t^4+4 t u^3+u^4\Big)+2 Q^2 s (8 s^3 (t+u)+9 s^2 (t+u)^2+4 s^4+s (13 t^2 u+4 t^3+13 t u^2+5 u^3)+\nonumber\\
        &(t+u)^2 (t^2+t u+u^2))+2 Q^{10} (18 s+10 t+9 u)+9 Q^{12}+s^2 (2 s^3 (t+u)+3 s^2 (t+u)^2+s^4+2 s (t+u)^3\nonumber\\
        &+(t^2+t u+u^2)^2)\Bigg)\Bigg]\Bigg\}
  \end{align}
    \begin{align}
\mathcal{B}_{0}^{^3S_1[1]}&=-\frac{512 Q^2}{y^2 \left(Q^2+s+t\right)^2 \left(2 Q^2+s+u\right)^2 \left(Q^2+t+u\right)^2}   (y-1) \Bigg\{-K_\perp^2 \left(Q^2+s\right) \Big(Q^2 \Big(13 s^2+20 s t+16 s u+7 t^2\nonumber\\
&+14 t u+3 u^2\Big)+2 Q^4 (9 s+7 t+5 u)+8 Q^6+(s+t+u) (u (3 s+4 t)+3 s (s+t))\Big)+2 K_\perp^4 \left(Q^2+s\right)^2(Q^2+s\nonumber\\
&+t+u)+t \Big(Q^4 (13 s^2+22 s t+34 s u+9 t^2+26 t u+13 u^2)+Q^2 \Big(u^2 (19 s+13 t)+u (s+t) (19 s+9 t)+(s+t)^2 \nonumber\\
&(3 s+t)+3 u^3\Big)+2 Q^6 (9 s+8 t+9 u)+8 Q^8+u (s+t+u)(u (3 s+2 t)+3 s (s+t))\Big)\Bigg\}
\end{align}
\begin{align}
\mathcal{B}_{0,2}^{^3S_1[8]}=\frac{15}{8}\mathcal{B}_{0,2}^{^3S_1[1]}
\end{align}
\begin{align}
    \mathcal{B}_0^{^1S_0[8]}&=-\frac{128 Q^2}{t^2 \Big(Q^2+s+t\Big)^2 \Big(2 Q^2+s+u\Big)^2 \Big(Q^2+t+u\Big)^4 y^2}  \Big\{4 \Big(Q^2+s\Big)^2 \Big(Q^2+u\Big) \Big(Q^2+t+u\Big)^2 \Big(Q^4+(t+u) Q^2\nonumber\\
    &+s t\Big) K_\perp^4+\Big(Q^2+s\Big) t \Big(28 Q^{12}+(64 s+87 t+96 u) Q^{10}+\Big(56 s^2+165 t s+108 t^2+124 u^2+40 (5 s+6 t) u\Big) Q^8\nonumber\\
    &+\Big(24 s^3+(135 t+152 u) s^2+2 \Big(87 t^2+193 u t+116 u^2\Big) s+70 t^3+72 u^3+238 t u^2+232 t^2 u\Big) Q^6+\Big(4 s^4+(69 t\nonumber\\
    &+56 u) s^3+4 \Big(29 t^2+60 u t+36 u^2\Big) s^2+4 \Big(26 t^3+75 u t^2+78 u^2 t+30 u^3\Big) s+8 (t+u) \Big(3 t^3+10 u t^2+10 u^2 t\nonumber\\
    &+2 u^3\Big)\Big) Q^4+\Big(8 (3 t+u) s^4+\Big(42 t^2+74 u t+40 u^2\Big) s^3+(t+u) \Big(51 t^2+83 u t+56 u^2\Big) s^2+2 (t+u) \Big(17 t^3\nonumber\\
    &+44 u t^2+35 u^2 t+12 u^3\Big) s+t (t+u)^2 \Big(3 t^2+10 u t+11 u^2\Big)\Big) Q^2+s \Big(4 t s^4+4 (t+u) (2 t+u) s^3+(t+u) \Big(9 t^2\nonumber\\
    &+9 u t+8 u^2\Big) s^2+2 (t+u)^2 \Big(5 t^2+7 u t+4 u^2\Big) s+t (t+u)^2 (3 t+u) (t+3 u)\Big)\Big) K_\perp^2-t \Big(Q^4+(s+t+u) Q^2\nonumber\\
    &+s u\Big) \Big(16 Q^{12}+(48 s+68 t+48 u) Q^{10}+\Big(52 s^2+4 (43 t+34 u) s+111 t^2+52 u^2+168 t u\Big) Q^8+2 \Big(12 s^3+\nonumber\\
    &(78 t+68 u) s^2+\Big(111 t^2+188 u t+68 u^2\Big) s+2 (t+u)^2 (25 t+6 u)\Big) Q^6+\Big(4 s^4+(60 t+56 u) s^3+\Big(169 t^2+\nonumber\\
    &288 u t+120 u^2\Big) s^2+2 \Big(69 t^3+173 u t^2+138 u^2 t+28 u^3\Big) +2 (t+u) \Big(31 t^3+57 u t^2+26 u^2 t+2 u^3\Big)\Big) Q^4+\nonumber\\
    &2 \Big(4 (t+u) s^4+\Big(30 t^2+44 u t+20 u^2\Big) s^3+\Big(37 t^3+91 u t^2+78 u^2 t+20 u^3\Big) s^2+(t+u) \Big(25 t^3+49 u t^2+36 u^2 t\nonumber\\
    &+4 u^3\Big) s+4 t (t+u)^2 \Big(3 t^2+5 u t+u^2\Big)\Big) Q^2+t^2 (t+u)^3 (3 t+7 u)+4 s^4 \Big(2 t^2+2 u t+u^2\Big)+4 s^3 (t+u) \Big(4 t^2\nonumber\\
        &+5 u t+2 u^2\Big)+2 s t (t+u)^2\Big(5 t^2+7 u t+4 u^2\Big)+s^2 (t+u) \Big(13 t^3+21 u t^2+20 u^2 t+4 u^3\Big)\Big)\Big\} (y-1)
\end{align}
\begin{align}
\mathcal{B}_{2}^{^3S_1[1]}&=\frac{512 K_\perp^2 Q^2}{y^2 \left(Q^2+s+t\right)^2 \left(2 Q^2+s+u\right)^2 \left(Q^2+t+u\right)^2}  \Bigg\{-Q^4 \Big(3 s^2 ((y-6) y+6)+4 s (y-2)^2 (t+u)\nonumber\\
&-t^2 ((y-10) y+10)
+2 t u ((y-2) y+2)+u^2 ((y-2) y+2)\Big)-Q^2 (s+t+u) \Big(s^2 (y (3 y-10)+10)\nonumber\\
&+2 s ((y-2) y+2) (t+u)
-2 t^2 ((y-6) y+6)\Big)-Q^6 (s ((y-14) y+14)+((y-6) y+6) (t+u))\nonumber\\
&+4 Q^8 (y-1)-s^2 ((y-2) y+2) (s+t+u)^2\Bigg\}
\end{align}

\begin{align}
        \mathcal{B}_2^{^1S_0 [8]}&=\frac{256  K_\perp^2 Q^2}{t^2 y^2 \Big(Q^2+s+t\Big)^2 \Big(2 Q^2+s+u\Big)^2 \Big(Q^2+t+u\Big)^2} \Big\{Q^8 \Big(13 s^2 ((y-2) y+2)+8 s t (y (2 y-7)+7)\nonumber\\
        &+34 s u ((y-2) y
        +2)+6 t^2 ((y-6) y+6)+20 t u (y-2)^2+13 u^2 ((y-2) y+2)\Big)+2 Q^6 \Big(s^2 (t (y (5 y-14)+14)\nonumber\\
        &+17 u ((y-2) y+2))
        +3 s^3 ((y-2) y+2)+s \Big(2 t^2 ((y-8) y+8)+t u (y (19 y-66)+66)\nonumber\\
        &+17 u^2 ((y-2) y+2)\Big)+(t+u) \Big(t^2 ((y-6) y+6)
        +t u (y (5 y-26)+26)+3 u^2 ((y-2) y+2)\Big)\Big)\nonumber\\
        &+Q^4 \Big(s^2 \Big(t^2 (2-y (3 y+2))+2 t u (y (11 y-30)+30)+30 u^2 ((y-2) y
        +2)\Big)+2 s^3 ((y-2) y+2) (t+7 u)\nonumber\\
        &+s^4 ((y-2) y+2)+2 s \Big(2 t^2 u (y (3 y-14)+14)+t^3 ((y-6) y+6)+2 t u^2 (y (7 y-24)+24)\nonumber\\
        &+7 u^3 ((y-2) y+2)\Big)+u (t+u)^2 (2 t ((y-6) y+6)+u ((y-2) y+2))\Big)+2 Q^2 s \Big(-s^2 ((y-2) y+2)\nonumber\\
        & \Big(2 t^2-2 t u-5 u^2\Big)+s^3 u ((y-2) y+2)+s u (t+u) \Big(2 t (y-2)^2+5 u ((y-2) y+2)\Big)+u \Big(2 t^2 u (y-2)^2+t^3 \nonumber\\
        &((y-6) y+6)+t u^2 (y (3 y-10)+10)+u^3 ((y-2) y+2)\Big)\Big)+4 Q^{10} \Big(3 s ((y-2) y+2)+2 t (y-2)^2\nonumber\\
        &+3 u ((y-2) y
        +2)\Big)+4 Q^{12} ((y-2) y+2)+s^2 ((y-2) y+2) (s+u) (t+u) (s (u-t)+u (t+u))\Big\}
    \end{align}   

\subsection{Quark channel}
\begin{align}
    \mathcal{A}_0^{^3S_1 [1]}=0
 \end{align}   
\begin{align}
    \mathcal{A}_0^{^3S_1 [8]}&=-\frac{4 Q^2}{3 M_\psi^3 s^2 u^2 y^2 \left(Q^2+s\right)^2} \Bigg\{Q^4 s (s^2 (t (6 y^2-28 y+28)+3 u ((y-2) y+2))+3 s^3 ((y-2) y+2)+s (4 t^2 ((y-6) y\nonumber\\
        &+6)+2 t u ((y-2) y+2)+3 u^2 ((y-2) y+2))+u ((y-2) y+2) (2 t^2+4 t u+3 u^2))+Q^2 s^2 (s^2 (t (y (3 y-14)\nonumber\\
        &+14)+3 u ((y-2) y+2))+s^3 ((y-2) y+2)+s (4 t^2 ((y-6) y+6)+4 t u ((y-2) y+2)+u^2 ((y-2) y+2))\nonumber\\
        &+4 t^2 u ((y-2) y+2)+2 t^3 ((y-6) y+6)+5 t u^2 ((y-2) y+2)+3 u^3 ((y-2) y+2))+Q^6 (s^2 (t (y (3 y-14)+14)\nonumber\\
        &+u ((y-2) y+2))+3 s^3 ((y-2) y+2)+3 s u^2 ((y-2) y+2)+u^2 ((y-2) y+2) (t+u))+Q^8 ((y-2) y+2)\nonumber\\
        & (s^2+u^2)+s^3 u ((y-2) y+2) (s^2+2 s t+2 t^2+2 t u+u^2)\Bigg\}
\end{align}

\begin{align}
        \mathcal{A}_0^{^1S_0 [8]}&=-\frac{8 Q^2\left(Q^2+s+t+u\right)}{M_\psi^3 t y^2 \left(Q^2+s\right)^2 \left(2 Q^2+s+u\right)^2} \Bigg\{Q^4 (7 s^2 ((y-2) y+2)+s \left(t \left(6 y^2-20 y+20\right)+4 u ((y-2) y+2)\right)\nonumber\\
        &+2 t^2 ((y-6) y+6)+2 t u ((y-6) y+6)+u^2 ((y-2) y+2))+2 Q^2 s (2 s^2 ((y-2) y+2)+s ((y-2) y+2)\nonumber\\
        & (t+u)+u (t ((y-6) y+6)+u ((y-2) y+2)))+2 Q^6 (3 s ((y-2) y+2)+2 t (y-2)^2+u ((y-2) y+2))\nonumber\\
        &+2 Q^8 ((y-2) y+2)+s^2 ((y-2) y+2) (s^2+u^2)\Bigg\}
\end{align}

\begin{align}
        \mathcal{A}_0^{^3P_J [8]}&=\frac{32 Q^2}{M_\psi^3 t y^2 \left(Q^2+s\right)^2 \left(2 Q^2+s+u\right)^4} \Bigg\{8 Q^2 (y-1) (Q^6 (26 s^2+12 s (3 t+2 u)+6 (6 t^2+6 t u+u^2))+Q^4 (s^2 (29 t\nonumber\\
        &+26 u)+22 s^3+2 s (32 t^2+33 t u+7 u^2)+56 t^2 u+32 t^3+29 t u^2+2 u^3)+Q^2 (s^2 (39 t^2+37 t u+10 u^2)+s^3 (11 t\nonumber\\
        &+12 u)+10 s^4+s (66 t^2 u+36 t^3+41 t u^2+4 u^3)+t (t+u) (8 t^2+12 t u+7 u^2))+8 Q^8 (2 (s+t)+u)+4 Q^{10}\nonumber\\
        &+s (s+u) (2 (s+t) (s^2+4 t^2)+u^2 (2 s+7 t)+t u (5 s+12 t)))-((y-2) y+2) (2 Q^6 (s^2 (180 t+149 u)+101 s^3\nonumber\\
        &+s (2 t+3 u) (50 t+29 u)+60 t^2 u+24 t^3+52 t u^2+15 u^3)+Q^4 (2 s^2 (67 t^2+191 t u+94 u^2)+222 s^3 (t+u)\nonumber\\
       & +125 s^4
        +2 s (82 t^2 u+20 t^3+101 t u^2+37 u^3)+(2 t^2+2 t u+u^2) (8 t^2+12 t u+7 u^2))+2 Q^2 s (s^2 (20 t^2+81 t u\nonumber\\
         &+43 u^2)       +s^3 (37 t+43 u)+22 s^4+s u (40 t^2+63 t u+29 u^2)+u (t+u) (8 t^2+12 t u+7 u^2))+2 Q^8 (97 s^2)\nonumber\\
        & +2 s (74 t+51 u
        +(2 t+u) (26 t+29 u))+8 Q^{10} (13 s+12 t+7 u)+24 Q^{12}+s^2 (s+u) (s^2 (12 t+7 u)+7 s^3\nonumber\\
        &+s (8 t^2+16 t u+7 u^2)
        +u (8 t^2+12 t u+7 u^2)))\Bigg\}
\end{align}
\bibliographystyle{apsrev}
\bibliography{jpsi_pion_ref}

\begin{thebibliography}{83}
\expandafter\ifx\csname natexlab\endcsname\relax\def\natexlab#1{#1}\fi
\expandafter\ifx\csname bibnamefont\endcsname\relax
  \def\bibnamefont#1{#1}\fi
\expandafter\ifx\csname bibfnamefont\endcsname\relax
  \def\bibfnamefont#1{#1}\fi
\expandafter\ifx\csname citenamefont\endcsname\relax
  \def\citenamefont#1{#1}\fi
\expandafter\ifx\csname url\endcsname\relax
  \def\url#1{\texttt{#1}}\fi
\expandafter\ifx\csname urlprefix\endcsname\relax\def\urlprefix{URL }\fi
\providecommand{\bibinfo}[2]{#2}
\providecommand{\eprint}[2][]{\url{#2}}

\bibitem[{\citenamefont{Mulders and Tangerman}(1996)}]{Mulders:1995dh}
\bibinfo{author}{\bibfnamefont{P.~J.} \bibnamefont{Mulders}} \bibnamefont{and}
  \bibinfo{author}{\bibfnamefont{R.~D.} \bibnamefont{Tangerman}},
  \bibinfo{journal}{Nucl. Phys. B} \textbf{\bibinfo{volume}{461}},
  \bibinfo{pages}{197} (\bibinfo{year}{1996}), \bibinfo{note}{[Erratum:
  Nucl.Phys.B 484, 538--540 (1997)]}, \eprint{hep-ph/9510301}.

\bibitem[{\citenamefont{Boer and Mulders}(1998)}]{Boer:1997nt}
\bibinfo{author}{\bibfnamefont{D.}~\bibnamefont{Boer}} \bibnamefont{and}
  \bibinfo{author}{\bibfnamefont{P.~J.} \bibnamefont{Mulders}},
  \bibinfo{journal}{Phys. Rev. D} \textbf{\bibinfo{volume}{57}},
  \bibinfo{pages}{5780} (\bibinfo{year}{1998}), \eprint{hep-ph/9711485}.

\bibitem[{\citenamefont{Bacchetta et~al.}(2007)\citenamefont{Bacchetta, Diehl,
  Goeke, Metz, Mulders, and Schlegel}}]{Bacchetta:2006tn}
\bibinfo{author}{\bibfnamefont{A.}~\bibnamefont{Bacchetta}},
  \bibinfo{author}{\bibfnamefont{M.}~\bibnamefont{Diehl}},
  \bibinfo{author}{\bibfnamefont{K.}~\bibnamefont{Goeke}},
  \bibinfo{author}{\bibfnamefont{A.}~\bibnamefont{Metz}},
  \bibinfo{author}{\bibfnamefont{P.~J.} \bibnamefont{Mulders}},
  \bibnamefont{and} \bibinfo{author}{\bibfnamefont{M.}~\bibnamefont{Schlegel}},
  \bibinfo{journal}{JHEP} \textbf{\bibinfo{volume}{02}}, \bibinfo{pages}{093}
  (\bibinfo{year}{2007}), \eprint{hep-ph/0611265}.

\bibitem[{\citenamefont{Tangerm~an and Mulders}(1995)}]{Tangerman:1994eh}
\bibinfo{author}{\bibfnamefont{R.~D.} \bibnamefont{Tangerm~an}}
  \bibnamefont{and} \bibinfo{author}{\bibfnamefont{P.~J.}
  \bibnamefont{Mulders}}, \bibinfo{journal}{Phys. Rev. D}
  \textbf{\bibinfo{volume}{51}}, \bibinfo{pages}{3357} (\bibinfo{year}{1995}),
  \eprint{hep-ph/9403227}.

\bibitem[{\citenamefont{Arnold et~al.}(2009)\citenamefont{Arnold, Metz, and
  Schlegel}}]{Arnold:2008kf}
\bibinfo{author}{\bibfnamefont{S.}~\bibnamefont{Arnold}},
  \bibinfo{author}{\bibfnamefont{A.}~\bibnamefont{Metz}}, \bibnamefont{and}
  \bibinfo{author}{\bibfnamefont{M.}~\bibnamefont{Schlegel}},
  \bibinfo{journal}{Phys. Rev. D} \textbf{\bibinfo{volume}{79}},
  \bibinfo{pages}{034005} (\bibinfo{year}{2009}), \eprint{0809.2262}.

\bibitem[{\citenamefont{Mulders and Rodrigues}(2001)}]{Mulders:2001pj}
\bibinfo{author}{\bibfnamefont{P.~J.} \bibnamefont{Mulders}} \bibnamefont{and}
  \bibinfo{author}{\bibfnamefont{J.}~\bibnamefont{Rodrigues}},
  \bibinfo{journal}{Phys. Rev. D} \textbf{\bibinfo{volume}{63}},
  \bibinfo{pages}{094021} (\bibinfo{year}{2001}),
  \urlprefix\url{https://link.aps.org/doi/10.1103/PhysRevD.63.094021}.

\bibitem[{\citenamefont{Sivers}(1990)}]{Sivers:1989cc}
\bibinfo{author}{\bibfnamefont{D.~W.} \bibnamefont{Sivers}},
  \bibinfo{journal}{Phys. Rev. D} \textbf{\bibinfo{volume}{41}},
  \bibinfo{pages}{83} (\bibinfo{year}{1990}).

\bibitem[{\citenamefont{Sivers}(1991)}]{Sivers:1990fh}
\bibinfo{author}{\bibfnamefont{D.~W.} \bibnamefont{Sivers}},
  \bibinfo{journal}{Phys. Rev. D} \textbf{\bibinfo{volume}{43}},
  \bibinfo{pages}{261} (\bibinfo{year}{1991}).

\bibitem[{\citenamefont{Anselmino et~al.}(2017)\citenamefont{Anselmino,
  Boglione, D'Alesio, Murgia, and Prokudin}}]{Anselmino:2016uie}
\bibinfo{author}{\bibfnamefont{M.}~\bibnamefont{Anselmino}},
  \bibinfo{author}{\bibfnamefont{M.}~\bibnamefont{Boglione}},
  \bibinfo{author}{\bibfnamefont{U.}~\bibnamefont{D'Alesio}},
  \bibinfo{author}{\bibfnamefont{F.}~\bibnamefont{Murgia}}, \bibnamefont{and}
  \bibinfo{author}{\bibfnamefont{A.}~\bibnamefont{Prokudin}},
  \bibinfo{journal}{JHEP} \textbf{\bibinfo{volume}{04}}, \bibinfo{pages}{046}
  (\bibinfo{year}{2017}), \eprint{1612.06413}.

\bibitem[{\citenamefont{Boglione et~al.}(2018)\citenamefont{Boglione, D'Alesio,
  Flore, and Gonzalez-Hernandez}}]{Boglione:2018dqd}
\bibinfo{author}{\bibfnamefont{M.}~\bibnamefont{Boglione}},
  \bibinfo{author}{\bibfnamefont{U.}~\bibnamefont{D'Alesio}},
  \bibinfo{author}{\bibfnamefont{C.}~\bibnamefont{Flore}}, \bibnamefont{and}
  \bibinfo{author}{\bibfnamefont{J.~O.} \bibnamefont{Gonzalez-Hernandez}},
  \bibinfo{journal}{JHEP} \textbf{\bibinfo{volume}{07}}, \bibinfo{pages}{148}
  (\bibinfo{year}{2018}), \eprint{1806.10645}.

\bibitem[{\citenamefont{Bury et~al.}(2021)\citenamefont{Bury, Prokudin, and
  Vladimirov}}]{Bury:2021sue}
\bibinfo{author}{\bibfnamefont{M.}~\bibnamefont{Bury}},
  \bibinfo{author}{\bibfnamefont{A.}~\bibnamefont{Prokudin}}, \bibnamefont{and}
  \bibinfo{author}{\bibfnamefont{A.}~\bibnamefont{Vladimirov}},
  \bibinfo{journal}{JHEP} \textbf{\bibinfo{volume}{05}}, \bibinfo{pages}{151}
  (\bibinfo{year}{2021}), \eprint{2103.03270}.

\bibitem[{\citenamefont{Mei\ss{}ner et~al.}(2007)\citenamefont{Mei\ss{}ner,
  Metz, and Goeke}}]{Meissner:2007}
\bibinfo{author}{\bibfnamefont{S.}~\bibnamefont{Mei\ss{}ner}},
  \bibinfo{author}{\bibfnamefont{A.}~\bibnamefont{Metz}}, \bibnamefont{and}
  \bibinfo{author}{\bibfnamefont{K.}~\bibnamefont{Goeke}},
  \bibinfo{journal}{Phys. Rev. D} \textbf{\bibinfo{volume}{76}},
  \bibinfo{pages}{034002} (\bibinfo{year}{2007}),
  \urlprefix\url{https://link.aps.org/doi/10.1103/PhysRevD.76.034002}.

\bibitem[{\citenamefont{Zheng et~al.}(2018)\citenamefont{Zheng, Aschenauer,
  Lee, Xiao, and Yin}}]{ElkePhysRevD.98.034011}
\bibinfo{author}{\bibfnamefont{L.}~\bibnamefont{Zheng}},
  \bibinfo{author}{\bibfnamefont{E.~C.} \bibnamefont{Aschenauer}},
  \bibinfo{author}{\bibfnamefont{J.~H.} \bibnamefont{Lee}},
  \bibinfo{author}{\bibfnamefont{B.-W.} \bibnamefont{Xiao}}, \bibnamefont{and}
  \bibinfo{author}{\bibfnamefont{Z.-B.} \bibnamefont{Yin}},
  \bibinfo{journal}{Phys. Rev. D} \textbf{\bibinfo{volume}{98}},
  \bibinfo{pages}{034011} (\bibinfo{year}{2018}),
  \urlprefix\url{https://link.aps.org/doi/10.1103/PhysRevD.98.034011}.

\bibitem[{\citenamefont{Boer et~al.}(2015)\citenamefont{Boer, Lorc{\'e},
  Pisano, and Zhou}}]{Boer2015Cristian}
\bibinfo{author}{\bibfnamefont{D.}~\bibnamefont{Boer}},
  \bibinfo{author}{\bibfnamefont{C.}~\bibnamefont{Lorc{\'e}}},
  \bibinfo{author}{\bibfnamefont{C.}~\bibnamefont{Pisano}}, \bibnamefont{and}
  \bibinfo{author}{\bibfnamefont{J.}~\bibnamefont{Zhou}},
  \bibinfo{journal}{Advances in High Energy Physics}
  \textbf{\bibinfo{volume}{2015}}, \bibinfo{pages}{371396}
  (\bibinfo{year}{2015}), ISSN \bibinfo{issn}{1687-7357},
  \urlprefix\url{https://doi.org/10.1155/2015/371396}.

\bibitem[{\citenamefont{Zeng et~al.}(2022)\citenamefont{Zeng, Liu, Sun, and
  Zhao}}]{ZengPhysRevD.106.094039}
\bibinfo{author}{\bibfnamefont{C.}~\bibnamefont{Zeng}},
  \bibinfo{author}{\bibfnamefont{T.}~\bibnamefont{Liu}},
  \bibinfo{author}{\bibfnamefont{P.}~\bibnamefont{Sun}}, \bibnamefont{and}
  \bibinfo{author}{\bibfnamefont{Y.}~\bibnamefont{Zhao}},
  \bibinfo{journal}{Phys. Rev. D} \textbf{\bibinfo{volume}{106}},
  \bibinfo{pages}{094039} (\bibinfo{year}{2022}),
  \urlprefix\url{https://link.aps.org/doi/10.1103/PhysRevD.106.094039}.

\bibitem[{\citenamefont{Agrawal et~al.}(2024)\citenamefont{Agrawal, Vasim, and
  Abir}}]{agrawal2024spinflip}
\bibinfo{author}{\bibfnamefont{S.}~\bibnamefont{Agrawal}},
  \bibinfo{author}{\bibfnamefont{N.}~\bibnamefont{Vasim}}, \bibnamefont{and}
  \bibinfo{author}{\bibfnamefont{R.}~\bibnamefont{Abir}},
  \emph{\bibinfo{title}{Spin-flip gluon gtmd $f_{1,2}$ at small-$x$}}
  (\bibinfo{year}{2024}), \eprint{2312.04132}.

\bibitem[{\citenamefont{Ji et~al.}(2003)\citenamefont{Ji, Ma, and
  Yuan}}]{Ji:2002xn}
\bibinfo{author}{\bibfnamefont{X.-d.} \bibnamefont{Ji}},
  \bibinfo{author}{\bibfnamefont{J.-P.} \bibnamefont{Ma}}, \bibnamefont{and}
  \bibinfo{author}{\bibfnamefont{F.}~\bibnamefont{Yuan}},
  \bibinfo{journal}{Nucl. Phys. B} \textbf{\bibinfo{volume}{652}},
  \bibinfo{pages}{383} (\bibinfo{year}{2003}), \eprint{hep-ph/0210430}.

\bibitem[{\citenamefont{Qiu and Sterman}(1999)}]{Qiu:1998ia}
\bibinfo{author}{\bibfnamefont{J.-w.} \bibnamefont{Qiu}} \bibnamefont{and}
  \bibinfo{author}{\bibfnamefont{G.~F.} \bibnamefont{Sterman}},
  \bibinfo{journal}{Phys. Rev. D} \textbf{\bibinfo{volume}{59}},
  \bibinfo{pages}{014004} (\bibinfo{year}{1999}), \eprint{hep-ph/9806356}.

\bibitem[{\citenamefont{Qiu and Sterman}(1991)}]{Qiu:1991pp}
\bibinfo{author}{\bibfnamefont{J.-w.} \bibnamefont{Qiu}} \bibnamefont{and}
  \bibinfo{author}{\bibfnamefont{G.~F.} \bibnamefont{Sterman}},
  \bibinfo{journal}{Phys. Rev. Lett.} \textbf{\bibinfo{volume}{67}},
  \bibinfo{pages}{2264} (\bibinfo{year}{1991}).

\bibitem[{\citenamefont{Collins}(2002)}]{Collins:2002kn}
\bibinfo{author}{\bibfnamefont{J.~C.} \bibnamefont{Collins}},
  \bibinfo{journal}{Phys. Lett. B} \textbf{\bibinfo{volume}{536}},
  \bibinfo{pages}{43} (\bibinfo{year}{2002}), \eprint{hep-ph/0204004}.

\bibitem[{\citenamefont{Gamberg et~al.}(2013)\citenamefont{Gamberg, Kang, and
  Prokudin}}]{GambergPhysRevLett.110.232301}
\bibinfo{author}{\bibfnamefont{L.}~\bibnamefont{Gamberg}},
  \bibinfo{author}{\bibfnamefont{Z.-B.} \bibnamefont{Kang}}, \bibnamefont{and}
  \bibinfo{author}{\bibfnamefont{A.}~\bibnamefont{Prokudin}},
  \bibinfo{journal}{Phys. Rev. Lett.} \textbf{\bibinfo{volume}{110}},
  \bibinfo{pages}{232301} (\bibinfo{year}{2013}),
  \urlprefix\url{https://link.aps.org/doi/10.1103/PhysRevLett.110.232301}.

\bibitem[{\citenamefont{Kang et~al.}(2011)\citenamefont{Kang, Qiu, Vogelsang,
  and Yuan}}]{KangPhysRevD.83.094001}
\bibinfo{author}{\bibfnamefont{Z.-B.} \bibnamefont{Kang}},
  \bibinfo{author}{\bibfnamefont{J.-W.} \bibnamefont{Qiu}},
  \bibinfo{author}{\bibfnamefont{W.}~\bibnamefont{Vogelsang}},
  \bibnamefont{and} \bibinfo{author}{\bibfnamefont{F.}~\bibnamefont{Yuan}},
  \bibinfo{journal}{Phys. Rev. D} \textbf{\bibinfo{volume}{83}},
  \bibinfo{pages}{094001} (\bibinfo{year}{2011}),
  \urlprefix\url{https://link.aps.org/doi/10.1103/PhysRevD.83.094001}.

\bibitem[{\citenamefont{Brodsky et~al.}(2002)\citenamefont{Brodsky, Hwang, and
  Schmidt}}]{Brodsky:2002cx}
\bibinfo{author}{\bibfnamefont{S.~J.} \bibnamefont{Brodsky}},
  \bibinfo{author}{\bibfnamefont{D.~S.} \bibnamefont{Hwang}}, \bibnamefont{and}
  \bibinfo{author}{\bibfnamefont{I.}~\bibnamefont{Schmidt}},
  \bibinfo{journal}{Phys. Lett. B} \textbf{\bibinfo{volume}{530}},
  \bibinfo{pages}{99} (\bibinfo{year}{2002}), \eprint{hep-ph/0201296}.

\bibitem[{\citenamefont{Belitsky et~al.}(2003)\citenamefont{Belitsky, Ji, and
  Yuan}}]{Belitsky:2002sm}
\bibinfo{author}{\bibfnamefont{A.~V.} \bibnamefont{Belitsky}},
  \bibinfo{author}{\bibfnamefont{X.}~\bibnamefont{Ji}}, \bibnamefont{and}
  \bibinfo{author}{\bibfnamefont{F.}~\bibnamefont{Yuan}},
  \bibinfo{journal}{Nucl. Phys. B} \textbf{\bibinfo{volume}{656}},
  \bibinfo{pages}{165} (\bibinfo{year}{2003}), \eprint{hep-ph/0208038}.

\bibitem[{\citenamefont{Ji and Yuan}(2002)}]{Ji:2002aa}
\bibinfo{author}{\bibfnamefont{X.-d.} \bibnamefont{Ji}} \bibnamefont{and}
  \bibinfo{author}{\bibfnamefont{F.}~\bibnamefont{Yuan}},
  \bibinfo{journal}{Phys. Lett. B} \textbf{\bibinfo{volume}{543}},
  \bibinfo{pages}{66} (\bibinfo{year}{2002}), \eprint{hep-ph/0206057}.

\bibitem[{\citenamefont{Boer et~al.}(2003)\citenamefont{Boer, Mulders, and
  Pijlman}}]{Boer:2003cm}
\bibinfo{author}{\bibfnamefont{D.}~\bibnamefont{Boer}},
  \bibinfo{author}{\bibfnamefont{P.~J.} \bibnamefont{Mulders}},
  \bibnamefont{and} \bibinfo{author}{\bibfnamefont{F.}~\bibnamefont{Pijlman}},
  \bibinfo{journal}{Nucl. Phys. B} \textbf{\bibinfo{volume}{667}},
  \bibinfo{pages}{201} (\bibinfo{year}{2003}), \eprint{hep-ph/0303034}.

\bibitem[{\citenamefont{Airapetian et~al.}(2005)\citenamefont{Airapetian,
  Akopov, Akopov, Amarian, Andrus, Aschenauer, Augustyniak, Avakian,
  Avetissian, Avetissian et~al.}}]{Airapetian:1994}
\bibinfo{author}{\bibfnamefont{A.}~\bibnamefont{Airapetian}},
  \bibinfo{author}{\bibfnamefont{N.}~\bibnamefont{Akopov}},
  \bibinfo{author}{\bibfnamefont{Z.}~\bibnamefont{Akopov}},
  \bibinfo{author}{\bibfnamefont{M.}~\bibnamefont{Amarian}},
  \bibinfo{author}{\bibfnamefont{A.}~\bibnamefont{Andrus}},
  \bibinfo{author}{\bibfnamefont{E.~C.} \bibnamefont{Aschenauer}},
  \bibinfo{author}{\bibfnamefont{W.}~\bibnamefont{Augustyniak}},
  \bibinfo{author}{\bibfnamefont{R.}~\bibnamefont{Avakian}},
  \bibinfo{author}{\bibfnamefont{A.}~\bibnamefont{Avetissian}},
  \bibinfo{author}{\bibfnamefont{E.}~\bibnamefont{Avetissian}},
  \bibnamefont{et~al.} (\bibinfo{collaboration}{The HERMES Collaboration}),
  \bibinfo{journal}{Phys. Rev. Lett.} \textbf{\bibinfo{volume}{94}},
  \bibinfo{pages}{012002} (\bibinfo{year}{2005}),
  \urlprefix\url{https://link.aps.org/doi/10.1103/PhysRevLett.94.012002}.

\bibitem[{\citenamefont{Airapetian et~al.}(2000)}]{HERMES:1999ryv}
\bibinfo{author}{\bibfnamefont{A.}~\bibnamefont{Airapetian}}
  \bibnamefont{et~al.} (\bibinfo{collaboration}{HERMES}),
  \bibinfo{journal}{Phys. Rev. Lett.} \textbf{\bibinfo{volume}{84}},
  \bibinfo{pages}{4047} (\bibinfo{year}{2000}), \eprint{hep-ex/9910062}.

\bibitem[{\citenamefont{Alexakhin et~al.}(2005)}]{COMPASS:2005csq}
\bibinfo{author}{\bibfnamefont{V.~Y.} \bibnamefont{Alexakhin}}
  \bibnamefont{et~al.} (\bibinfo{collaboration}{COMPASS}),
  \bibinfo{journal}{Phys. Rev. Lett.} \textbf{\bibinfo{volume}{94}},
  \bibinfo{pages}{202002} (\bibinfo{year}{2005}), \eprint{hep-ex/0503002}.

\bibitem[{\citenamefont{Adolph et~al.}(2012)\citenamefont{Adolph, Alekseev,
  Alexakhin, Alexandrov, Alexeev, Amoroso, Antonov, Austregesilo, Badełek,
  Balestra et~al.}}]{ADOLPH2012383}
\bibinfo{author}{\bibfnamefont{C.}~\bibnamefont{Adolph}},
  \bibinfo{author}{\bibfnamefont{M.}~\bibnamefont{Alekseev}},
  \bibinfo{author}{\bibfnamefont{V.}~\bibnamefont{Alexakhin}},
  \bibinfo{author}{\bibfnamefont{Y.}~\bibnamefont{Alexandrov}},
  \bibinfo{author}{\bibfnamefont{G.}~\bibnamefont{Alexeev}},
  \bibinfo{author}{\bibfnamefont{A.}~\bibnamefont{Amoroso}},
  \bibinfo{author}{\bibfnamefont{A.}~\bibnamefont{Antonov}},
  \bibinfo{author}{\bibfnamefont{A.}~\bibnamefont{Austregesilo}},
  \bibinfo{author}{\bibfnamefont{B.}~\bibnamefont{Badełek}},
  \bibinfo{author}{\bibfnamefont{F.}~\bibnamefont{Balestra}},
  \bibnamefont{et~al.}, \bibinfo{journal}{Physics Letters B}
  \textbf{\bibinfo{volume}{717}}, \bibinfo{pages}{383} (\bibinfo{year}{2012}),
  ISSN \bibinfo{issn}{0370-2693},
  \urlprefix\url{https://www.sciencedirect.com/science/article/pii/S0370269312010039}.

\bibitem[{\citenamefont{D'Alesio et~al.}(2015)\citenamefont{D'Alesio, Murgia,
  and Pisano}}]{DAlesio:2015fwo}
\bibinfo{author}{\bibfnamefont{U.}~\bibnamefont{D'Alesio}},
  \bibinfo{author}{\bibfnamefont{F.}~\bibnamefont{Murgia}}, \bibnamefont{and}
  \bibinfo{author}{\bibfnamefont{C.}~\bibnamefont{Pisano}},
  \bibinfo{journal}{JHEP} \textbf{\bibinfo{volume}{09}}, \bibinfo{pages}{119}
  (\bibinfo{year}{2015}), \eprint{1506.03078}.

\bibitem[{\citenamefont{D'Alesio
  et~al.}(2019{\natexlab{a}})\citenamefont{D'Alesio, Flore, Murgia, Pisano, and
  Taels}}]{DAlesio:2018rnv}
\bibinfo{author}{\bibfnamefont{U.}~\bibnamefont{D'Alesio}},
  \bibinfo{author}{\bibfnamefont{C.}~\bibnamefont{Flore}},
  \bibinfo{author}{\bibfnamefont{F.}~\bibnamefont{Murgia}},
  \bibinfo{author}{\bibfnamefont{C.}~\bibnamefont{Pisano}}, \bibnamefont{and}
  \bibinfo{author}{\bibfnamefont{P.}~\bibnamefont{Taels}},
  \bibinfo{journal}{Phys. Rev. D} \textbf{\bibinfo{volume}{99}},
  \bibinfo{pages}{036013} (\bibinfo{year}{2019}{\natexlab{a}}),
  \eprint{1811.02970}.

\bibitem[{\citenamefont{Adare et~al.}(2014)}]{PHENIX:2013wle}
\bibinfo{author}{\bibfnamefont{A.}~\bibnamefont{Adare}} \bibnamefont{et~al.}
  (\bibinfo{collaboration}{PHENIX}), \bibinfo{journal}{Phys. Rev. D}
  \textbf{\bibinfo{volume}{90}}, \bibinfo{pages}{012006}
  (\bibinfo{year}{2014}), \eprint{1312.1995}.

\bibitem[{\citenamefont{Braaten et~al.}(1996)\citenamefont{Braaten, Fleming,
  and Yuan}}]{Braaten_1996}
\bibinfo{author}{\bibfnamefont{E.}~\bibnamefont{Braaten}},
  \bibinfo{author}{\bibfnamefont{S.}~\bibnamefont{Fleming}}, \bibnamefont{and}
  \bibinfo{author}{\bibfnamefont{T.~C.} \bibnamefont{Yuan}},
  \bibinfo{journal}{Annual Review of Nuclear and Particle Science}
  \textbf{\bibinfo{volume}{46}}, \bibinfo{pages}{197–235}
  (\bibinfo{year}{1996}), ISSN \bibinfo{issn}{1545-4134},
  \urlprefix\url{http://dx.doi.org/10.1146/annurev.nucl.46.1.197}.

\bibitem[{\citenamefont{Bodwin et~al.}(1995)\citenamefont{Bodwin, Braaten, and
  Lepage}}]{BodwinPhysRevD.51.1125}
\bibinfo{author}{\bibfnamefont{G.~T.} \bibnamefont{Bodwin}},
  \bibinfo{author}{\bibfnamefont{E.}~\bibnamefont{Braaten}}, \bibnamefont{and}
  \bibinfo{author}{\bibfnamefont{G.~P.} \bibnamefont{Lepage}},
  \bibinfo{journal}{Phys. Rev. D} \textbf{\bibinfo{volume}{51}},
  \bibinfo{pages}{1125} (\bibinfo{year}{1995}),
  \urlprefix\url{https://link.aps.org/doi/10.1103/PhysRevD.51.1125}.

\bibitem[{\citenamefont{Amundson et~al.}(1997)\citenamefont{Amundson, Éboli,
  Gregores, and Halzen}}]{AMUNDSON1997323}
\bibinfo{author}{\bibfnamefont{J.}~\bibnamefont{Amundson}},
  \bibinfo{author}{\bibfnamefont{O.}~\bibnamefont{Éboli}},
  \bibinfo{author}{\bibfnamefont{E.}~\bibnamefont{Gregores}}, \bibnamefont{and}
  \bibinfo{author}{\bibfnamefont{F.}~\bibnamefont{Halzen}},
  \bibinfo{journal}{Physics Letters B} \textbf{\bibinfo{volume}{390}},
  \bibinfo{pages}{323} (\bibinfo{year}{1997}), ISSN \bibinfo{issn}{0370-2693},
  \urlprefix\url{https://www.sciencedirect.com/science/article/pii/S0370269396014177}.

\bibitem[{\citenamefont{Butensch\"on and
  Kniehl}(2011)}]{KniehlPhysRevLett.106.022003}
\bibinfo{author}{\bibfnamefont{M.}~\bibnamefont{Butensch\"on}}
  \bibnamefont{and} \bibinfo{author}{\bibfnamefont{B.~A.}
  \bibnamefont{Kniehl}}, \bibinfo{journal}{Phys. Rev. Lett.}
  \textbf{\bibinfo{volume}{106}}, \bibinfo{pages}{022003}
  (\bibinfo{year}{2011}),
  \urlprefix\url{https://link.aps.org/doi/10.1103/PhysRevLett.106.022003}.

\bibitem[{\citenamefont{Godbole et~al.}(2012)\citenamefont{Godbole, Misra,
  Mukherjee, and Rawoot}}]{GRMAMARVPhysRevD.85.094013}
\bibinfo{author}{\bibfnamefont{R.~M.} \bibnamefont{Godbole}},
  \bibinfo{author}{\bibfnamefont{A.}~\bibnamefont{Misra}},
  \bibinfo{author}{\bibfnamefont{A.}~\bibnamefont{Mukherjee}},
  \bibnamefont{and} \bibinfo{author}{\bibfnamefont{V.~S.}
  \bibnamefont{Rawoot}}, \bibinfo{journal}{Phys. Rev. D}
  \textbf{\bibinfo{volume}{85}}, \bibinfo{pages}{094013}
  (\bibinfo{year}{2012}),
  \urlprefix\url{https://link.aps.org/doi/10.1103/PhysRevD.85.094013}.

\bibitem[{\citenamefont{Rajesh et~al.}(2018{\natexlab{a}})\citenamefont{Rajesh,
  Kishore, and Mukherjee}}]{RSKRMAPhysRevD.98.014007}
\bibinfo{author}{\bibfnamefont{S.}~\bibnamefont{Rajesh}},
  \bibinfo{author}{\bibfnamefont{R.}~\bibnamefont{Kishore}}, \bibnamefont{and}
  \bibinfo{author}{\bibfnamefont{A.}~\bibnamefont{Mukherjee}},
  \bibinfo{journal}{Phys. Rev. D} \textbf{\bibinfo{volume}{98}},
  \bibinfo{pages}{014007} (\bibinfo{year}{2018}{\natexlab{a}}),
  \urlprefix\url{https://link.aps.org/doi/10.1103/PhysRevD.98.014007}.

\bibitem[{\citenamefont{Mukherjee and
  Rajesh}(2017{\natexlab{a}})}]{Mukherjee2017}
\bibinfo{author}{\bibfnamefont{A.}~\bibnamefont{Mukherjee}} \bibnamefont{and}
  \bibinfo{author}{\bibfnamefont{S.}~\bibnamefont{Rajesh}},
  \bibinfo{journal}{The European Physical Journal C}
  \textbf{\bibinfo{volume}{77}}, \bibinfo{pages}{854}
  (\bibinfo{year}{2017}{\natexlab{a}}), ISSN \bibinfo{issn}{1434-6052},
  \urlprefix\url{https://doi.org/10.1140/epjc/s10052-017-5406-4}.

\bibitem[{\citenamefont{Fleming et~al.}(2020)\citenamefont{Fleming, Makris, and
  Mehen}}]{Fleming2020}
\bibinfo{author}{\bibfnamefont{S.}~\bibnamefont{Fleming}},
  \bibinfo{author}{\bibfnamefont{Y.}~\bibnamefont{Makris}}, \bibnamefont{and}
  \bibinfo{author}{\bibfnamefont{T.}~\bibnamefont{Mehen}},
  \bibinfo{journal}{Journal of High Energy Physics}
  \textbf{\bibinfo{volume}{4}}, \bibinfo{pages}{122} (\bibinfo{year}{2020}),
  ISSN \bibinfo{issn}{1029-8479},
  \urlprefix\url{https://doi.org/10.1007/JHEP04(2020)122}.

\bibitem[{\citenamefont{Echevarria}(2019)}]{Echevarria2019}
\bibinfo{author}{\bibfnamefont{M.~G.} \bibnamefont{Echevarria}},
  \bibinfo{journal}{Journal of High Energy Physics}
  \textbf{\bibinfo{volume}{10}}, \bibinfo{pages}{144} (\bibinfo{year}{2019}),
  ISSN \bibinfo{issn}{1029-8479},
  \urlprefix\url{https://doi.org/10.1007/JHEP10(2019)144}.

\bibitem[{\citenamefont{Boer et~al.}(2023)\citenamefont{Boer, Bor, Maxia,
  Pisano, and Yuan}}]{Boer2023}
\bibinfo{author}{\bibfnamefont{D.}~\bibnamefont{Boer}},
  \bibinfo{author}{\bibfnamefont{J.}~\bibnamefont{Bor}},
  \bibinfo{author}{\bibfnamefont{L.}~\bibnamefont{Maxia}},
  \bibinfo{author}{\bibfnamefont{C.}~\bibnamefont{Pisano}}, \bibnamefont{and}
  \bibinfo{author}{\bibfnamefont{F.}~\bibnamefont{Yuan}},
  \bibinfo{journal}{Journal of High Energy Physics}
  \textbf{\bibinfo{volume}{8}}, \bibinfo{pages}{105} (\bibinfo{year}{2023}),
  ISSN \bibinfo{issn}{1029-8479},
  \urlprefix\url{https://doi.org/10.1007/JHEP08(2023)105}.

\bibitem[{\citenamefont{Echevarria et~al.}(2024)\citenamefont{Echevarria,
  Romera, and Taels}}]{Echevarria:2024idp}
\bibinfo{author}{\bibfnamefont{M.~G.} \bibnamefont{Echevarria}},
  \bibinfo{author}{\bibfnamefont{S.~F.} \bibnamefont{Romera}},
  \bibnamefont{and} \bibinfo{author}{\bibfnamefont{P.}~\bibnamefont{Taels}}
  (\bibinfo{year}{2024}), \eprint{2407.04793}.

\bibitem[{\citenamefont{Chakrabarti
  et~al.}(2023{\natexlab{a}})\citenamefont{Chakrabarti, Kishore, Mukherjee, and
  Rajesh}}]{Chakrabarti:2022rjr}
\bibinfo{author}{\bibfnamefont{D.}~\bibnamefont{Chakrabarti}},
  \bibinfo{author}{\bibfnamefont{R.}~\bibnamefont{Kishore}},
  \bibinfo{author}{\bibfnamefont{A.}~\bibnamefont{Mukherjee}},
  \bibnamefont{and} \bibinfo{author}{\bibfnamefont{S.}~\bibnamefont{Rajesh}},
  \bibinfo{journal}{Phys. Rev. D} \textbf{\bibinfo{volume}{107}},
  \bibinfo{pages}{014008} (\bibinfo{year}{2023}{\natexlab{a}}),
  \eprint{2211.08709}.

\bibitem[{\citenamefont{D'Alesio
  et~al.}(2019{\natexlab{b}})\citenamefont{D'Alesio, Murgia, Pisano, and
  Taels}}]{DAlesio:2019qpk}
\bibinfo{author}{\bibfnamefont{U.}~\bibnamefont{D'Alesio}},
  \bibinfo{author}{\bibfnamefont{F.}~\bibnamefont{Murgia}},
  \bibinfo{author}{\bibfnamefont{C.}~\bibnamefont{Pisano}}, \bibnamefont{and}
  \bibinfo{author}{\bibfnamefont{P.}~\bibnamefont{Taels}},
  \bibinfo{journal}{Phys. Rev. D} \textbf{\bibinfo{volume}{100}},
  \bibinfo{pages}{094016} (\bibinfo{year}{2019}{\natexlab{b}}),
  \eprint{1908.00446}.

\bibitem[{\citenamefont{Kishore et~al.}(2022)\citenamefont{Kishore, Mukherjee,
  Pawar, and Siddiqah}}]{Kishore:2022ddb}
\bibinfo{author}{\bibfnamefont{R.}~\bibnamefont{Kishore}},
  \bibinfo{author}{\bibfnamefont{A.}~\bibnamefont{Mukherjee}},
  \bibinfo{author}{\bibfnamefont{A.}~\bibnamefont{Pawar}}, \bibnamefont{and}
  \bibinfo{author}{\bibfnamefont{M.}~\bibnamefont{Siddiqah}},
  \bibinfo{journal}{Phys. Rev. D} \textbf{\bibinfo{volume}{106}},
  \bibinfo{pages}{034009} (\bibinfo{year}{2022}), \eprint{2203.13516}.

\bibitem[{\citenamefont{Kishore et~al.}(2020)\citenamefont{Kishore, Mukherjee,
  and Rajesh}}]{Kishore:2019fzb}
\bibinfo{author}{\bibfnamefont{R.}~\bibnamefont{Kishore}},
  \bibinfo{author}{\bibfnamefont{A.}~\bibnamefont{Mukherjee}},
  \bibnamefont{and} \bibinfo{author}{\bibfnamefont{S.}~\bibnamefont{Rajesh}},
  \bibinfo{journal}{Phys. Rev. D} \textbf{\bibinfo{volume}{101}},
  \bibinfo{pages}{054003} (\bibinfo{year}{2020}), \eprint{1908.03698}.

\bibitem[{\citenamefont{Maxia and Yuan}(2024)}]{maxia2024azimuthal}
\bibinfo{author}{\bibfnamefont{L.}~\bibnamefont{Maxia}} \bibnamefont{and}
  \bibinfo{author}{\bibfnamefont{F.}~\bibnamefont{Yuan}},
  \emph{\bibinfo{title}{Azimuthal angular correlation of $j/\psi$ plus jet
  production at the eic}} (\bibinfo{year}{2024}), \eprint{2403.02097}.

\bibitem[{\citenamefont{Boer et~al.}(2011)\citenamefont{Boer, Brodsky, Mulders,
  and Pisano}}]{Boer:2011}
\bibinfo{author}{\bibfnamefont{D.}~\bibnamefont{Boer}},
  \bibinfo{author}{\bibfnamefont{S.~J.} \bibnamefont{Brodsky}},
  \bibinfo{author}{\bibfnamefont{P.~J.} \bibnamefont{Mulders}},
  \bibnamefont{and} \bibinfo{author}{\bibfnamefont{C.}~\bibnamefont{Pisano}},
  \bibinfo{journal}{Phys. Rev. Lett.} \textbf{\bibinfo{volume}{106}},
  \bibinfo{pages}{132001} (\bibinfo{year}{2011}),
  \urlprefix\url{https://link.aps.org/doi/10.1103/PhysRevLett.106.132001}.

\bibitem[{\citenamefont{Pisano et~al.}(2013)\citenamefont{Pisano, Boer,
  Brodsky, Buffing, and Mulders}}]{Pisano:2013cya}
\bibinfo{author}{\bibfnamefont{C.}~\bibnamefont{Pisano}},
  \bibinfo{author}{\bibfnamefont{D.}~\bibnamefont{Boer}},
  \bibinfo{author}{\bibfnamefont{S.~J.} \bibnamefont{Brodsky}},
  \bibinfo{author}{\bibfnamefont{M.~G.~A.} \bibnamefont{Buffing}},
  \bibnamefont{and} \bibinfo{author}{\bibfnamefont{P.~J.}
  \bibnamefont{Mulders}}, \bibinfo{journal}{JHEP}
  \textbf{\bibinfo{volume}{10}}, \bibinfo{pages}{024} (\bibinfo{year}{2013}),
  \eprint{1307.3417}.

\bibitem[{\citenamefont{Efremov et~al.}(2018)\citenamefont{Efremov, Ivanov, and
  Teryaev}}]{Efremov:2017iwh}
\bibinfo{author}{\bibfnamefont{A.~V.} \bibnamefont{Efremov}},
  \bibinfo{author}{\bibfnamefont{N.~Y.} \bibnamefont{Ivanov}},
  \bibnamefont{and} \bibinfo{author}{\bibfnamefont{O.~V.}
  \bibnamefont{Teryaev}}, \bibinfo{journal}{Phys. Lett. B}
  \textbf{\bibinfo{volume}{777}}, \bibinfo{pages}{435} (\bibinfo{year}{2018}),
  \eprint{1711.05221}.

\bibitem[{\citenamefont{Kishore and Mukherjee}(2019)}]{Kishore:2018ugo}
\bibinfo{author}{\bibfnamefont{R.}~\bibnamefont{Kishore}} \bibnamefont{and}
  \bibinfo{author}{\bibfnamefont{A.}~\bibnamefont{Mukherjee}},
  \bibinfo{journal}{Phys. Rev. D} \textbf{\bibinfo{volume}{99}},
  \bibinfo{pages}{054012} (\bibinfo{year}{2019}), \eprint{1811.07495}.

\bibitem[{\citenamefont{Kishore et~al.}(2021)\citenamefont{Kishore, Mukherjee,
  and Siddiqah}}]{Kishore:2021vsm}
\bibinfo{author}{\bibfnamefont{R.}~\bibnamefont{Kishore}},
  \bibinfo{author}{\bibfnamefont{A.}~\bibnamefont{Mukherjee}},
  \bibnamefont{and} \bibinfo{author}{\bibfnamefont{M.}~\bibnamefont{Siddiqah}},
  \bibinfo{journal}{Phys. Rev. D} \textbf{\bibinfo{volume}{104}},
  \bibinfo{pages}{094015} (\bibinfo{year}{2021}), \eprint{2103.09070}.

\bibitem[{\citenamefont{D'Alesio et~al.}(2022)\citenamefont{D'Alesio, Maxia,
  Murgia, Pisano, and Rajesh}}]{DAlesio:2021yws}
\bibinfo{author}{\bibfnamefont{U.}~\bibnamefont{D'Alesio}},
  \bibinfo{author}{\bibfnamefont{L.}~\bibnamefont{Maxia}},
  \bibinfo{author}{\bibfnamefont{F.}~\bibnamefont{Murgia}},
  \bibinfo{author}{\bibfnamefont{C.}~\bibnamefont{Pisano}}, \bibnamefont{and}
  \bibinfo{author}{\bibfnamefont{S.}~\bibnamefont{Rajesh}},
  \bibinfo{journal}{JHEP} \textbf{\bibinfo{volume}{03}}, \bibinfo{pages}{037}
  (\bibinfo{year}{2022}), \eprint{2110.07529}.

\bibitem[{\citenamefont{D'Alesio et~al.}(2020)\citenamefont{D'Alesio, Maxia,
  Murgia, Pisano, and Rajesh}}]{DAlesio:2020eqo}
\bibinfo{author}{\bibfnamefont{U.}~\bibnamefont{D'Alesio}},
  \bibinfo{author}{\bibfnamefont{L.}~\bibnamefont{Maxia}},
  \bibinfo{author}{\bibfnamefont{F.}~\bibnamefont{Murgia}},
  \bibinfo{author}{\bibfnamefont{C.}~\bibnamefont{Pisano}}, \bibnamefont{and}
  \bibinfo{author}{\bibfnamefont{S.}~\bibnamefont{Rajesh}},
  \bibinfo{journal}{Phys. Rev. D} \textbf{\bibinfo{volume}{102}},
  \bibinfo{pages}{094011} (\bibinfo{year}{2020}), \eprint{2007.03353}.

\bibitem[{\citenamefont{Rajesh et~al.}(2018{\natexlab{b}})\citenamefont{Rajesh,
  Kishore, and Mukherjee}}]{Rajesh:2018qks}
\bibinfo{author}{\bibfnamefont{S.}~\bibnamefont{Rajesh}},
  \bibinfo{author}{\bibfnamefont{R.}~\bibnamefont{Kishore}}, \bibnamefont{and}
  \bibinfo{author}{\bibfnamefont{A.}~\bibnamefont{Mukherjee}},
  \bibinfo{journal}{Phys. Rev. D} \textbf{\bibinfo{volume}{98}},
  \bibinfo{pages}{014007} (\bibinfo{year}{2018}{\natexlab{b}}),
  \eprint{1802.10359}.

\bibitem[{\citenamefont{Qiu et~al.}(2011)\citenamefont{Qiu, Schlegel, and
  Vogelsang}}]{Qiu:2011}
\bibinfo{author}{\bibfnamefont{J.-W.} \bibnamefont{Qiu}},
  \bibinfo{author}{\bibfnamefont{M.}~\bibnamefont{Schlegel}}, \bibnamefont{and}
  \bibinfo{author}{\bibfnamefont{W.}~\bibnamefont{Vogelsang}},
  \bibinfo{journal}{Phys. Rev. Lett.} \textbf{\bibinfo{volume}{107}},
  \bibinfo{pages}{062001} (\bibinfo{year}{2011}),
  \urlprefix\url{https://link.aps.org/doi/10.1103/PhysRevLett.107.062001}.

\bibitem[{\citenamefont{Boer and Pisano}(2015)}]{Boer:2014lka}
\bibinfo{author}{\bibfnamefont{D.}~\bibnamefont{Boer}} \bibnamefont{and}
  \bibinfo{author}{\bibfnamefont{C.}~\bibnamefont{Pisano}},
  \bibinfo{journal}{Phys. Rev. D} \textbf{\bibinfo{volume}{91}},
  \bibinfo{pages}{074024} (\bibinfo{year}{2015}), \eprint{1412.5556}.

\bibitem[{\citenamefont{del Castillo et~al.}(2021)\citenamefont{del Castillo,
  Echevarria, Makris, and Scimemi}}]{delCastillo:2020omr}
\bibinfo{author}{\bibfnamefont{R.~F.} \bibnamefont{del Castillo}},
  \bibinfo{author}{\bibfnamefont{M.~G.} \bibnamefont{Echevarria}},
  \bibinfo{author}{\bibfnamefont{Y.}~\bibnamefont{Makris}}, \bibnamefont{and}
  \bibinfo{author}{\bibfnamefont{I.}~\bibnamefont{Scimemi}},
  \bibinfo{journal}{JHEP} \textbf{\bibinfo{volume}{01}}, \bibinfo{pages}{088}
  (\bibinfo{year}{2021}), \eprint{2008.07531}.

\bibitem[{\citenamefont{Kang et~al.}(2021)\citenamefont{Kang, Reiten, Shao, and
  Terry}}]{Kang:2020xgk}
\bibinfo{author}{\bibfnamefont{Z.-B.} \bibnamefont{Kang}},
  \bibinfo{author}{\bibfnamefont{J.}~\bibnamefont{Reiten}},
  \bibinfo{author}{\bibfnamefont{D.~Y.} \bibnamefont{Shao}}, \bibnamefont{and}
  \bibinfo{author}{\bibfnamefont{J.}~\bibnamefont{Terry}},
  \bibinfo{journal}{JHEP} \textbf{\bibinfo{volume}{05}}, \bibinfo{pages}{286}
  (\bibinfo{year}{2021}), \eprint{2012.01756}.

\bibitem[{\citenamefont{Kniehl and Kramer}(1999)}]{Kniehl1999}
\bibinfo{author}{\bibfnamefont{B.~A.} \bibnamefont{Kniehl}} \bibnamefont{and}
  \bibinfo{author}{\bibfnamefont{G.}~\bibnamefont{Kramer}},
  \bibinfo{journal}{The European Physical Journal C - Particles and Fields}
  \textbf{\bibinfo{volume}{6}}, \bibinfo{pages}{493} (\bibinfo{year}{1999}),
  ISSN \bibinfo{issn}{1434-6052},
  \urlprefix\url{https://doi.org/10.1007/s100529800921}.

\bibitem[{\citenamefont{Bacchetta
  et~al.}(2020{\natexlab{a}})\citenamefont{Bacchetta, Boer, Pisano, and
  Taels}}]{Bacchetta2020}
\bibinfo{author}{\bibfnamefont{A.}~\bibnamefont{Bacchetta}},
  \bibinfo{author}{\bibfnamefont{D.}~\bibnamefont{Boer}},
  \bibinfo{author}{\bibfnamefont{C.}~\bibnamefont{Pisano}}, \bibnamefont{and}
  \bibinfo{author}{\bibfnamefont{P.}~\bibnamefont{Taels}},
  \bibinfo{journal}{The European Physical Journal C}
  \textbf{\bibinfo{volume}{80}}, \bibinfo{pages}{72}
  (\bibinfo{year}{2020}{\natexlab{a}}), ISSN \bibinfo{issn}{1434-6052},
  \urlprefix\url{https://doi.org/10.1140/epjc/s10052-020-7620-8}.

\bibitem[{\citenamefont{Koike et~al.}(2011)\citenamefont{Koike, Tanaka, and
  Yoshida}}]{Koike:2011ns}
\bibinfo{author}{\bibfnamefont{Y.}~\bibnamefont{Koike}},
  \bibinfo{author}{\bibfnamefont{K.}~\bibnamefont{Tanaka}}, \bibnamefont{and}
  \bibinfo{author}{\bibfnamefont{S.}~\bibnamefont{Yoshida}},
  \bibinfo{journal}{Phys. Rev. D} \textbf{\bibinfo{volume}{83}},
  \bibinfo{pages}{114014} (\bibinfo{year}{2011}), \eprint{1104.0798}.

\bibitem[{\citenamefont{Banu et~al.}(2023)\citenamefont{Banu, Mukherjee, Pawar,
  and Rajesh}}]{BKMAPARSPhysRevD.108.034005}
\bibinfo{author}{\bibfnamefont{K.}~\bibnamefont{Banu}},
  \bibinfo{author}{\bibfnamefont{A.}~\bibnamefont{Mukherjee}},
  \bibinfo{author}{\bibfnamefont{A.}~\bibnamefont{Pawar}}, \bibnamefont{and}
  \bibinfo{author}{\bibfnamefont{S.}~\bibnamefont{Rajesh}},
  \bibinfo{journal}{Phys. Rev. D} \textbf{\bibinfo{volume}{108}},
  \bibinfo{pages}{034005} (\bibinfo{year}{2023}),
  \urlprefix\url{https://link.aps.org/doi/10.1103/PhysRevD.108.034005}.

\bibitem[{\citenamefont{Mukherjee and
  Rajesh}(2017{\natexlab{b}})}]{Mukherjee:2016qxa}
\bibinfo{author}{\bibfnamefont{A.}~\bibnamefont{Mukherjee}} \bibnamefont{and}
  \bibinfo{author}{\bibfnamefont{S.}~\bibnamefont{Rajesh}},
  \bibinfo{journal}{Eur. Phys. J. C} \textbf{\bibinfo{volume}{77}},
  \bibinfo{pages}{854} (\bibinfo{year}{2017}{\natexlab{b}}),
  \eprint{1609.05596}.

\bibitem[{\citenamefont{Boer et~al.}(2016)\citenamefont{Boer, Mulders, Pisano,
  and Zhou}}]{Boer:2016fqd}
\bibinfo{author}{\bibfnamefont{D.}~\bibnamefont{Boer}},
  \bibinfo{author}{\bibfnamefont{P.~J.} \bibnamefont{Mulders}},
  \bibinfo{author}{\bibfnamefont{C.}~\bibnamefont{Pisano}}, \bibnamefont{and}
  \bibinfo{author}{\bibfnamefont{J.}~\bibnamefont{Zhou}},
  \bibinfo{journal}{JHEP} \textbf{\bibinfo{volume}{08}}, \bibinfo{pages}{001}
  (\bibinfo{year}{2016}), \eprint{1605.07934}.

\bibitem[{\citenamefont{Bacchetta et~al.}(2000)\citenamefont{Bacchetta,
  Boglione, Henneman, and Mulders}}]{Bacchetta:1999kz}
\bibinfo{author}{\bibfnamefont{A.}~\bibnamefont{Bacchetta}},
  \bibinfo{author}{\bibfnamefont{M.}~\bibnamefont{Boglione}},
  \bibinfo{author}{\bibfnamefont{A.}~\bibnamefont{Henneman}}, \bibnamefont{and}
  \bibinfo{author}{\bibfnamefont{P.~J.} \bibnamefont{Mulders}},
  \bibinfo{journal}{Phys. Rev. Lett.} \textbf{\bibinfo{volume}{85}},
  \bibinfo{pages}{712} (\bibinfo{year}{2000}), \eprint{hep-ph/9912490}.

\bibitem[{\citenamefont{Kniehl and Kramer}(2006)}]{Kniehl:2006mw}
\bibinfo{author}{\bibfnamefont{B.~A.} \bibnamefont{Kniehl}} \bibnamefont{and}
  \bibinfo{author}{\bibfnamefont{G.}~\bibnamefont{Kramer}},
  \bibinfo{journal}{Phys. Rev. D} \textbf{\bibinfo{volume}{74}},
  \bibinfo{pages}{037502} (\bibinfo{year}{2006}), \eprint{hep-ph/0607306}.

\bibitem[{\citenamefont{D'Alesio et~al.}(2017)\citenamefont{D'Alesio, Murgia,
  Pisano, and Taels}}]{Alesio:2017um}
\bibinfo{author}{\bibfnamefont{U.}~\bibnamefont{D'Alesio}},
  \bibinfo{author}{\bibfnamefont{F.}~\bibnamefont{Murgia}},
  \bibinfo{author}{\bibfnamefont{C.}~\bibnamefont{Pisano}}, \bibnamefont{and}
  \bibinfo{author}{\bibfnamefont{P.}~\bibnamefont{Taels}},
  \bibinfo{journal}{Phys. Rev. D} \textbf{\bibinfo{volume}{96}},
  \bibinfo{pages}{036011} (\bibinfo{year}{2017}),
  \urlprefix\url{https://link.aps.org/doi/10.1103/PhysRevD.96.036011}.

\bibitem[{\citenamefont{Boer et~al.}(2012)\citenamefont{Boer, den Dunnen,
  Pisano, Schlegel, and Vogelsang}}]{Boer:2011kf}
\bibinfo{author}{\bibfnamefont{D.}~\bibnamefont{Boer}},
  \bibinfo{author}{\bibfnamefont{W.~J.} \bibnamefont{den Dunnen}},
  \bibinfo{author}{\bibfnamefont{C.}~\bibnamefont{Pisano}},
  \bibinfo{author}{\bibfnamefont{M.}~\bibnamefont{Schlegel}}, \bibnamefont{and}
  \bibinfo{author}{\bibfnamefont{W.}~\bibnamefont{Vogelsang}},
  \bibinfo{journal}{Phys. Rev. Lett.} \textbf{\bibinfo{volume}{108}},
  \bibinfo{pages}{032002} (\bibinfo{year}{2012}), \eprint{1109.1444}.

\bibitem[{\citenamefont{Boer and Pisano}(2012)}]{Boer:2012bt}
\bibinfo{author}{\bibfnamefont{D.}~\bibnamefont{Boer}} \bibnamefont{and}
  \bibinfo{author}{\bibfnamefont{C.}~\bibnamefont{Pisano}},
  \bibinfo{journal}{Phys. Rev. D} \textbf{\bibinfo{volume}{86}},
  \bibinfo{pages}{094007} (\bibinfo{year}{2012}), \eprint{1208.3642}.

\bibitem[{\citenamefont{Bacchetta et~al.}(2004)\citenamefont{Bacchetta,
  D'Alesio, Diehl, and Miller}}]{Bacchetta:2004jz}
\bibinfo{author}{\bibfnamefont{A.}~\bibnamefont{Bacchetta}},
  \bibinfo{author}{\bibfnamefont{U.}~\bibnamefont{D'Alesio}},
  \bibinfo{author}{\bibfnamefont{M.}~\bibnamefont{Diehl}}, \bibnamefont{and}
  \bibinfo{author}{\bibfnamefont{C.~A.} \bibnamefont{Miller}},
  \bibinfo{journal}{Phys. Rev. D} \textbf{\bibinfo{volume}{70}},
  \bibinfo{pages}{117504} (\bibinfo{year}{2004}), \eprint{hep-ph/0410050}.

\bibitem[{\citenamefont{Anselmino et~al.}(2005)\citenamefont{Anselmino,
  Boglione, D'Alesio, Kotzinian, Murgia, and Prokudin}}]{Anselmino:2005ea}
\bibinfo{author}{\bibfnamefont{M.}~\bibnamefont{Anselmino}},
  \bibinfo{author}{\bibfnamefont{M.}~\bibnamefont{Boglione}},
  \bibinfo{author}{\bibfnamefont{U.}~\bibnamefont{D'Alesio}},
  \bibinfo{author}{\bibfnamefont{A.}~\bibnamefont{Kotzinian}},
  \bibinfo{author}{\bibfnamefont{F.}~\bibnamefont{Murgia}}, \bibnamefont{and}
  \bibinfo{author}{\bibfnamefont{A.}~\bibnamefont{Prokudin}},
  \bibinfo{journal}{Phys. Rev. D} \textbf{\bibinfo{volume}{72}},
  \bibinfo{pages}{094007} (\bibinfo{year}{2005}), \bibinfo{note}{[Erratum:
  Phys.Rev.D 72, 099903 (2005)]}, \eprint{hep-ph/0507181}.

\bibitem[{\citenamefont{Boglione et~al.}(2021)\citenamefont{Boglione, D'Alesio,
  Flore, Gonzalez-Hernandez, Murgia, and Prokudin}}]{Boglione:2021aha}
\bibinfo{author}{\bibfnamefont{M.}~\bibnamefont{Boglione}},
  \bibinfo{author}{\bibfnamefont{U.}~\bibnamefont{D'Alesio}},
  \bibinfo{author}{\bibfnamefont{C.}~\bibnamefont{Flore}},
  \bibinfo{author}{\bibfnamefont{J.~O.} \bibnamefont{Gonzalez-Hernandez}},
  \bibinfo{author}{\bibfnamefont{F.}~\bibnamefont{Murgia}}, \bibnamefont{and}
  \bibinfo{author}{\bibfnamefont{A.}~\bibnamefont{Prokudin}},
  \bibinfo{journal}{Phys. Lett. B} \textbf{\bibinfo{volume}{815}},
  \bibinfo{pages}{136135} (\bibinfo{year}{2021}), \eprint{2101.03955}.

\bibitem[{\citenamefont{Adam et~al.}(2021)\citenamefont{Adam, Adamczyk, Adams,
  Adkins, Agakishiev, Aggarwal, Ahammed, Alekseev, Anderson, Aparin
  et~al.}}]{adam2021measurement}
\bibinfo{author}{\bibfnamefont{J.}~\bibnamefont{Adam}},
  \bibinfo{author}{\bibfnamefont{L.}~\bibnamefont{Adamczyk}},
  \bibinfo{author}{\bibfnamefont{J.}~\bibnamefont{Adams}},
  \bibinfo{author}{\bibfnamefont{J.}~\bibnamefont{Adkins}},
  \bibinfo{author}{\bibfnamefont{G.}~\bibnamefont{Agakishiev}},
  \bibinfo{author}{\bibfnamefont{M.}~\bibnamefont{Aggarwal}},
  \bibinfo{author}{\bibfnamefont{Z.}~\bibnamefont{Ahammed}},
  \bibinfo{author}{\bibfnamefont{I.}~\bibnamefont{Alekseev}},
  \bibinfo{author}{\bibfnamefont{D.}~\bibnamefont{Anderson}},
  \bibinfo{author}{\bibfnamefont{A.}~\bibnamefont{Aparin}},
  \bibnamefont{et~al.}, \bibinfo{journal}{Physical Review D}
  \textbf{\bibinfo{volume}{103}}, \bibinfo{pages}{092009}
  (\bibinfo{year}{2021}).

\bibitem[{\citenamefont{Bacchetta
  et~al.}(2020{\natexlab{b}})\citenamefont{Bacchetta, Celiberto, Radici, and
  Taels}}]{Bacchetta:2020vty}
\bibinfo{author}{\bibfnamefont{A.}~\bibnamefont{Bacchetta}},
  \bibinfo{author}{\bibfnamefont{F.~G.} \bibnamefont{Celiberto}},
  \bibinfo{author}{\bibfnamefont{M.}~\bibnamefont{Radici}}, \bibnamefont{and}
  \bibinfo{author}{\bibfnamefont{P.}~\bibnamefont{Taels}},
  \bibinfo{journal}{Eur. Phys. J. C} \textbf{\bibinfo{volume}{80}},
  \bibinfo{pages}{733} (\bibinfo{year}{2020}{\natexlab{b}}),
  \eprint{2005.02288}.

\bibitem[{\citenamefont{Chakrabarti
  et~al.}(2023{\natexlab{b}})\citenamefont{Chakrabarti, Choudhary, Gurjar,
  Kishore, Maji, Mondal, and Mukherjee}}]{Chakrabarti:2023djs}
\bibinfo{author}{\bibfnamefont{D.}~\bibnamefont{Chakrabarti}},
  \bibinfo{author}{\bibfnamefont{P.}~\bibnamefont{Choudhary}},
  \bibinfo{author}{\bibfnamefont{B.}~\bibnamefont{Gurjar}},
  \bibinfo{author}{\bibfnamefont{R.}~\bibnamefont{Kishore}},
  \bibinfo{author}{\bibfnamefont{T.}~\bibnamefont{Maji}},
  \bibinfo{author}{\bibfnamefont{C.}~\bibnamefont{Mondal}}, \bibnamefont{and}
  \bibinfo{author}{\bibfnamefont{A.}~\bibnamefont{Mukherjee}},
  \bibinfo{journal}{Phys. Rev. D} \textbf{\bibinfo{volume}{108}},
  \bibinfo{pages}{014009} (\bibinfo{year}{2023}{\natexlab{b}}),
  \eprint{2304.09908}.

\bibitem[{\citenamefont{Chao et~al.}(2012)\citenamefont{Chao, Ma, Shao, Wang,
  and Zhang}}]{Chao:2012iv}
\bibinfo{author}{\bibfnamefont{K.-T.} \bibnamefont{Chao}},
  \bibinfo{author}{\bibfnamefont{Y.-Q.} \bibnamefont{Ma}},
  \bibinfo{author}{\bibfnamefont{H.-S.} \bibnamefont{Shao}},
  \bibinfo{author}{\bibfnamefont{K.}~\bibnamefont{Wang}}, \bibnamefont{and}
  \bibinfo{author}{\bibfnamefont{Y.-J.} \bibnamefont{Zhang}},
  \bibinfo{journal}{Phys. Rev. Lett.} \textbf{\bibinfo{volume}{108}},
  \bibinfo{pages}{242004} (\bibinfo{year}{2012}), \eprint{1201.2675}.

\bibitem[{\citenamefont{Sharma and Vitev}(2013)}]{Sharma:2012dy}
\bibinfo{author}{\bibfnamefont{R.}~\bibnamefont{Sharma}} \bibnamefont{and}
  \bibinfo{author}{\bibfnamefont{I.}~\bibnamefont{Vitev}},
  \bibinfo{journal}{Phys. Rev. C} \textbf{\bibinfo{volume}{87}},
  \bibinfo{pages}{044905} (\bibinfo{year}{2013}), \eprint{1203.0329}.

\bibitem[{\citenamefont{Bertone et~al.}(2017)\citenamefont{Bertone, Carrazza,
  Hartland, Nocera, and Rojo}}]{Bertone:2017tyb}
\bibinfo{author}{\bibfnamefont{V.}~\bibnamefont{Bertone}},
  \bibinfo{author}{\bibfnamefont{S.}~\bibnamefont{Carrazza}},
  \bibinfo{author}{\bibfnamefont{N.~P.} \bibnamefont{Hartland}},
  \bibinfo{author}{\bibfnamefont{E.~R.} \bibnamefont{Nocera}},
  \bibnamefont{and} \bibinfo{author}{\bibfnamefont{J.}~\bibnamefont{Rojo}}
  (\bibinfo{collaboration}{NNPDF}), \bibinfo{journal}{Eur. Phys. J. C}
  \textbf{\bibinfo{volume}{77}}, \bibinfo{pages}{516} (\bibinfo{year}{2017}),
  \eprint{1706.07049}.

\bibitem[{\citenamefont{Hou et~al.}(2021)\citenamefont{Hou, Gao, Hobbs, Xie,
  Dulat, Guzzi, Huston, Nadolsky, Pumplin, Schmidt
  et~al.}}]{HouPhysRevD.103.014013}
\bibinfo{author}{\bibfnamefont{T.-J.} \bibnamefont{Hou}},
  \bibinfo{author}{\bibfnamefont{J.}~\bibnamefont{Gao}},
  \bibinfo{author}{\bibfnamefont{T.~J.} \bibnamefont{Hobbs}},
  \bibinfo{author}{\bibfnamefont{K.}~\bibnamefont{Xie}},
  \bibinfo{author}{\bibfnamefont{S.}~\bibnamefont{Dulat}},
  \bibinfo{author}{\bibfnamefont{M.}~\bibnamefont{Guzzi}},
  \bibinfo{author}{\bibfnamefont{J.}~\bibnamefont{Huston}},
  \bibinfo{author}{\bibfnamefont{P.}~\bibnamefont{Nadolsky}},
  \bibinfo{author}{\bibfnamefont{J.}~\bibnamefont{Pumplin}},
  \bibinfo{author}{\bibfnamefont{C.}~\bibnamefont{Schmidt}},
  \bibnamefont{et~al.}, \bibinfo{journal}{Phys. Rev. D}
  \textbf{\bibinfo{volume}{103}}, \bibinfo{pages}{014013}
  (\bibinfo{year}{2021}),
  \urlprefix\url{https://link.aps.org/doi/10.1103/PhysRevD.103.014013}.

\bibitem[{\citenamefont{Chakrabarti
  et~al.}(2023{\natexlab{c}})\citenamefont{Chakrabarti, Choudhary, Gurjar,
  Kishore, Maji, Mondal, and Mukherjee}}]{ChakrabartiPhysRevD.108.014009}
\bibinfo{author}{\bibfnamefont{D.}~\bibnamefont{Chakrabarti}},
  \bibinfo{author}{\bibfnamefont{P.}~\bibnamefont{Choudhary}},
  \bibinfo{author}{\bibfnamefont{B.}~\bibnamefont{Gurjar}},
  \bibinfo{author}{\bibfnamefont{R.}~\bibnamefont{Kishore}},
  \bibinfo{author}{\bibfnamefont{T.}~\bibnamefont{Maji}},
  \bibinfo{author}{\bibfnamefont{C.}~\bibnamefont{Mondal}}, \bibnamefont{and}
  \bibinfo{author}{\bibfnamefont{A.}~\bibnamefont{Mukherjee}},
  \bibinfo{journal}{Phys. Rev. D} \textbf{\bibinfo{volume}{108}},
  \bibinfo{pages}{014009} (\bibinfo{year}{2023}{\natexlab{c}}),
  \urlprefix\url{https://link.aps.org/doi/10.1103/PhysRevD.108.014009}.

\end{thebibliography}

\end{document}